%% file: paper.tex
\documentclass[preprint,12pt]{elsarticle}

\usepackage{graphicx}

\usepackage{amssymb}

\usepackage{amsmath}
\usepackage[hang,nooneline]{subfigure}
\usepackage{algorithm}
\usepackage{algorithmic}
\usepackage{url}
\usepackage{color}
\usepackage{colortbl}
\usepackage{multirow}
\usepackage{array}

\definecolor{darkgrey}{rgb}{0.8,0.8,0.8}
\definecolor{lightgray}{rgb}{0.95,0.95,0.95}
\definecolor{lightblue}{rgb}{0.0,0.56,0.78}
\definecolor{darkblue}{rgb}{0.18,0.35,0.48}

\newcommand{\Walberla}{\textsc{waLBerla}}
\DeclareMathAlphabet{\mathpzc}{OT1}{pzc}{m}{it}
\newcommand{\pe}{$\mathpzc{pe}$}

\journal{Journal of Computational Science}

\begin{document}

\begin{frontmatter}



\title{All good things come in threes - Three beads learn to swim with lattice Boltzmann and a rigid body solver}


\author[LSS,EAM]{Kristina Pickl\corref{cor1}}
\ead{kristina.pickl@informatik.uni-erlangen.de}
\cortext[cor1]{Corresponding author}
\author[LSS]{Jan G{\"o}tz}
\author[ZISC]{Klaus Iglberger}
\author[EAM,ThPhy]{Jayant Pande}
\author[EAM,ZISC,ThPhy]{\\Klaus Mecke}
\author[EAM,ZISC,ThPhy]{Ana-Sun$\check{\text{c}}$ana Smith}
\author[LSS,EAM,ZISC]{Ulrich R{\"u}de}

\address[LSS]{Lehrstuhl f\"{u}r Systemsimulation, Friedrich-Alexander Universit\"{a}t Erlangen-N\"{u}rnberg, D-91058 Erlangen, Germany}
\address[EAM]{Cluster of Excellence: Engineering of Advanced Materials, Friedrich-Alexander Universit\"{a}t Erlangen-N\"{u}rnberg,  D-91052 Erlangen, Germany}
\address[ZISC]{Zentralinstitut f\"{u}r Scientific Computing, Friedrich-Alexander Universit\"{a}t Erlangen-N\"{u}rnberg, D-91058 Erlangen, Germany}
\address[ThPhy]{Institut f\"{u}r Theoretische Physik, Friedrich-Alexander Universit\"{a}t Erlangen-N\"{u}rnberg,  D-91058 Erlangen, Germany}

\begin{abstract}
\input{abstract}
\end{abstract}

\begin{keyword}
Stokes flow  \sep self-propelled microorganism \sep lattice Boltzmann method \sep numerical simulation



\end{keyword}

\end{frontmatter}


\input{introduction}

\input{numerical_method}

\input{setup}

\input{vacuum_validation}

\input{results}

\input{conclusion}

\section*{Acknowledgements}
The work was supported by the Kompetenznetz\-werk f\"ur Technisch-Wissen\-schaftliches Hoch- und H\"ochstleistungsrechnen in Bayern (\textsc{konwihr}) under project \textsc{waLBerlaMC} and the Bundesministerium f\"ur Bildung und Forschung under the \textsc{SKALB} project no. 01IH08003A. Moreover, the authors gratefully acknowledge the support of the Cluster of Excellence 'Engineering of Advanced Materials' at the University of Erlangen-Nuremberg, which is funded by the German Research Foundation (DFG) within the framework of its 'Excellence Initiative'.

\clearpage
 \appendix
\input{appendix}


\clearpage
\bibliographystyle{model1b-num-names}
\bibliography{bibfile}







\end{document}

%% file: abstract.tex
We simulate the self-propulsion of devices in a fluid in the regime of low
Reynolds numbers. Each device consists of three bodies (spheres or
capsules) connected with two damped harmonic springs. Sinusoidal driving
forces compress the springs which are resolved within a rigid body physics
engine. The latter is consistently coupled to a 3D lattice Boltzmann
framework for the fluid dynamics. In simulations of three-sphere devices,
we find that the propulsion velocity agrees well with theoretical
predictions. In simulations where some or all spheres are replaced by
capsules, we find that the asymmetry of the design strongly affects the
propelling efficiency.

%% file: introduction.tex
\section{Introduction}
\label{sec:introduction}

Engineered micro-devices, developed in such a way that they are able to move alone through a
fluid and, simultaneously, emit a signal, can be of crucial use in various fields of science.
A few engineered self-propulsive devices have been developed over the last
decade. In one approach, the device consisted of magnetically actuated superparamagnetic particles linked together by DNA double strands~\cite{Dreyfus:2005:MAS}.
In another approach,  miniature
semiconductor diodes floating in water were used. When powered by an external alternating electric field, the voltage induced between the diode electrodes changed,
which resulted in particle-localized electro-osmotic flow. In the latter case, the devices could respond and emit light, thus providing a signal that could be used as
a carrier~\cite{Chang:2005:MAS}.
A third example is a swimmer based on the creation of a traveling wave along a piezoelectric layered beam divided into several
segments with a voltage with the same frequency but different phases and amplitudes applied to each segment~\cite{Kosa:2007:PMSM}.

Propulsion through a fluid induces the flow of the surrounding fluid, which in turn affects the propulsion of the device. Most generally
the fluid motion is described by the Navier-Stokes equation
\begin{equation}\label{eq:NavierStokes}
\frac{\partial \vec{u}}{\partial t}+(\vec{u} \cdot \nabla)\vec{u}=-\frac{1}{\rho}\nabla p +\frac{\mu}{\rho}\nabla^2\vec{u}+\vec{f}.
\end{equation}

Here $\vec{u}$ is the velocity, $\rho$ is the density, $p$ is the pressure, $\mu$ is the dynamic viscosity of the fluid, and $\vec{f}$ is the applied force in the fluid. The first term on
the left hand side describes the acceleration of the fluid, whereas the second term accounts for non-linear inertial effects that may give rise to effects such as
turbulence. On the right hand side the first term is the driving pressure gradient, while the second term takes into account the viscous dissipation.

In the case of low Reynolds numbers, the inertia terms on the left side of equation~(\ref{eq:NavierStokes}) can be dropped and one is left with 

\begin{equation}\label{eq:Stokes}
0=-\frac{1}{\rho}\nabla p +\frac{\mu}{\rho}\nabla^2\vec{u}+\vec{f},
\end{equation}
which together with the incompressibility equation $\nabla \cdot \vec{u}=0$ describes the so-called Stokes flow. The main characteristics of the Stokes flow emerge from the
domination of viscous forces. Consequently, the flow is always laminar (no turbulence or vortex shedding) and characterized by a small momentum. Since the flow is
proportional to the forces applied, linear superposition of solutions is valid. Furthermore, the Stokes flow is instantaneous, which means that the time-development is
given solely by the effects of the boundaries. Finally, there is no coasting, and the flow is time-reversible, which has important implications for swimming strategies
at low Reynolds numbers~\cite{Taylor:1951:ASMO}.

The propulsion strategy in this regime must involve a non-time-reversible motion, thus involving more than one degree of freedom~\cite{Purcell:1977:LALRN}. Such a
requirement is not limiting if a large number of degrees of freedom is available. However, if one tries to design a self-propelled device (e.g.~a swimmer) with only a
few degrees of freedom, a balance between simplicity and functionality has to be found.

A number of analytic and numerical studies have been performed to understand the behavior of a single or a couple of swimmers under
various conditions~\cite{Golestanian:2011:HSLRe,Lauga:2009:HSM}. However, most of these types of approaches are limited to simple geometries of swimmers and fail when numerous swimmers are
involved leading to a collective behavior. In addition to the effective treatment of many body interactions, the regimes in which inertia starts to play a role, as
well as the problem of motion in confined geometry or transport in turbulent flow, are inaccessible. In these circumstances extensive simulations become essential, but
difficult to perform efficiently~\cite{Pooley:2008:LBSTSMS, Yang:2010:SBSRSF}.

\begin{figure}[hbt]
\begin{center}
\includegraphics[width=6.5cm]{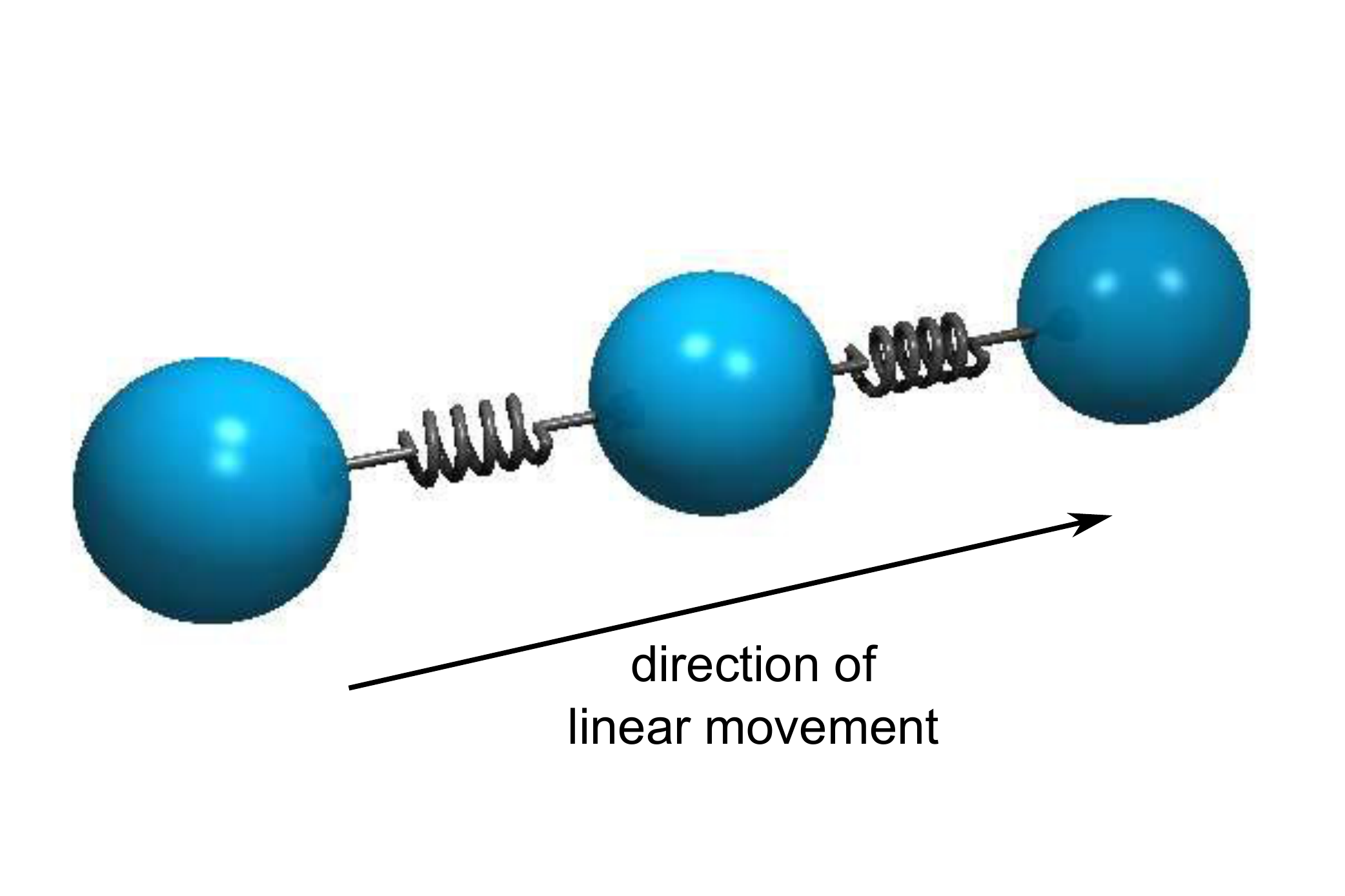}
\end{center}
\caption{A three-sphere swimming device with two translational degrees of freedom.}
\label{fig:exampleswimmer}
\end{figure}
In order to meet these challenges, we augmented our already existing massively parallel lattice Boltzmann simulation framework \Walberla{}, \textit{widely applicable Lattice
Boltzmann solver from Erlangen}~\cite{Feichtinger:2011:HPCsd}, with swimmers. The motion of the latter is included into \Walberla{} by coupling it with the \pe{} rigid multibody physics engine, a framework for 
the simulation of rigid, fully resolved bodies of arbitrary shape~\cite{iglberger:phd:10}.
This allows us to resolve both, the driven motion of a swimmer within the fluid, and the induced motion of the fluid in a consistent manner.
For the swimming device we choose the simplest 	
possible design, consisting of three rigid bodies connected with two springs~(Figure~\ref{fig:exampleswimmer}).
This design has been studied extensively in the three spheres geometry by analytical methods~\cite{Golestanian:2008:ARS}. 

This paper starts with the elucidation of the background of the numerical methods, which form the basis of our simulation, in
Section~\ref{sec:numerical}. Here, the lattice Boltzmann method (LBM) of the \Walberla{} framework, the mechanics of the \pe{} framework and, finally, the coupling
of both with special focus on the swimmers are briefly explained. In Section~\ref{sec:setup}, our new implementation approach for the integration of the swimmers is
considered in detail, including the cycling strategy. Moreover, our design variants are presented.
Afterwards, in Section~\ref{sec:validationVacuum}, a quantitative validation of the simulation results of the \pe{} engine is conducted.
In Section~\ref{sec:results}, validation criteria for the swimmer are defined, and asymmetric and symmetric designs are simulated and analyzed in detail to demonstrate the 
capability of our extension of the framework. Finally, in Section~\ref{sec:concAfw} a conclusion of the achieved results is given, and propositions for future work are made.

%% file: numerical_method.tex
\section{Numerical methods}
\label{sec:numerical}
\subsection{The lattice Boltzmann method}
\input{lbm}
\subsection{The rigid body physics engine \pe}
\input{pe}
\subsection{Coupling lattice Boltzmann to the \pe{} for swimmers}
\input{coupling}

%% file: lbm.tex
\label{sec:lbm}
The LBM can be used as an alternative to the classical Navier--Stokes (NS)
solvers to simulate fluid flows~\cite{Chen:98:LBM_For_Fluid_Flows,Succi:01:LBM_For_Fluid_Dynamics}
The fluid is modeled
as a set of imaginary, discretized particles positioned on an
equidistant grid of lattice cells.
Moreover, the particles are only allowed to move
in fixed, pre-defined directions given by the finite set of velocity vectors, resulting
from the discretization of the velocity space.
Unlike other methods, which rely on the automatic generation of a body fitted mesh during the simulation, the 
\Walberla{} framework uses the LBM on the same (block-)structured mesh, even when the objects are moving. This
eliminates the need for dynamic re-meshing, and hence offers a significant advantage with respect to performance.

In this study we use the common three-dimensional D3Q19 velocity phase discretization model
originally developed by Qian, d'Humi\`{e}res and Lallemand~\cite{Qian:1992:LBGK} with $N=19$
particle distribution functions (PDFs) $f_\alpha: \Omega \times T \rightarrow [0,1)$,
where $\Omega \subset \mathbb{R}^3$ and $T \subset \mathbb{R}^+_0$ are the spatial and the time
domain, respectively.
\begin{figure}[htb]
   \centering
   \includegraphics[width=.45\textwidth]{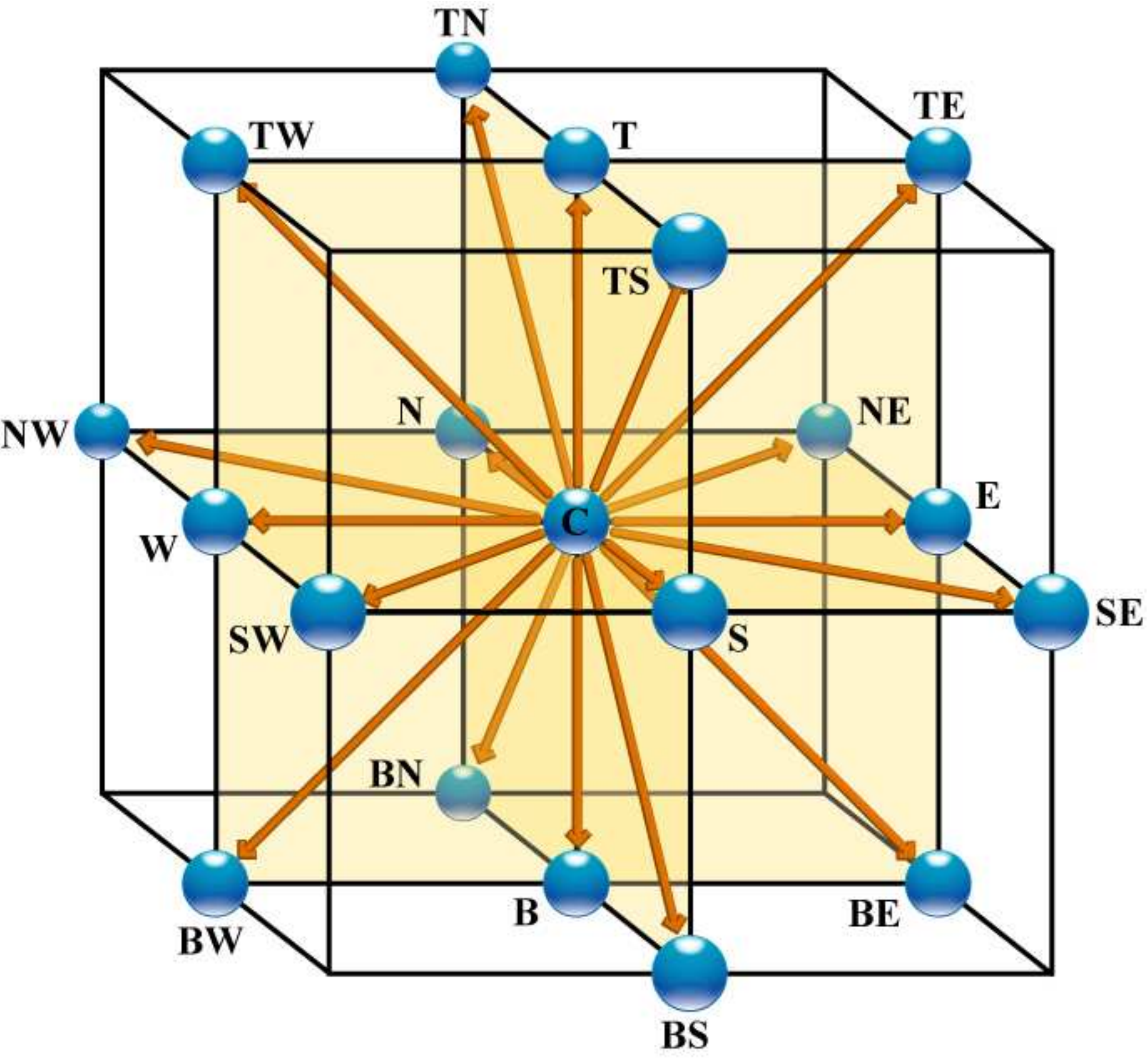}
   \caption{\label{fig:d3q19}
      The D3Q19 velocity phase discretization model.
   }
\end{figure}
Figure~\ref{fig:d3q19} illustrates the 19 directions.
The corresponding dimensionless discrete
velocity set is denoted by $\{\vec{e}_\alpha|\alpha=0,\ldots,N-1\}$. This model has been
shown to be both stable and efficient~\cite{Wei:2002:LBM3D}. For the work presented in this
paper, we adopt a lattice Boltzmann collision scheme proposed by Bhatnagar, Gross and Krook, called
\textsc{lbgk}~\cite{Bhatnagar:1954:Collision_In_Gases,Qian:1992:LBGK}
\begin{equation} \label{eq:lbgk}
   f_{\alpha} ( \vec{x}_i+\vec{e}_{\alpha} \Delta t , t+\Delta t ) = f_{\alpha}(\vec{x}_i,t) - \frac{1}{\tau} [f_{\alpha}(\vec{x}_i,t) - f_{\alpha}^{(\operatorname{eq})}(\vec{x}_i,t)]\,,
\end{equation}
where $\vec{x}_i$ is a cell in the discretized simulation domain, $t$ is the current time step, $t+\Delta t$ is the next time step, $\tau$ is the relaxation time in units of time step $\Delta t$ (which is set to be 1 here) and
$f_{\alpha}^{(\operatorname{eq})}$ represents the equilibrium distribution.

The time evolution of the distribution function, given by equation~(\ref{eq:lbgk}), is usually solved in two steps, known as the collision step and the
streaming step, respectively:
\begin{equation} \label{eq:collide}
   \tilde{f}_{\alpha} ( \vec{x}_i , t ) = f_{\alpha}(\vec{x}_i,t) - \frac{1}{\tau} [f_{\alpha}(\vec{x}_i,t) - f_{\alpha}^{(\operatorname{eq})}(\vec{x}_i,t)]\,,
\end{equation}
\begin{equation} \label{eq:streaming}
   f_{\alpha} (\vec{x}_i+\vec{e}_{\alpha} \Delta t , t+\Delta t) =  \tilde{f}_{\alpha} ( \vec{x}_i , t )\,,
\end{equation}
where $\tilde{f}_{\alpha}$ denotes the post-collision state of the distribution function. The
collision step is a local single-time relaxation towards equilibrium. The streaming step advects all PDFs except for~$f_0$ to their neighboring lattice
site depending on the velocity. Thus, for each time step only the information of the next neighbors is needed.

As a first-order no-slip boundary condition a simple bounce-back scheme is used, where
distribution functions pointing to a neighboring wall are just reflected such that both normal
and tangential velocities vanish:
\begin{equation}
   f_{\bar\alpha}( \vec{x}_f , t) = \tilde{f}_{\alpha}(\vec{x}_f,t)\label{lbm:bounce_back}\,,
\end{equation}
\noindent with $\bar\alpha$ representing the index of the opposite direction of
$\alpha$, $\vec{e}_{\bar\alpha}=-\vec{e}_\alpha$, and $\vec{x}_f$ explicitly denoting
the fluid cell. 

Due to the locality of the cell updates, the LBM can be implemented extremely efficiently
(see for instance~\cite{dimePPL03,2005:LBM:Wellein}). For the same reason, the parallelization of LBM 
is comparatively straightforward~\cite{Koerner:2005:PLBM}. It is based on a
subdomain partitioning that is realized in the~\Walberla{} framework by a patch data structure as
described in Feichtinger et~al.~\cite{Feichtinger:2011:HPCsd}. 

%% file: pe.tex
\label{sec:pe}
Simulating the dynamics of rigid bodies involves both the treatment of movement by a
discretization of Newton's equations of motion, as well as handling collisions between
rigid objects. 
The forces involved may be external, such as gravity, or between bodies, such as springs. Furthermore, a number
of velocity constraints are available through which hinges, sliders, ball joints and the like are implemented~\cite{Pickl:2009:MA-Pickl}.
The collisions between rigid bodies are either treated directly by the application of the
restitution laws in the form of a linear complementarity problem (LCP)~\cite{Anitescu:1997:LCP} or by applying
Hertzian contact mechanics (as for instance in the DEM approach)~\cite{Cundall:1979:DEM}. 

In our work, we fully resolve the geometry of the rigid bodies. In contrast to mass-point-based approaches,
this for instance enables us to easily exchange the geometries of the swimmers
in our simulations (see Subsection~\ref{subsec:alternativeDesign}).
The rigid body framework used for this is the \pe{} rigid multibody physics
engine~\cite{iglberger:phd:10}. Due to its highly flexible, modular and
massively parallel implementation, the framework allows for a direct selection of the
time discretization scheme and collision treatment. By this it can easily be adjusted to various
simulation scenarios. For instance, it has already been successfully used to simulate
large-scale granular flow scenarios~\cite{iglberger:2010:CSRD} and can be efficiently coupled
for simulations of particulate flows~\cite{Goetz:2010:SC10} (see
Subsection~\ref{sec:coupling}).

The springs of the \pe{} are modeled as damped harmonic oscillators in which the bodies are subject to the harmonic and the damping force. The first one is
\begin{equation}\label{eq:spring_i}
\vec{F}_\text{spring}^{S_n}=-k \,\Delta \vec{x}^{S_n} \hspace{0.5cm} \text{for all springs } n,
\end{equation}
where
$k$ is the force constant and $\Delta \vec{x}^{S_n} = (\vec{x}^{B_i}-\vec{x}^{B_j})- \vec{l}_0 $ is the deformation of the spring for $i\neq j$. The deformation is defined as the difference between the rest length $\vec{l}_0$ of the spring and its current length, which equals the current distance between the connected bodies $B_i$ and $B_j$, whose current positions are denoted by  $\vec{x}^{B_i}$ and $\vec{x}^{B_j}$, respectively.
In order to account for the dissipation in the spring, a damping force, proportional to the velocity $\Delta \vec{u}^{S_n}$ at which the spring is contracting is implemented as in the following
\begin{equation}\label{eq:damp_i}
\vec{F}_\text{damp}^{S_n}=- \gamma \, \Delta \vec{u}^{S_n} \hspace{0.5cm} \text{for all springs } n.
\end{equation}
Here 
$\gamma$ is the damping coefficient and 
$ \Delta \vec{u}^{S_n} = \vec{u}^{B_i}- \vec{u}^{B_j}$ for $i\neq j$. Here, $\vec{u}^{B_i}$ and $\vec{u}^{B_j}$ are the velocities of the bodies $B_i$ and $B_j$. This damping force hence models the internal friction resulting from the deformation 
of the spring.
Consequently, each of the two bodies $B_i$ and $B_j$ connected by a spring $S_n$ feels the force

\begin{align}
  \vec{F}_\text{osci}^{B_i}  &=  - k \, \Delta \vec{x}^{S_n} - \gamma \, \Delta \vec{u}^{S_n}, \label{eq:osci_i}\\  
  \vec{F}_\text{osci}^{B_j}  &=  \;\;\; k \, \Delta \vec{x}^{S_n} + \gamma \, \Delta \vec{u}^{S_n}\label{eq:osci_j},
\end{align}
respectively.

A standard way to classify a damped harmonic oscillator is by its damping ratio 
\begin{equation}\label{eq:dampingFactor}
   D = \frac {\gamma}{2 \sqrt{m \, k}}.
\end{equation}
If $D>1$ the system is overdamped and upon an excitation the system returns exponentially to equilibrium without oscillating. If $D\ll1$ the system is underdamped. It will oscillate
with a re-normalized frequency decreasing the amplitude to zero until the equilibrium state is reached. We choose the parameters
of the springs of the swimmer in such a way that the latter regime is achieved.

\begin{algorithm}[h!t]
   \caption{Rigid Body Time Step}
   \label{alg:pe}
   \begin{algorithmic}[1]
      \STATE{//Collision detection}
      \FOR{\textit{each rigid body ${B_i}$}}
        \STATE{\textit{Detect all contacts k}}
	\STATE{\textit{Add them to set of constraints c}}
      \ENDFOR
      \STATE{}
      \STATE{//Constraint resolution}
      \FOR{\textit{each constraint c}}
	\STATE{\textit{Determine acting constraint forces $\vec{F}^{B_i}_\text{c}$}}
      \ENDFOR
      \STATE{}
      \STATE{//Time integration}
      \FOR{\textit{each rigid body $B_i$}}
	\STATE{\textit{Apply forces $\vec{F}^{B_i}_\text{tot}$: update position and velocity}}
      \ENDFOR
   \end{algorithmic}
\end{algorithm}
Algorithm~\ref{alg:pe} gives an overview of the necessary steps within a single time step of
the \pe{} framework.
During each time step, collisions between rigid bodies have to be detected
and resolved appropriately in order to keep rigid bodies from interpenetrating.
Therefore, initially, all contacts $k$ between pairs of colliding rigid bodies are detected and
added to the set of constraints $c$, which already includes the potentially existing velocity constraints.
In the subsequent constraint resolution step, the acting constraining forces
$\vec{F}^{B_i}_\text{c}$, including the forces of the harmonic potentials $\vec{F}^{B_i}_\text{osci}$,
are determined. Moreover, the detected collisions are resolved depending on the
selected collision response algorithm. It is important to note that interpenetration cannot be completely prevented due to the
discrete time stepping, but small penetrations can be corrected in the collision response phase.
In the final step all rigid bodies ${B_i}$ are moved forward in time according to Newton's laws of
motion, depending on their current velocity and the total forces $\vec{F}^{B_i}_\text{tot}$ acting at the given instance of time.

In the simulations of swimmers, the size of the time step was chosen such that rigid bodies do not 
interpenetrate.
Another limitation on the size of the time step originates from the force constant of the harmonic oscillators and the frequency of the driving forces. In all cases
the time step must be small enough to enable all bodies to respond sufficiently quickly to changes in the acting forces.

%% file: coupling.tex
\label{sec:coupling}

\begin{figure*}[htbp]
   \centering
   \includegraphics[width=\textwidth]{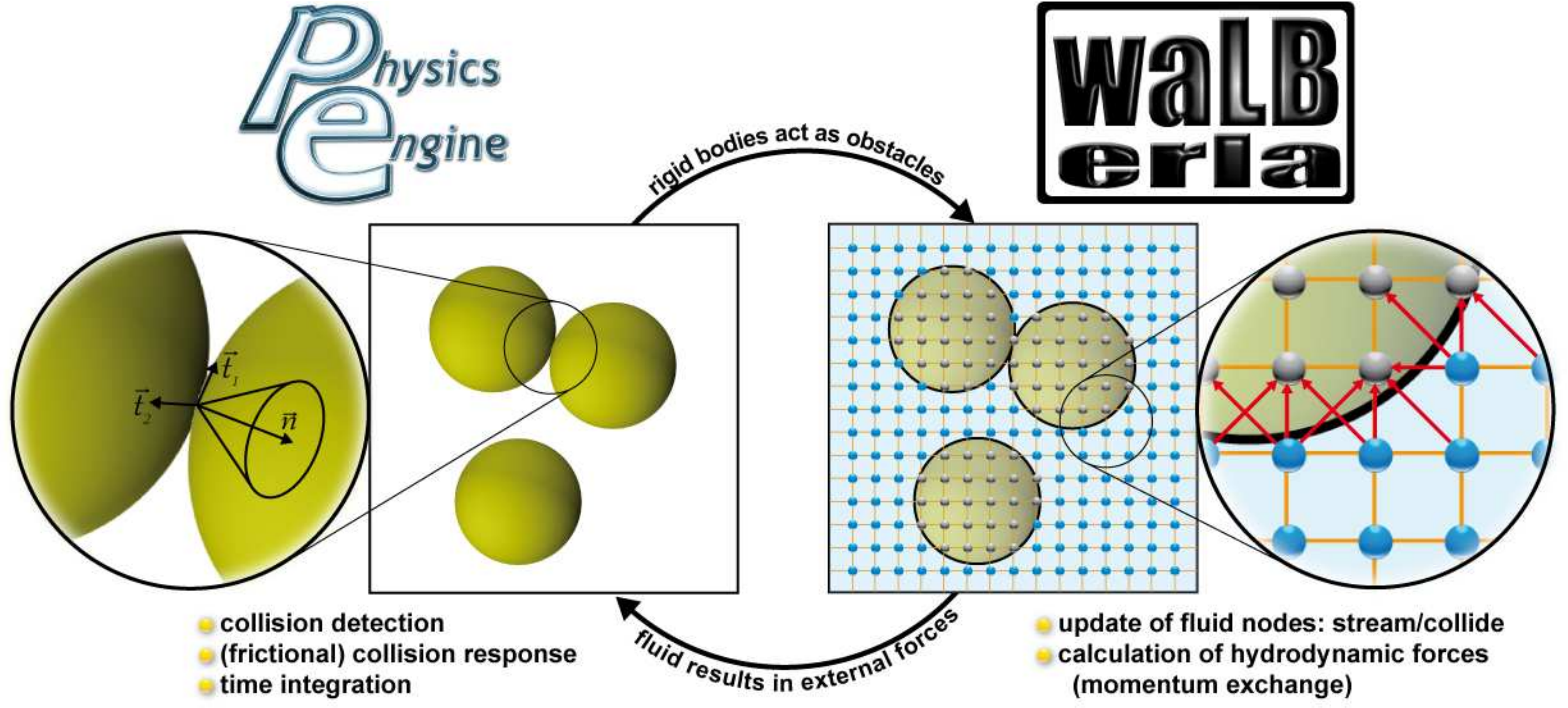}
   \caption{\label{fig:coupling}
      Illustration of the two-way coupling of~\Walberla{} fluid solver and \pe{} rigid body dynamics solver.
   }
\end{figure*}
\begin{figure*}[htbp]
   \centering
   \subfigure[Initial setup: The velocities $\vec{u}$ of the object cells $\vec{x}_b$
      are set to the velocity $\vec{u}_w(\vec{x}_b)$ of the object.~In this example the
      object only has a translational velocity component.~Fluid cells are marked
      with $\vec{x}_f$.]
   {
      \label{fig:ob_treat}
      \includegraphics[height=0.2\textheight]{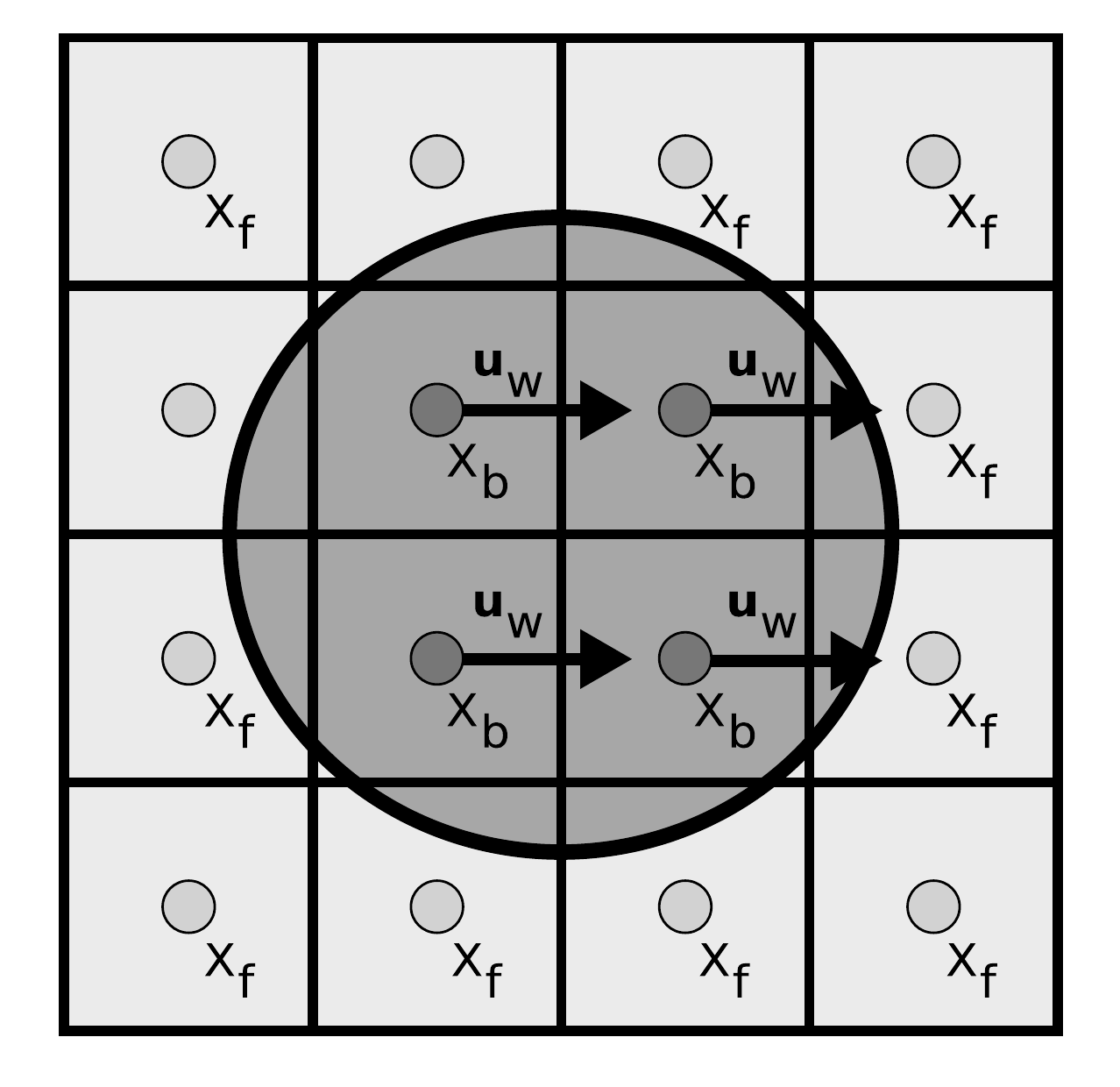}
   }
   \hspace{1cm}
   \subfigure[Updated setup: Two fluid cells have to be transformed to object cells, and for
      two object cells the PDFs have to be reconstructed.]
   {
      \label{fig:ob_reconstr}
      \includegraphics[height=0.2\textheight]{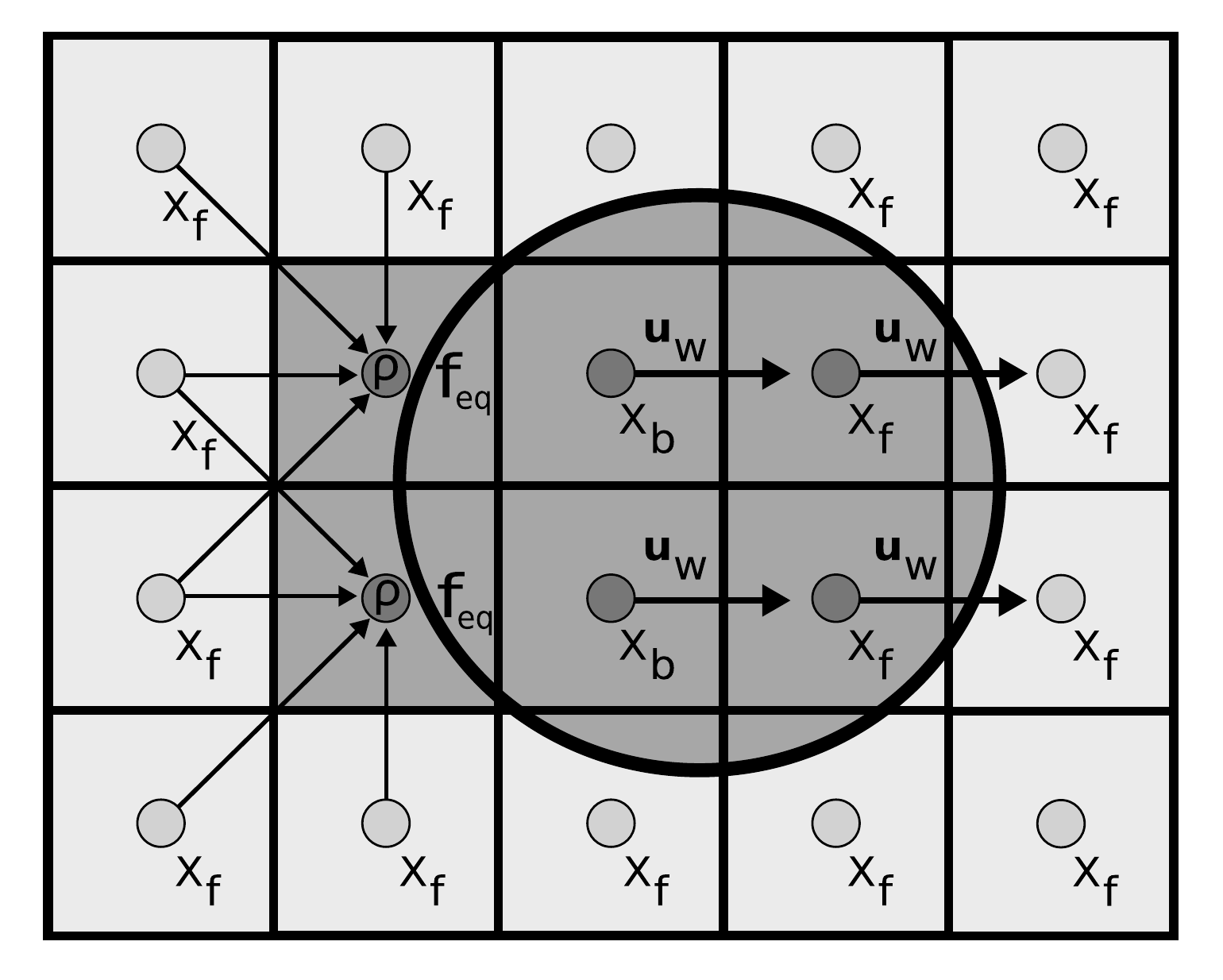}
   }
   \caption{2D mapping example \cite{Goetz:2010:ParComp}.}
\end{figure*}
The coupling between the lattice Boltzmann flow solver and the rigid body dynamics simulation
has to be two-way: rigid bodies have to be represented as (moving) boundaries in the flow
simulation, whereas the flow corresponds to hydrodynamic forces acting on the rigid bodies (Figure~\ref{fig:coupling}). We use an explicit coupling algorithm, shown in Algorithm~\ref{alg:LBM}, which we adopt for the swimmers. The additional steps are 
highlighted. Here, a driving force acts on the corresponding bodies according to a carefully defined protocol (see Subsection \ref{subsec:cycling}).

\begin{algorithm}[htb]
   \caption{Coupled LBM-PE solver for the swimmers}
   \label{alg:LBM}
   \begin{algorithmic}[1]
      \FOR{\textit{each swimmer S}}
         \STATE{\textit{Map S to lattice grid}}
      \ENDFOR
      \STATE{}
      \FOR{\textit{each lattice cell}}
         \STATE{\textit{Stream and collide}}
      \ENDFOR
      \STATE{}
      \FOR{\textit{each surface cell}}
         \STATE{\textit{Add forces $\vec{F}^{B_i}_\text{hydro}$ from fluid to rigid objects ${B_i}$}}
      \ENDFOR
      \STATE{}
      \color{lightblue}
      \FOR{\textit{each body ${B_i}$ of the swimmers}}
         \STATE{\textit{Add external forces $\vec{F}^{B_i}_\text{dri}$}}
      \ENDFOR
      \color{black}
      \STATE{}
      \STATE{//Time step in the rigid body simulation (see Algorithm~\ref{alg:pe})}
      \FOR{\textit{each rigid body $B_i$}}
	\STATE{\textit{$\hspace{0.4cm}\vec{F}^{B_i}_\text{tot} = \vec{F}^{B_i}_\text{hydro} + \vec{F}^{B_i}_\text{dri} + \vec{F}^{B_i}_\text{c}$}}
      \ENDFOR
   \end{algorithmic}
\end{algorithm}

The first step in the coupled algorithm is the mapping of all rigid bodies onto the lattice
Boltzmann grid (see Figure~\ref{fig:ob_treat} for an example). Objects are thus represented
as flag fields for the flow solver. In our implementation, each lattice node with a cell center
inside an object is treated as a moving boundary. For these cells, we apply the boundary
condition
\begin{eqnarray}
f_{\bar\alpha}(\vec{x}_f,t) &=& \tilde
f_{\alpha}(\vec{x}_f,t) + 6 w_{\alpha} \rho_w(\vec{x}_f,t)
\vec{e}_{\bar \alpha} \cdot \vec{u}_w(\vec{x}_f+\vec{e}_{\alpha},t)\,,
\end{eqnarray}
which is a variation of the standard no-slip boundary conditions for moving walls~\cite{2003:LBM:Yu}.
Here $\rho_w(\vec{x}_f,t)$ is the fluid density close to the wall, $w_{\alpha}$ is the weight in the LBM depending on the
stencil and the direction, and the velocity $\vec{u}_w(\vec{x}_f+\vec{e}_{\alpha},t)$ of
each object cell corresponds to the velocity of the object at the cell's position at time $t$. In this way, the fluid at the surface of an object
is given the same velocity as the object's surface which depends on the rotational and the translational velocities of the object.

The final part of the first step is to account for the flag changes due to the movement of the objects.
Here,
two cases can occur (see Figure~\ref{fig:ob_reconstr}): the fluid cells $\vec{x}_f$ can turn
into object cells, resulting in a conversion of the cell to a moving boundary. In the
reverse case, i.e.~a boundary cell turns into a fluid cell, the missing PDFs
have to be reconstructed. In our implementation, the missing PDFs are set to the equilibrium
distributions $f^{(eq)}_\alpha(\rho,\vec{u})$, where the macroscopic velocity $\vec{u}$ is
given by the velocity $\vec{u}_w(\vec{x}_f)$ of the object cell, and the density $\rho$ is computed as an
average of the surrounding fluid cells. Further details of this procedure have been discussed previously by us in Iglberger et
al.~\cite{2008:LBM:Iglberger}. There are more formal ways for the initialization of the missing distribution functions. One of them is described in the paper by Mei et al.~\cite{2006:Mei:Initial_conditions}. 

During the subsequent stream and collide step, the fluid flow acts through hydrodynamic forces on the
rigid objects: Fluid particles stream from their cells to neighboring cells and, in case they
enter a cell occupied by a moving object, are reversed, causing a momentum exchange between
the fluid and the particles. The total hydrodynamic force $\vec{F}^{B_i}_\text{hydro}$ on each object ${B_i}$ resulting from this momentum
exchange~\cite{2003:LBM:Yu} can be easily evaluated due to the kinetic origin of the LBM 
by
\begin{eqnarray}
\label{equ:hydroF}
\vec{F}^{B_i}_\text{hydro} = \sum_{\vec{x}_b} \sum_{\alpha=1}^{19} \vec{e}_\alpha \left[ 2\tilde
f_\alpha(\vec{x}_f,t) + 6 w_{\alpha} \rho_w(\vec{x}_f,t)
\vec{e}_{\bar \alpha} \cdot \vec{u}_w(\vec{x}_f+\vec{e}_{\alpha},t)
 \right] \frac{\Delta x}{\Delta t}\,,
\end{eqnarray}
where $\vec{x}_b$ are all obstacle cells of the object that neighbor at least one fluid cell. A comparison of different approaches for the momentum exchange is given in Lorenz et al.~\cite{Lorenz:2009}.

In the third step, the driving force $\vec{F}^{B_i}_\text{dri}$, resulting from the cycling strategy, is added to the previously calculated hydrodynamic force on each body ${B_i}$. Together with the constraint force $\vec{F}^{B_i}_\text{c}$, this results in a total force
\begin{equation}
\label{equ:totalF}
\vec{F}^{B_i}_\text{tot} = \vec{F}^{B_i}_\text{hydro} + \vec{F}^{B_i}_\text{dri} + \vec{F}^{B_i}_\text{c}
\end{equation}
on each object.

The fourth and final step of the coupling algorithm consists of determining the rigid body movements of the objects due to the influence of $\vec{F}^{B_i}_\text{tot}$ in the rigid body framework. This results in a position change of the objects, which then are mapped again to the LBM grid in the next time step.
A validation of this method is described in Binder et al.~\cite{Binder:2006:SHDAP}, and Iglberger et al.~\cite{2008:LBM:Iglberger}.

%% file: setup.tex
\section{Integration of the swimming device}
\label{sec:setup}

The modeling of a three-sphere swimmer always involves a similar physical setup. On the one hand, the connections between the objects need to be modeled. Here, stiff rods \cite{Najafi:2004:SSLR, Putz:2009:HSMM} or springs \cite{Felderhof:2006:SwimAni} are  most commonly used. On the other hand, the cycling strategy, responsible for the characteristic movement of the swimmer, needs to be defined.

\subsection{Elementary design}
\label{subsec:elementaryDesign}
\begin{figure}[htbp]
\begin{center}
\includegraphics[width=10cm]{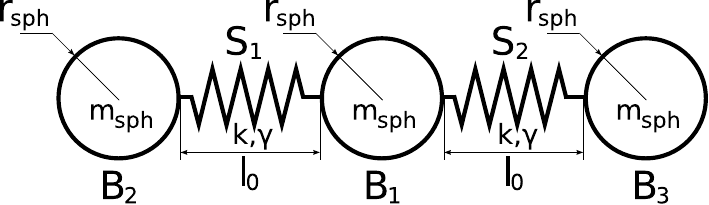}
\end{center}
\caption{A three-sphere swimming device. The three spheres have the same mass $m_\text{sph}$ and radius $r_\text{sph}$. Similarly, the two damped harmonic oscillators have identical force constant $k$, damping parameter $\gamma$, and rest length~$l_0$. }
\label{fig:swimmer}
\end{figure}
Our elementary design (Figure~\ref{fig:swimmer}), inspired by the analytical modeling in Golestanian and
Ajdari~\cite{Golestanian:2008:ARS}, and in Felderhof~\cite{Felderhof:2006:SwimAni},
consists of three spheres ($B_1, B_2, B_3$) with identical radius $r_\text{sph}$ and mass $m_\text{sph}$.
The connections between the three rigid objects of our swimmer are realized by two damped harmonic springs $S_1$ and $S_2$ with equal
force constant $k$ and damping parameter $\gamma$.
The forces are applied on the spheres along the main axis of the swimmer, in our case the z-axis, resulting in a translation of the swimmer in this direction.

In accordance with the equations (\ref{eq:osci_i}) and (\ref{eq:osci_j}), the different objects feel the force
\begin{align}
  \vec{F}_\text{osci}^{B_2} & =  - k \, \Delta \vec{x}^{S_1} - \gamma \, \Delta \vec{u}^{S_1}, \label{eq:osci_b1} \\  
  \vec{F}_\text{osci}^{B_1} & =  \;\;\,k \, \left( \Delta \vec{x}^{S_1} + \Delta \vec{x}^{S_2}\right) + \gamma \, \left( \Delta \vec{u}^{S_1}+ \Delta \vec{u}^{S_2}\right), \label{eq:osci_b2}\\
  \vec{F}_\text{osci}^{B_3} & =  - k \, \Delta \vec{x}^{S_2} - \gamma \,\Delta \vec{u}^{S_2},\label{eq:osci_b3}
\end{align}
respectively.

Since the bodies of the swimmer do not collide with each other,
the total force acting on each body $B_i$ of the swimmer, given in equation~(\ref{equ:totalF}), can be rewritten as
\begin{equation}
\label{equ:totalFSwimmer}
\vec{F}^{B_i}_\text{tot} = \vec{F}^{B_i}_\text{hydro} + \vec{F}^{B_i}_\text{dri} + \vec{F}^{B_i}_\text{osci}.
\end{equation}

\subsection{Alternative designs}
\label{subsec:alternativeDesign}
\begin{figure}[htbp]
\begin{center}
\includegraphics[width=12cm]{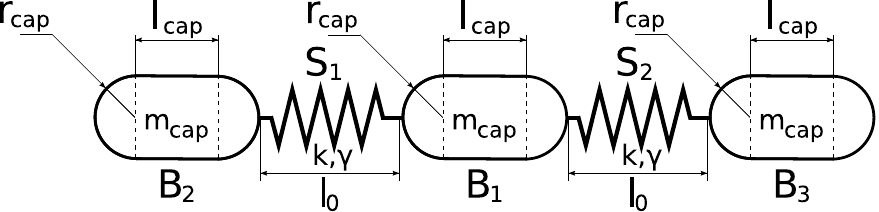}
\end{center}
\caption{A three-capsule swimming device. The three capsules have the same mass $m_\text{cap}$, length $l_\text{cap}$ and radius $r_\text{cap}$. Similarly, the two damped harmonic oscillators have identical force constant $k$, damping parameter $\gamma$, and rest length~$l_0$. }
\label{fig:swimmerCapsule}
\end{figure}

\begin{figure}[htbp]
\centering
  \subfigure[]{
    \includegraphics[height=45pt]{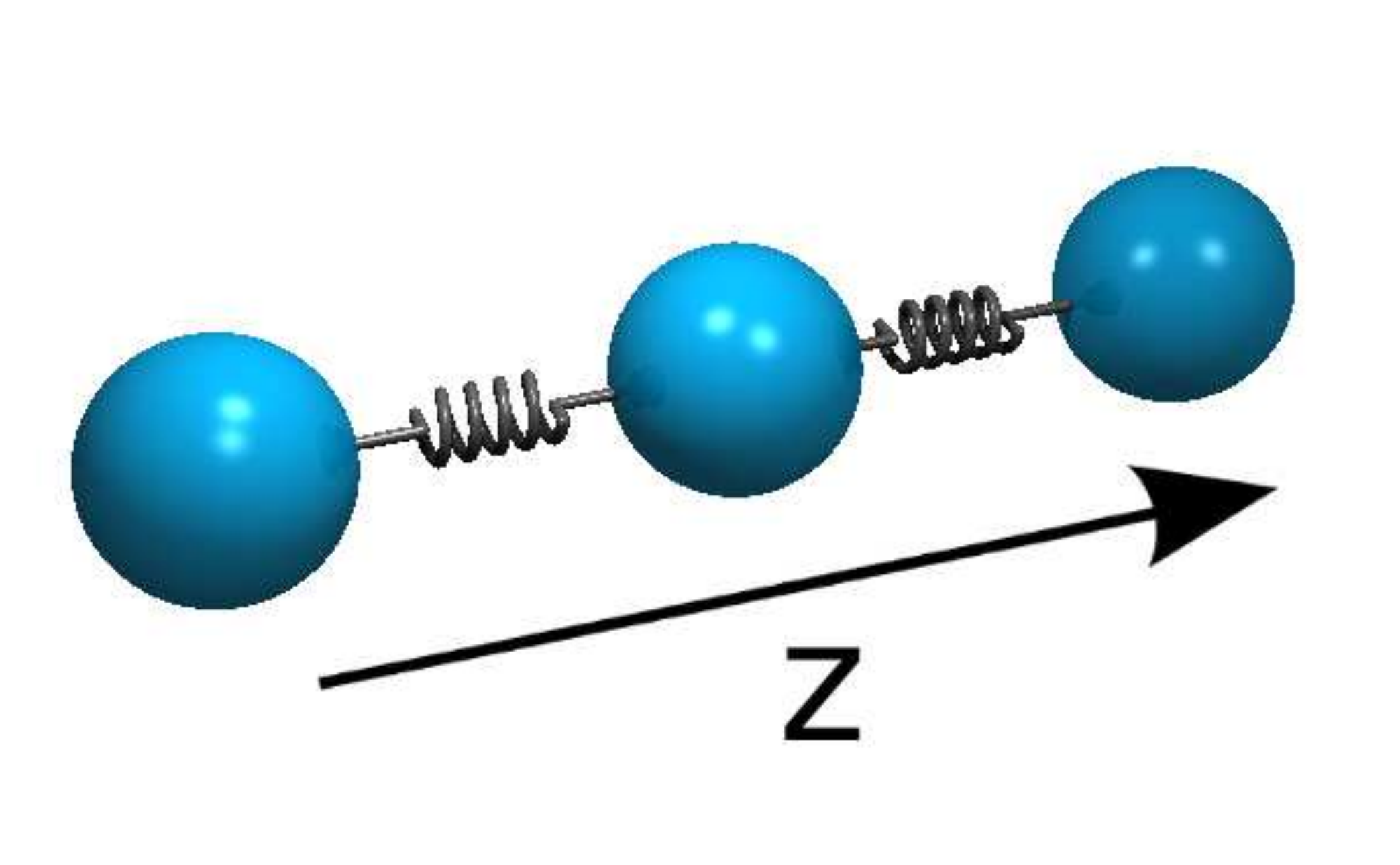}
    \label{fig:designA}
  }
  \hfill
  \subfigure[]{
    \includegraphics[height=45pt]{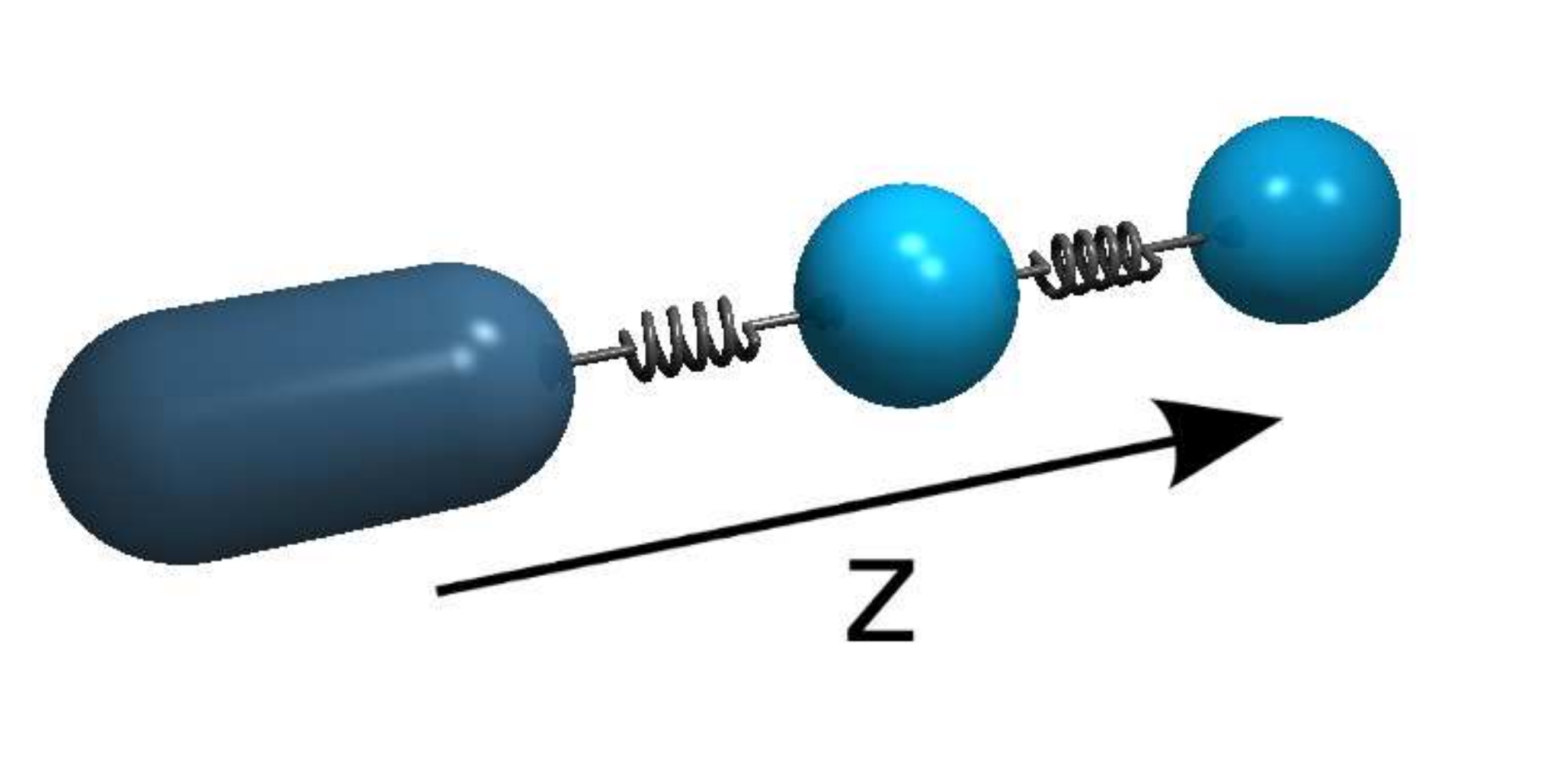}
    \label{fig:designB}
  }
  \hfill
  \subfigure[]{
    \includegraphics[height=45pt]{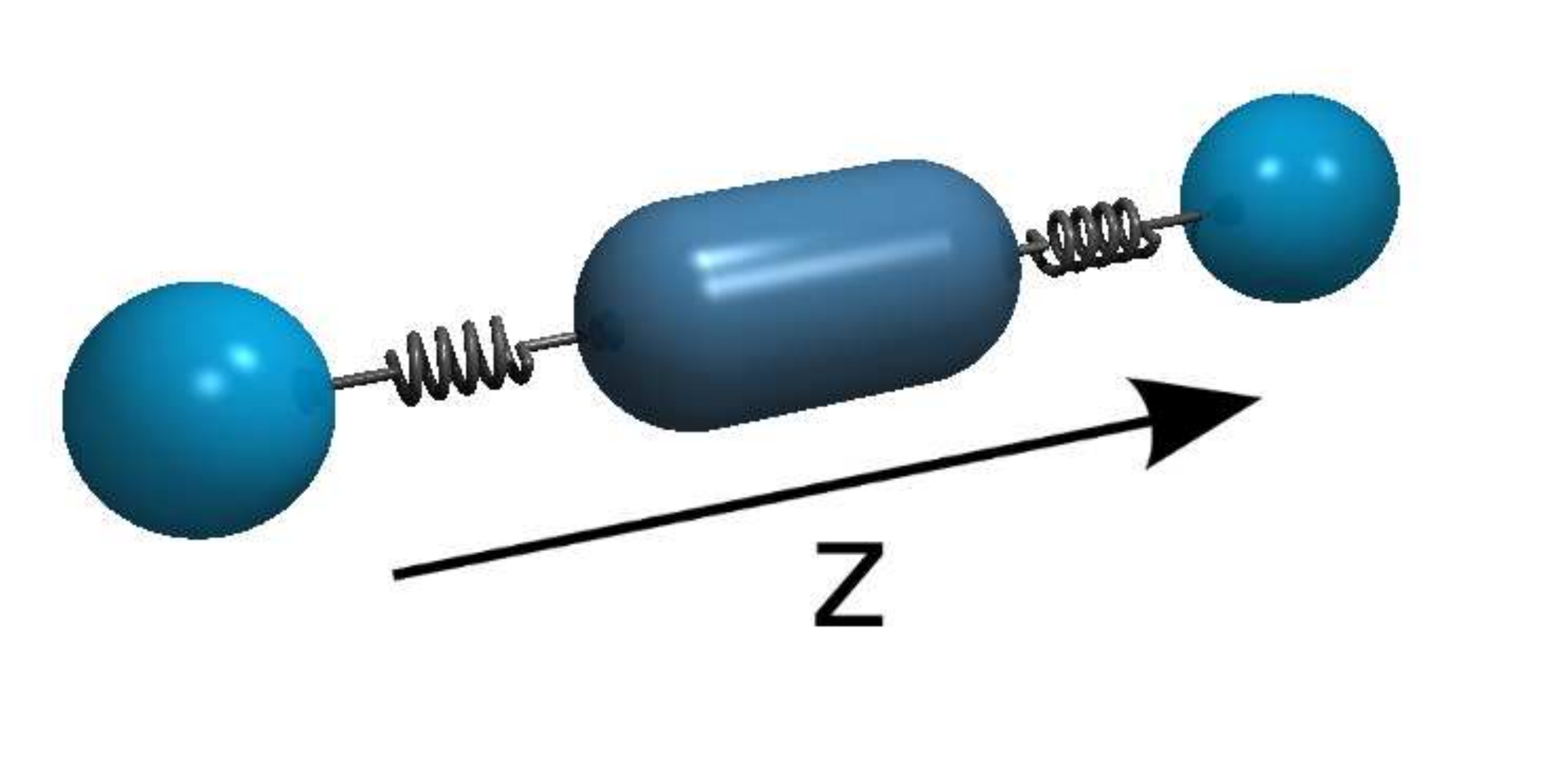}
    \label{fig:designC}
  }
  \hfill
  \subfigure[]{
    \includegraphics[height=45pt]{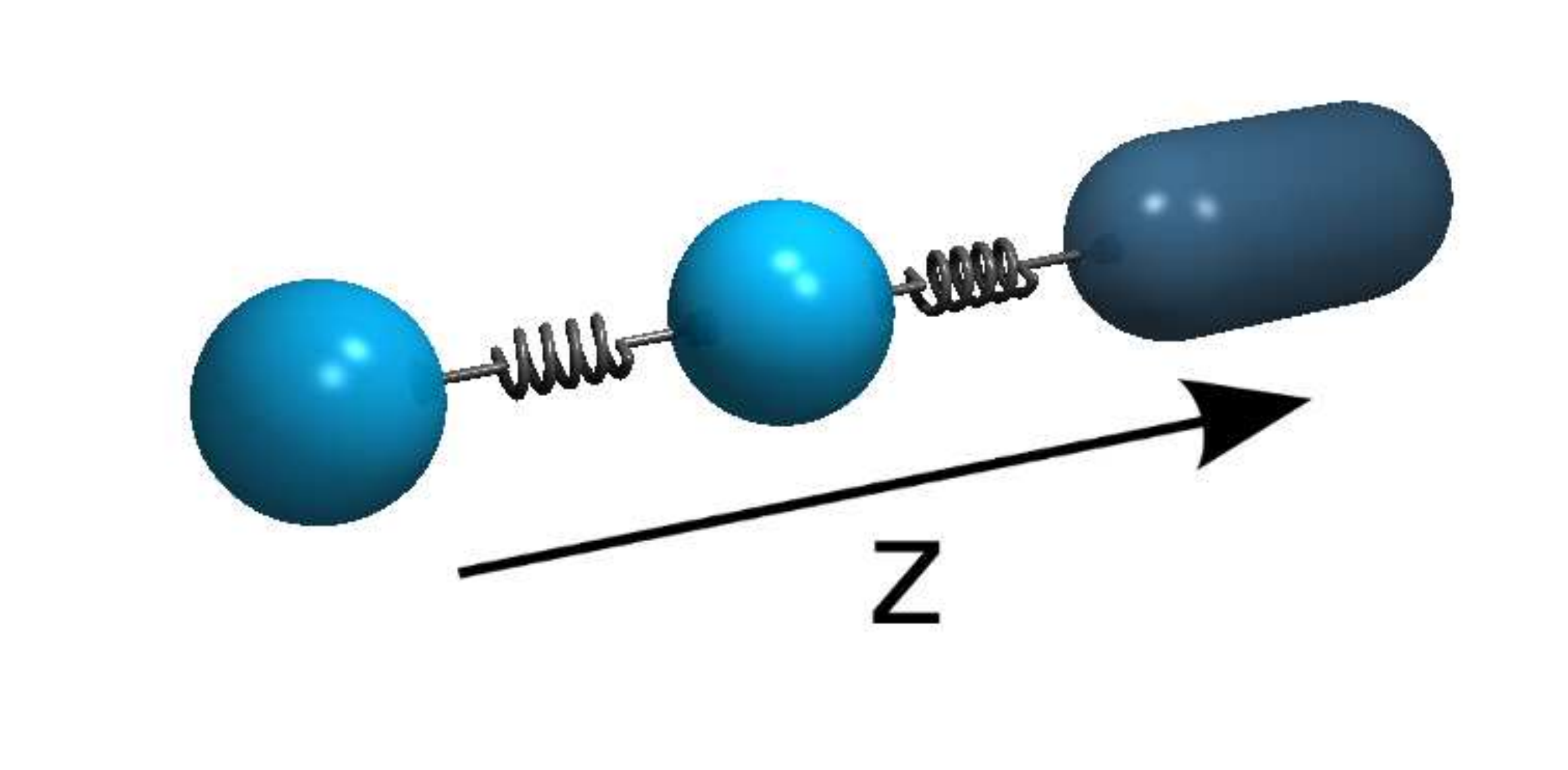}
    \label{fig:designD}
  }
  \hfill
  \subfigure[]{
    \includegraphics[height=45pt]{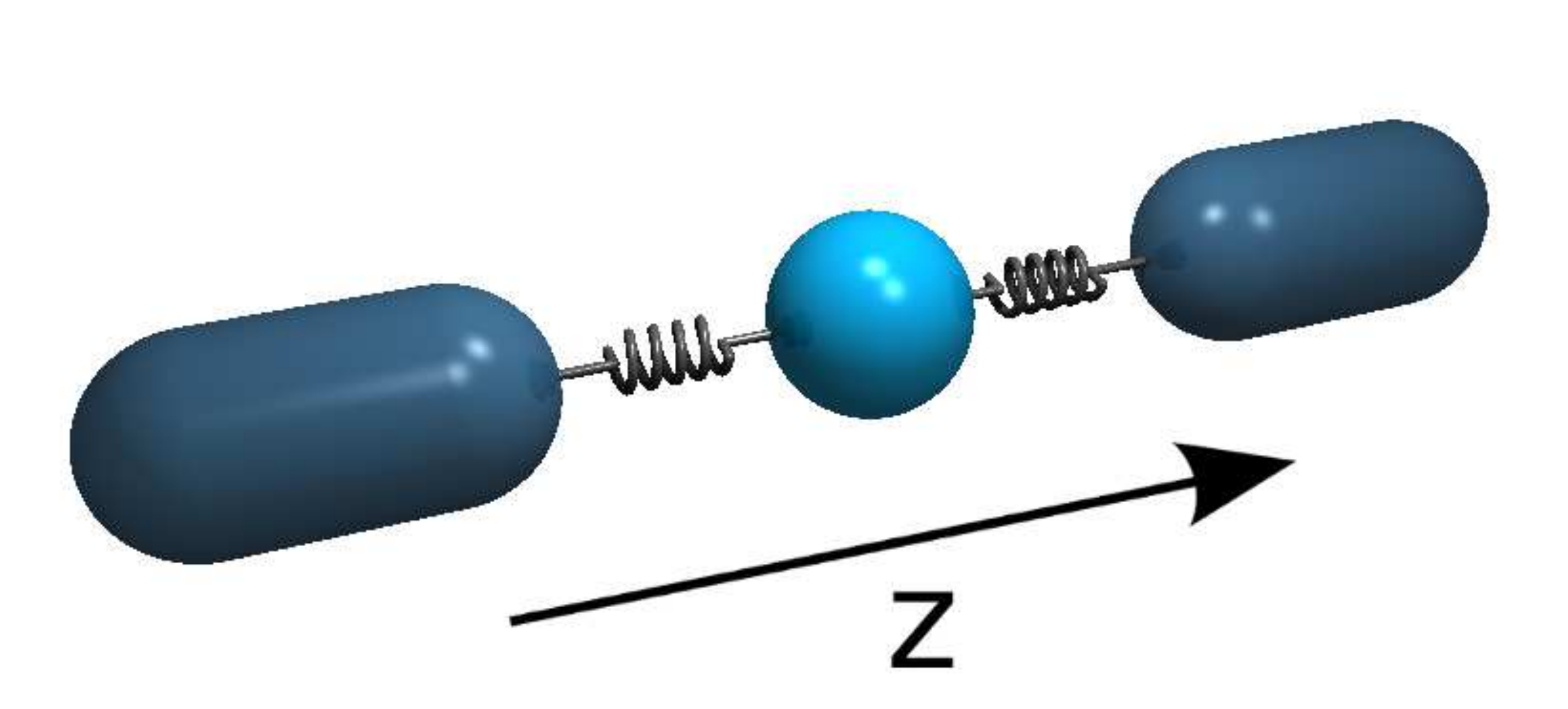}
    \label{fig:designE}
  }
  \hfill
  \subfigure[]{
    \includegraphics[height=45pt]{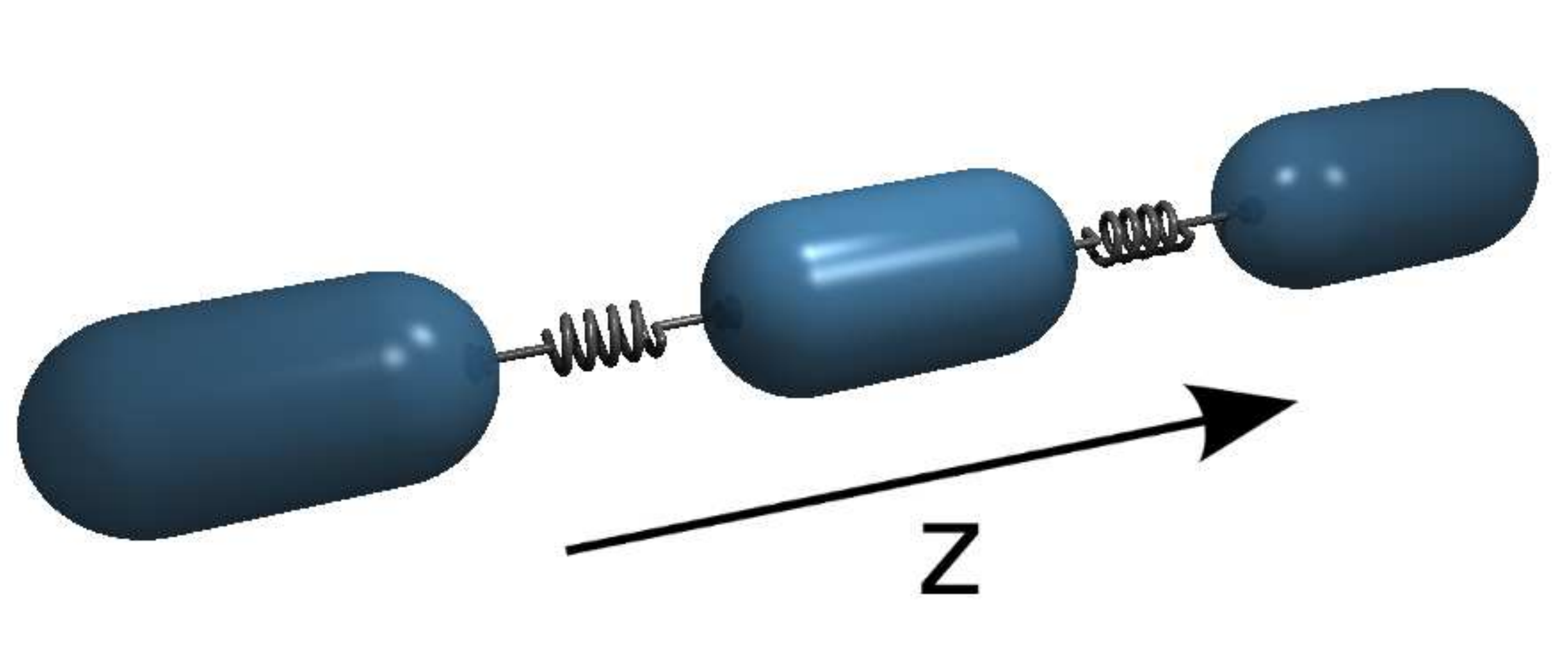}
    \label{fig:designF}
  }
  \hfill
  \subfigure[]{
    \includegraphics[height=45pt]{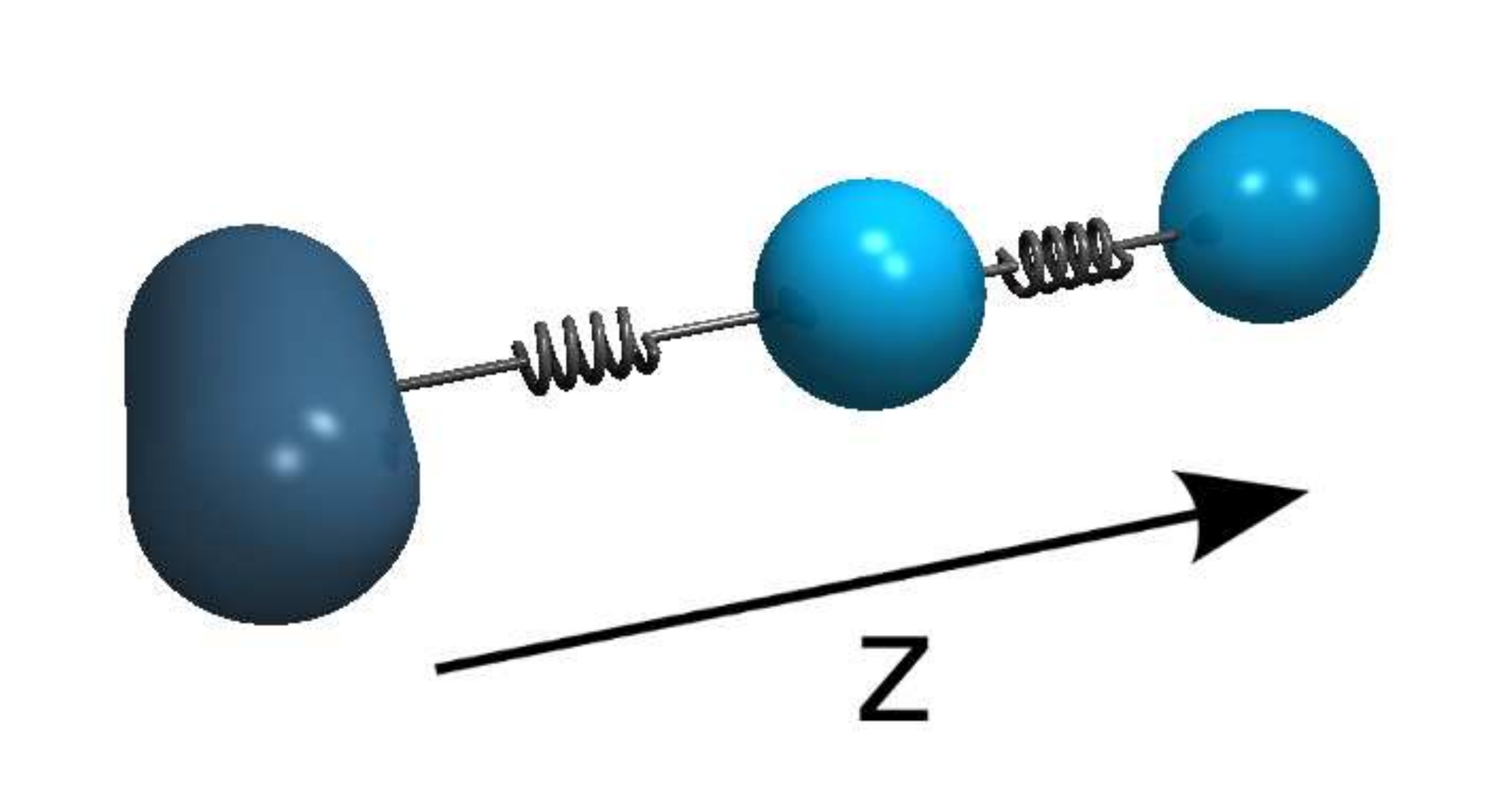}
    \label{fig:designG}
  }
  \hfill
  \subfigure[]{
    \includegraphics[height=45pt]{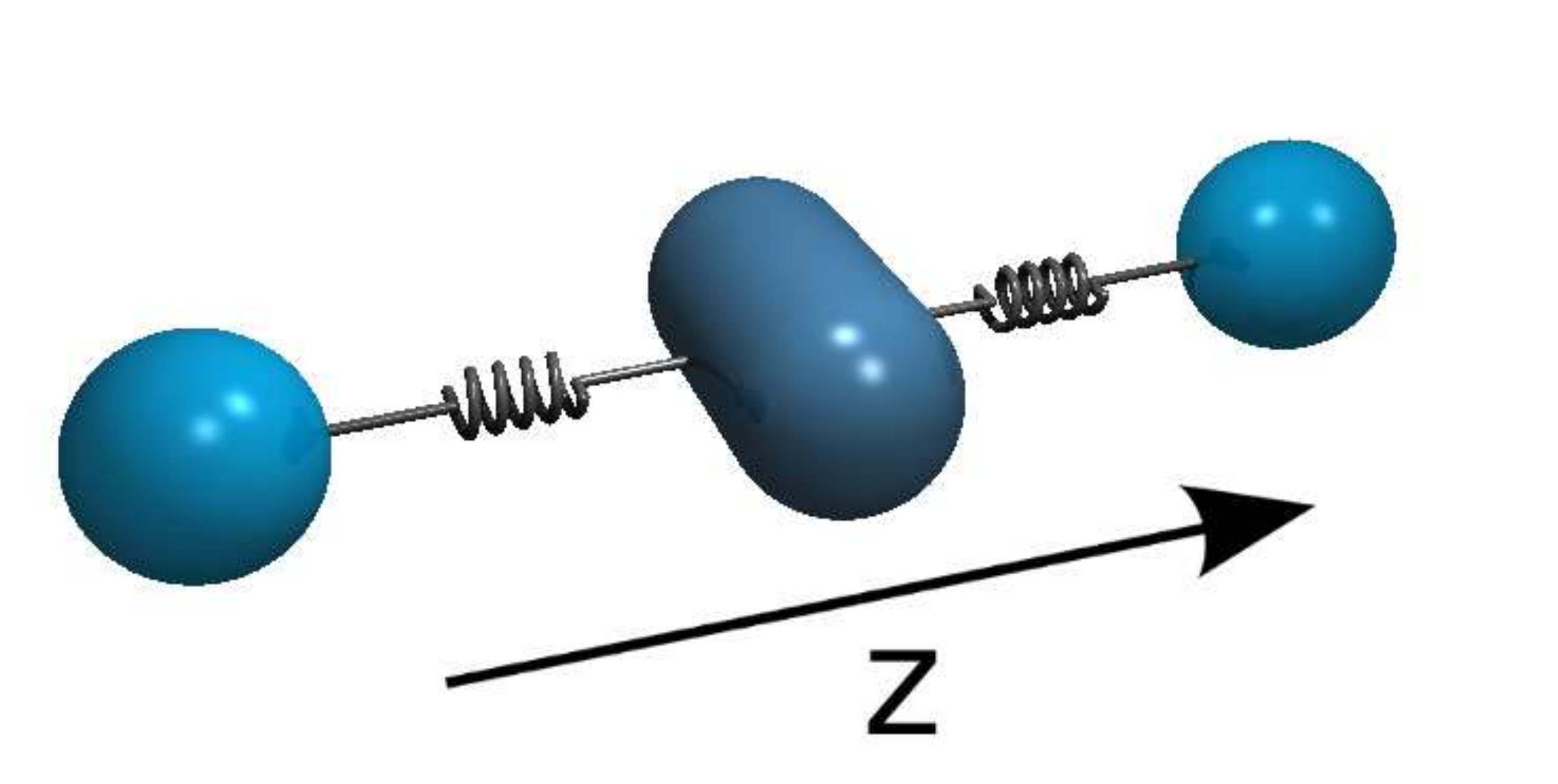}
    \label{fig:designH}
  }
  \hfill
  \subfigure[]{
    \includegraphics[height=45pt]{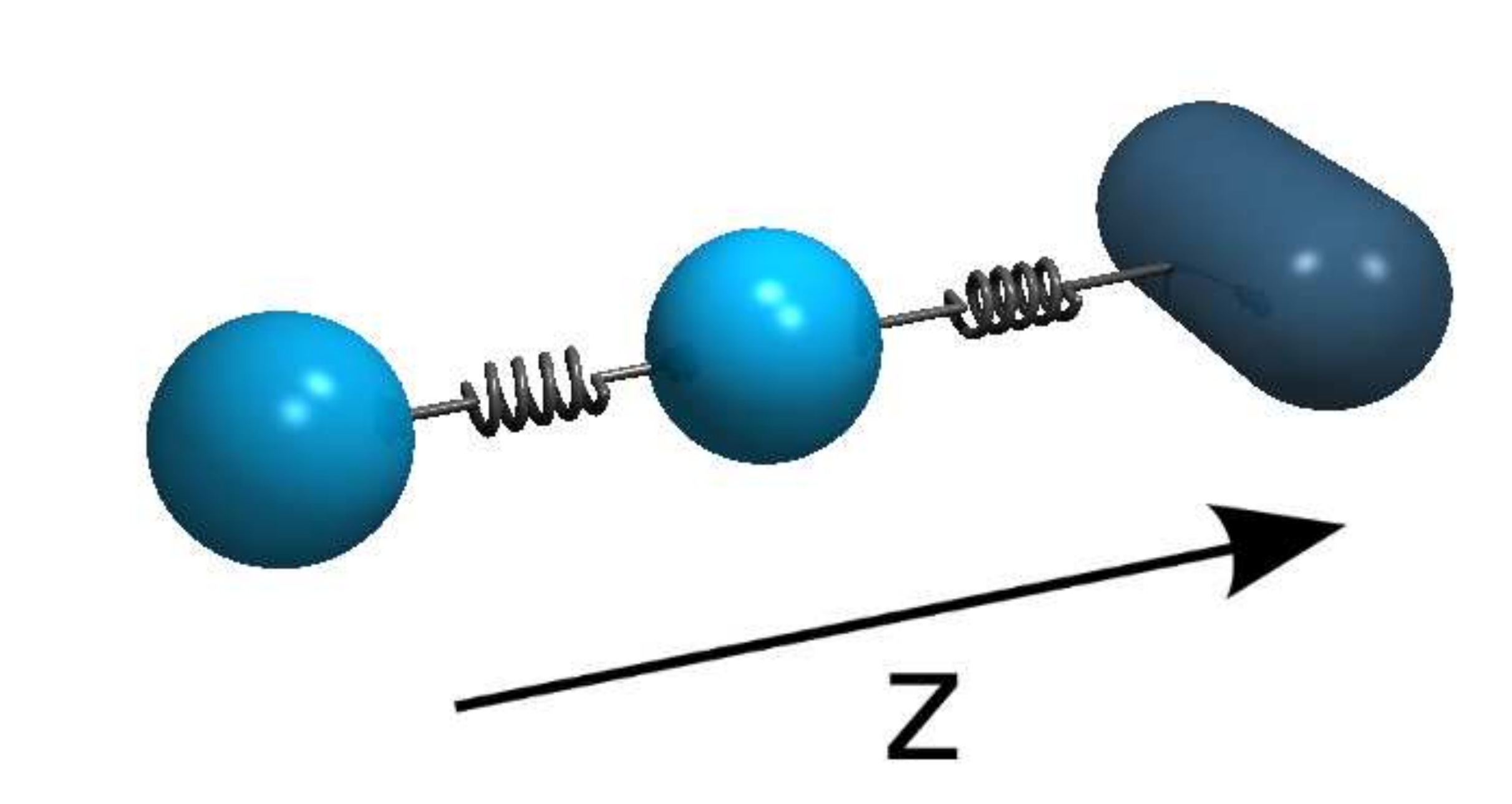}
    \label{fig:designI}
  }
  \hfill
  \subfigure[]{
    \includegraphics[height=45pt]{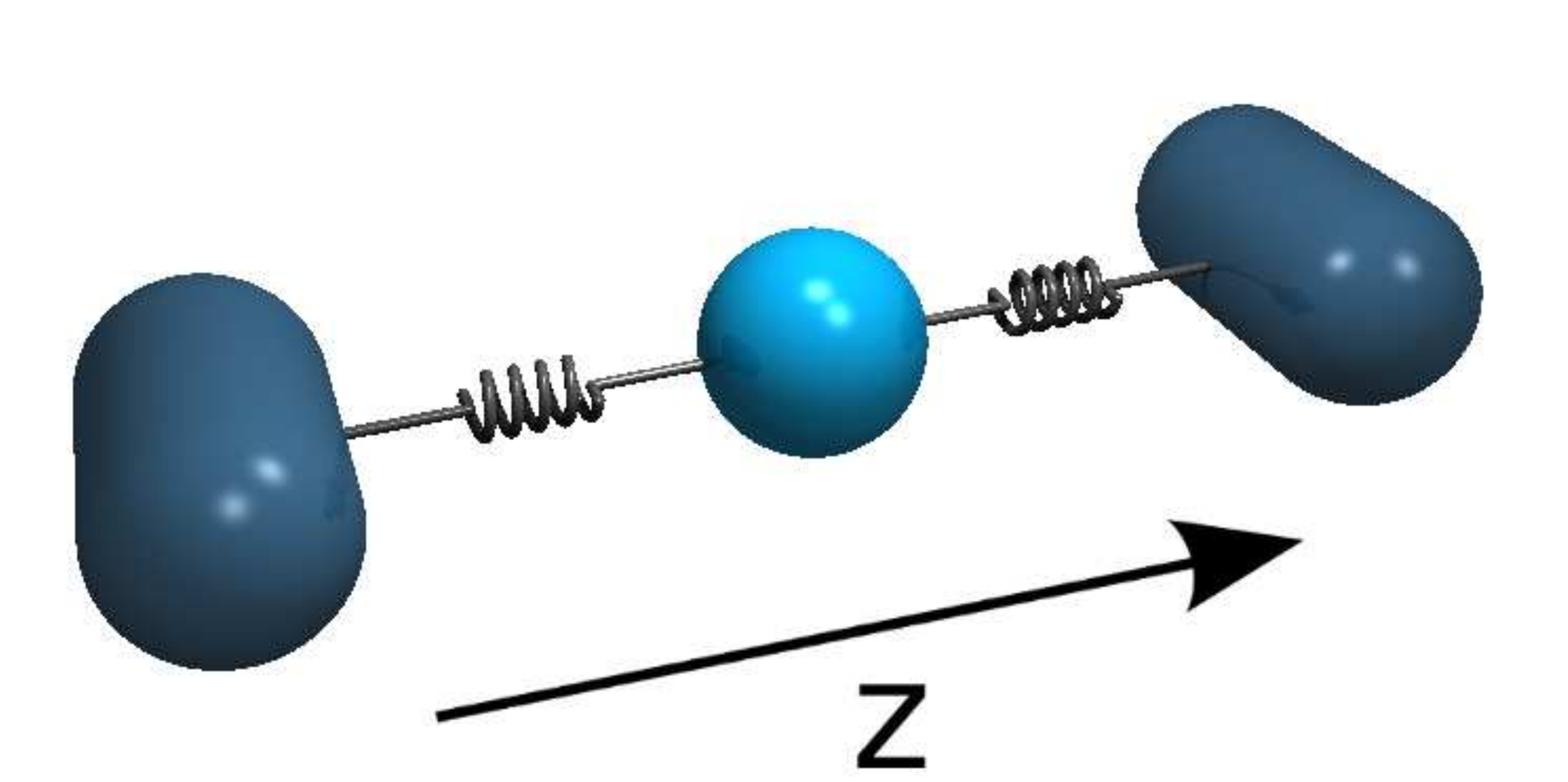}
    \label{fig:designJ}
  }
  \hfill
  \subfigure[]{
    \includegraphics[height=45pt]{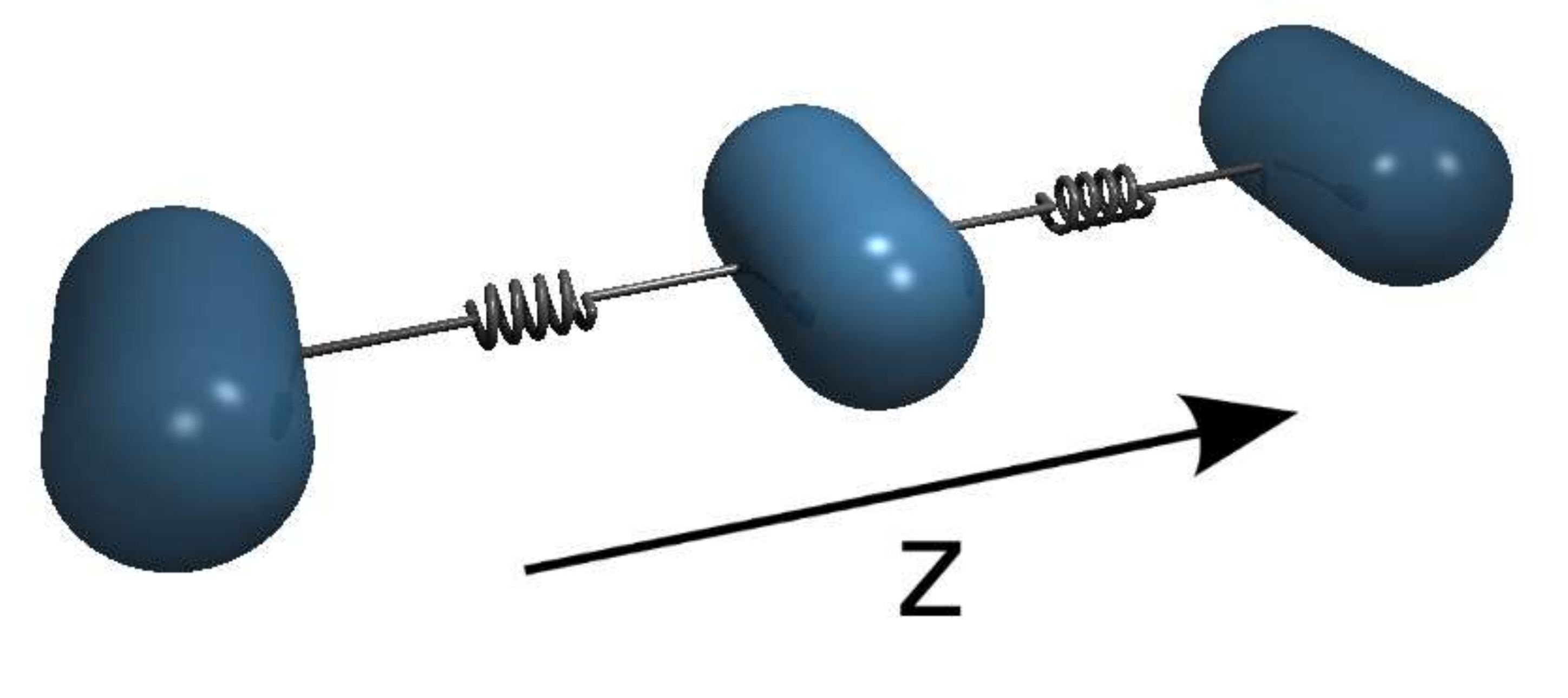}
    \label{fig:designK}
  }
  \hfill
  \subfigure[]{
    \includegraphics[height=45pt]{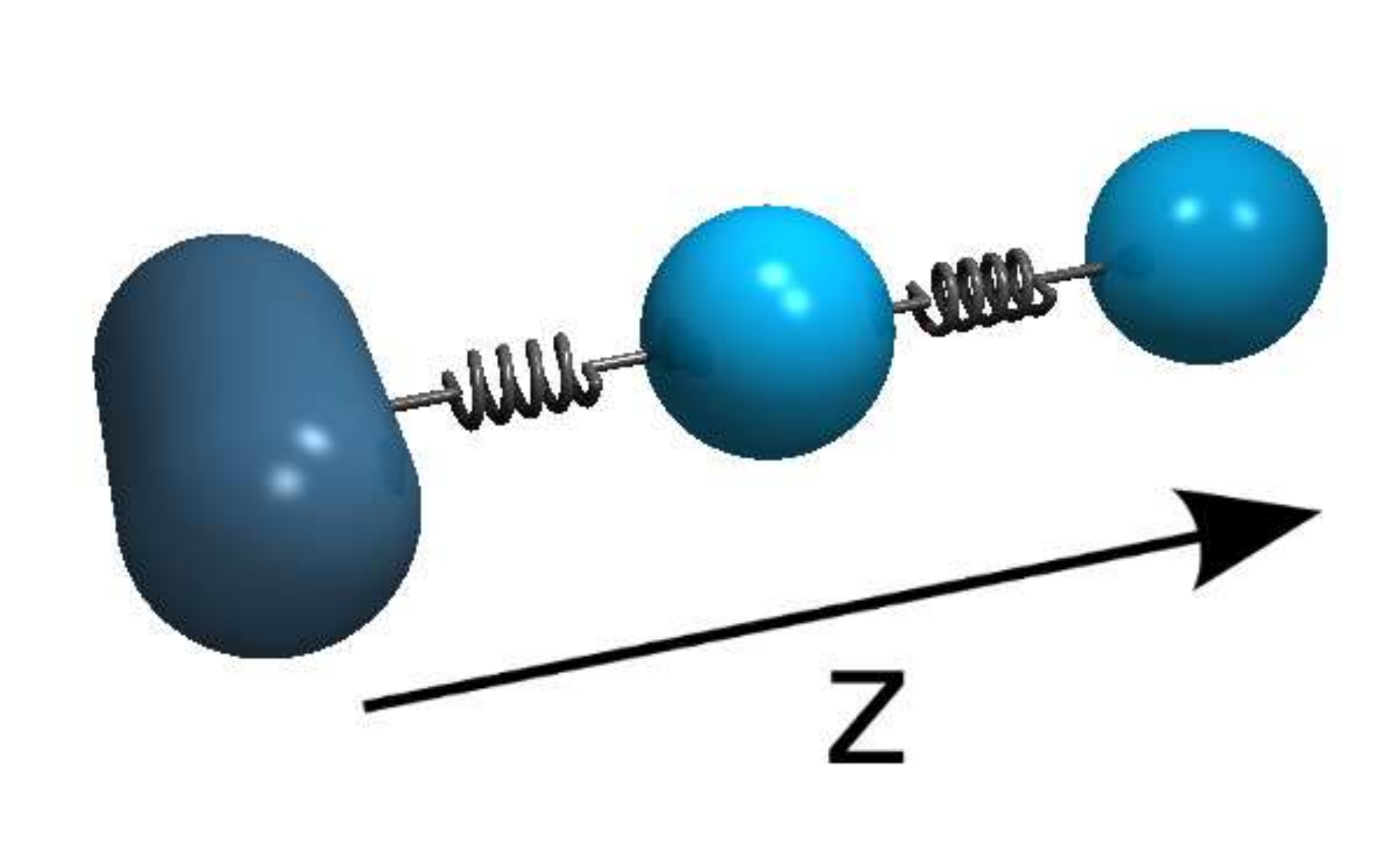}
    \label{fig:designL}
  }
  \hfill
  \subfigure[]{
    \includegraphics[height=45pt]{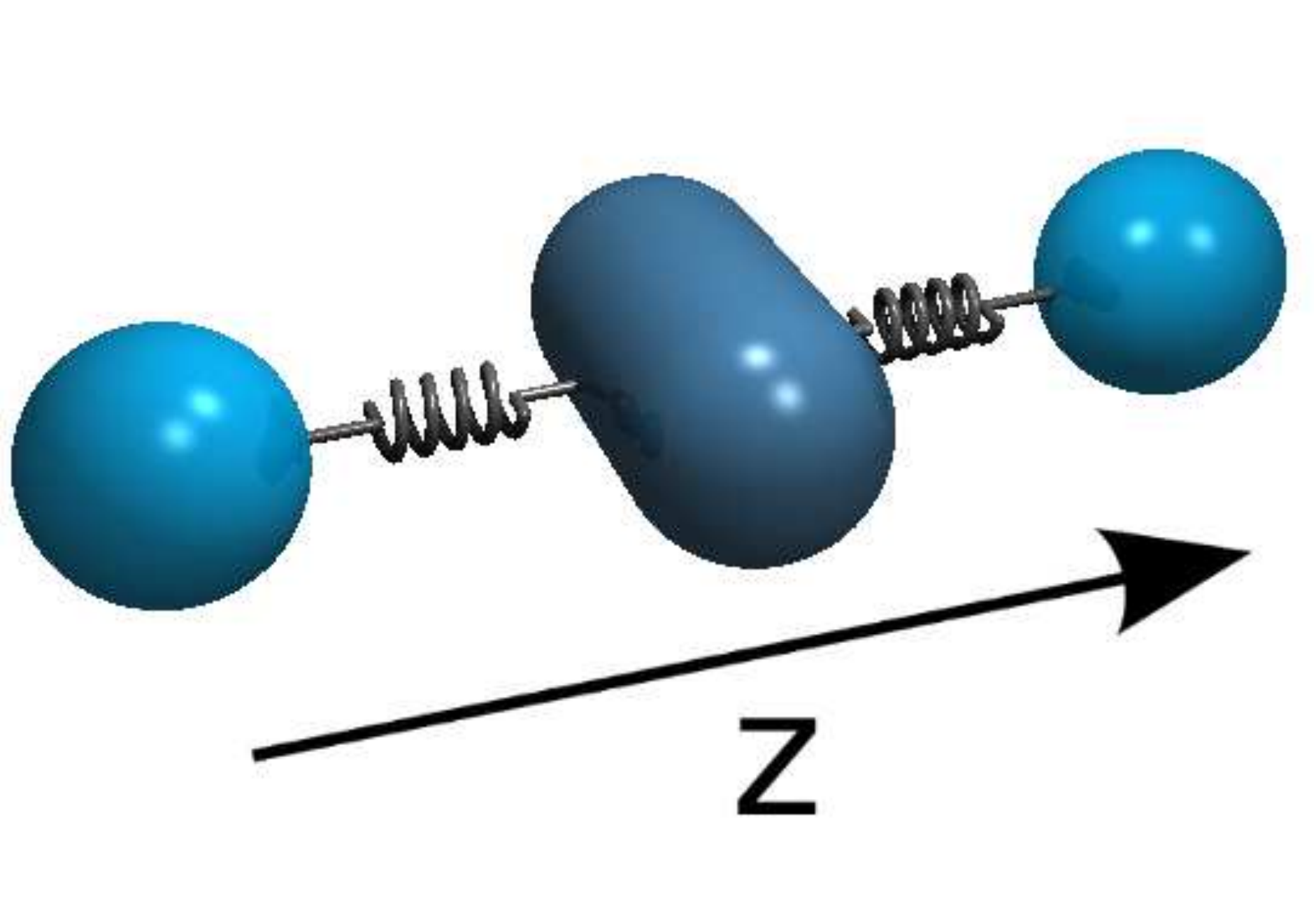}
    \label{fig:designM}
  }
  \hfill
  \subfigure[]{
    \includegraphics[height=45pt]{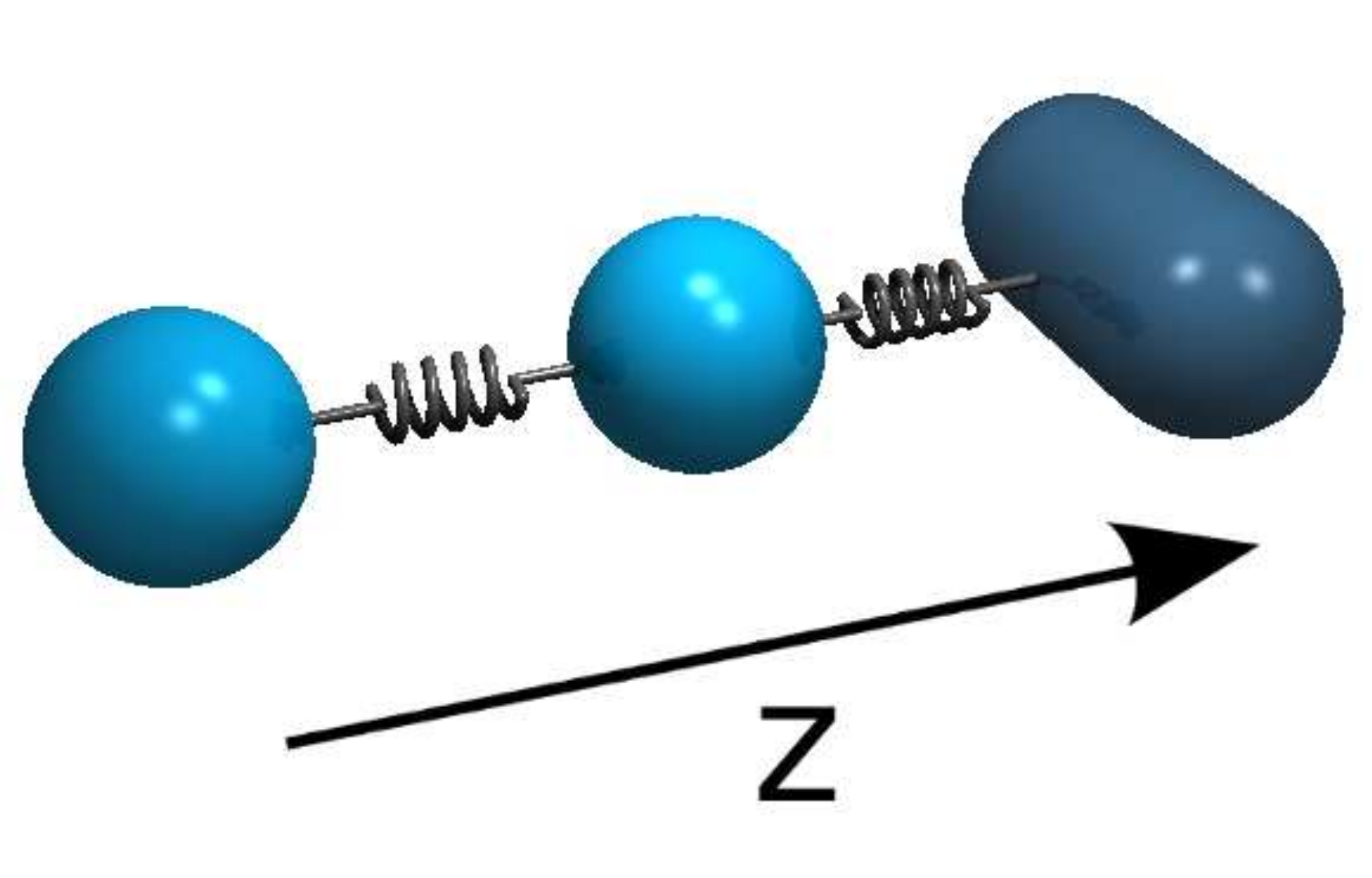}
    \label{fig:designN}
  }
  \hfill
  \subfigure[]{
    \includegraphics[height=45pt]{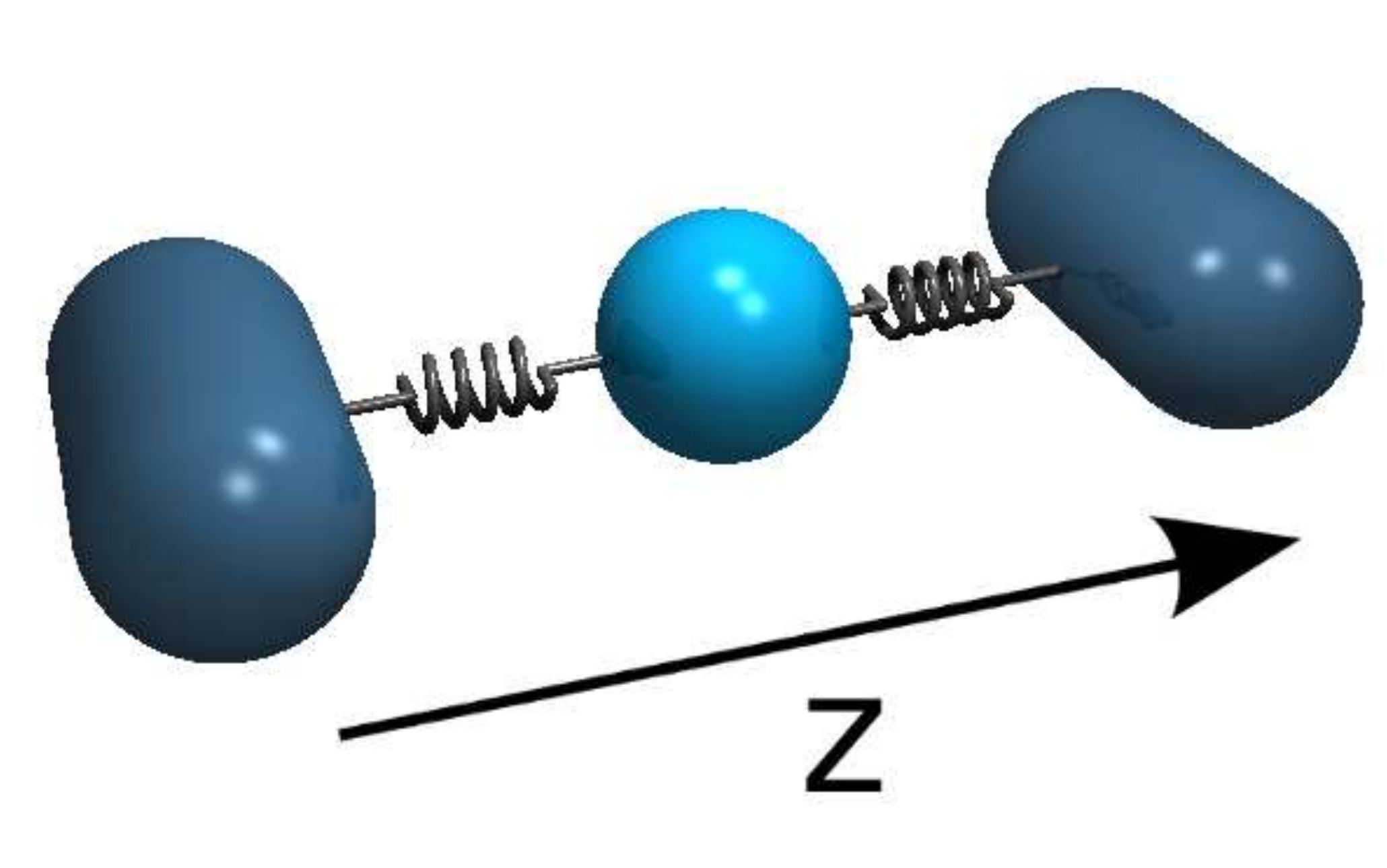}
    \label{fig:designO}
  }
  \hfill
  \subfigure[]{
    \includegraphics[height=45pt]{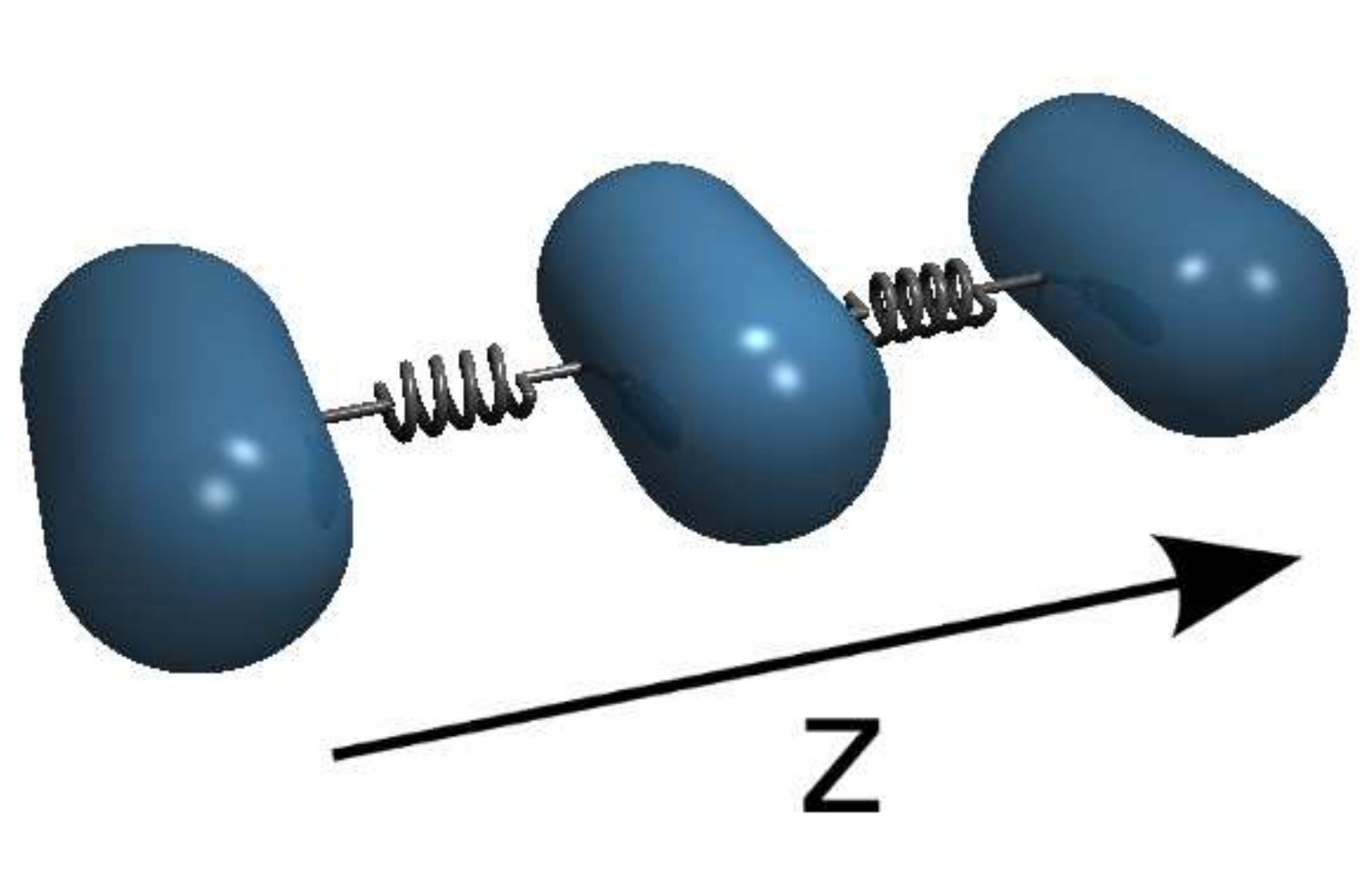}
    \label{fig:designP}
  }
  \hfill
  \subfigure[]{
    \includegraphics[height=45pt]{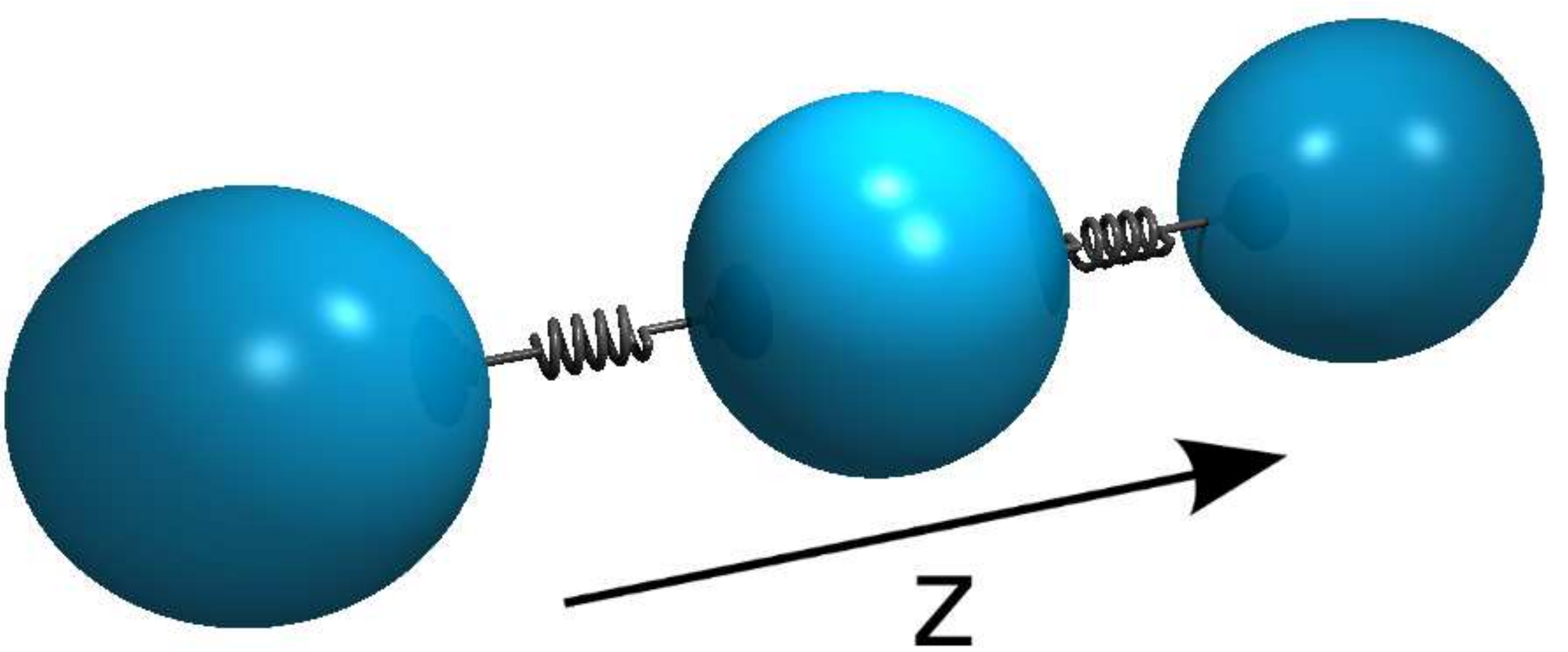}
    \label{fig:designQ}
  }
  \caption[Design options~\subref{fig:designA}--\subref{fig:designQ} of our swimming device. The capsules in the designs~\subref{fig:designG}--\subref{fig:designP} are rotated by $\pi/2$. Further, the spheres in design~\subref{fig:designQ} have larger radii, equal to the lengths of the capsules in~\subref{fig:designF} and~\subref{fig:designK}.]
  {\label{fig:designOptions}Design options~\subref{fig:designA}--\subref{fig:designQ} of our swimming device. The capsules in the designs~\subref{fig:designG}--\subref{fig:designP} are rotated by $\pi/2$. Further, the spheres in design~\subref{fig:designQ} have larger radii, equal to the lengths of the capsules in~\subref{fig:designF} and~\subref{fig:designK}.}%
\end{figure}

Since we want to investigate the effect of different geometries on the swimming behavior of our
self-propelled device, we will replace the spheres of the elementary three-sphere swimmer design by capsules.
For example, Figure~\ref{fig:swimmerCapsule} depicts the characteristic parameters of a three-capsule swimming device.
The comparably smooth edges of a capsule, and thus its resulting smooth behavior in a fluid, substantiate our choice of this particular geometry. 
Figure~\ref{fig:designOptions} shows our assembly variants.

\subsection{Cycling strategy}
\label{subsec:cycling}
Purcell's Scallop theorem~\cite{Purcell:1977:LALRN} necessitates the motion of a swimmer to be non-time-reversible at low Reynolds numbers. Additionally, the total applied force on a swimmer should vanish over one cycle, which means that the displacement of the center of mass of a swimmer over one cycle should be zero in the absence of a fluid.

\begin{figure}[htb]
\begin{center}
\includegraphics[width=12cm]{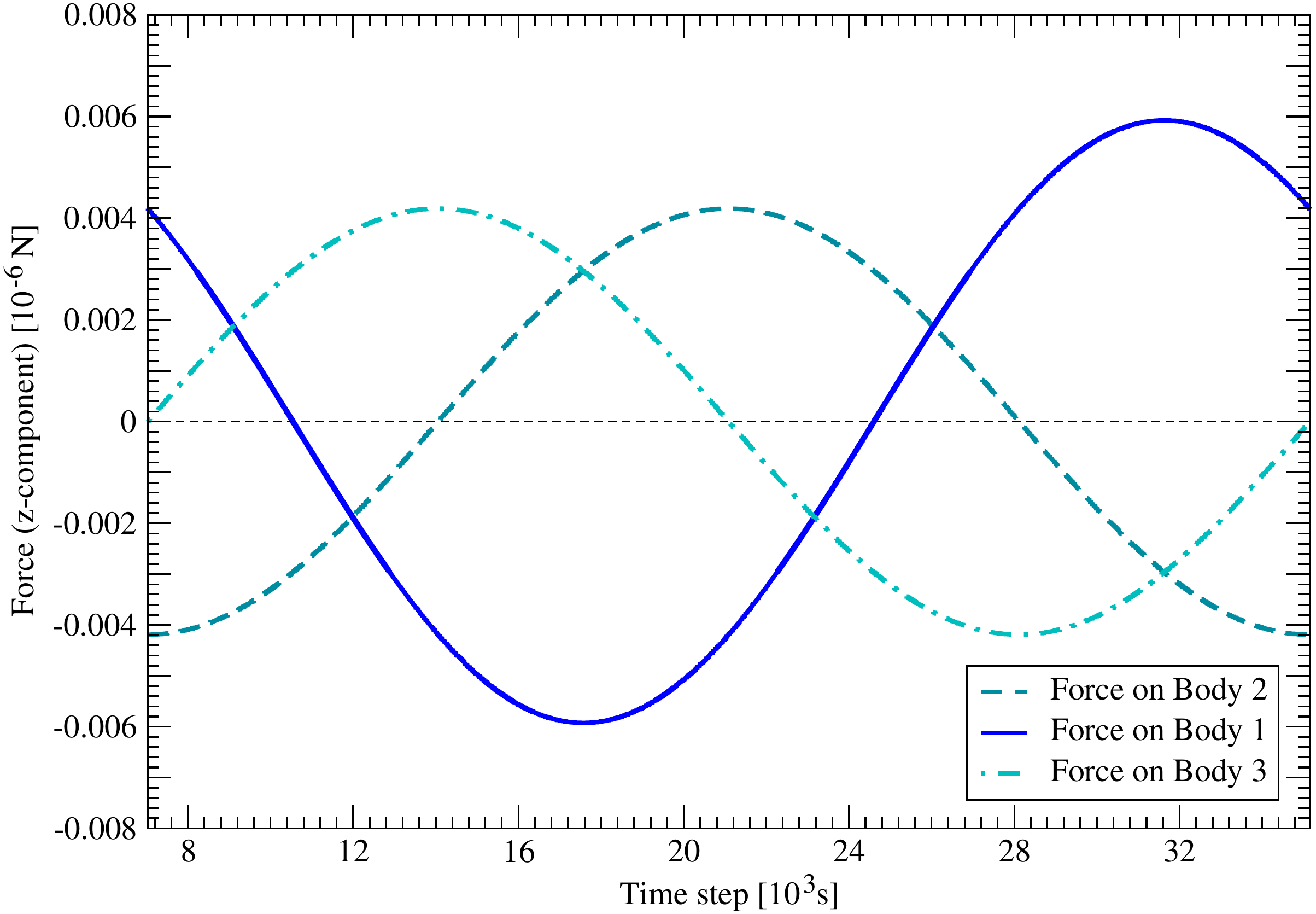}
\end{center}
\caption{Fragments of the z-components of the driving forces on all three bodies.}
\label{fig:forces}
\end{figure}

In contrast to the common approach of imposing known velocities on the constituent bodies of a swimmer, used, e.g., by Earl et al.~\cite{Earl:2007:MMSLR} and Najafi and Golestanian~\cite{Najafi:2004:SSLR}, our driving protocol imposes known forces on the objects. These forces, only applied along the main axis of the swimmer on the center of mass of each body, are given by

\begin{align}\label{eq:FB2}
   {\left( \vec{F}_{\text{dri}}^{B_2}\right)}_z &= - a  \cdot \text{sin} \left( \omega t \right) = - a \cdot \text{sin} \left( \frac{2 \pi t}{T} \right), \\
\label{eq:FB3}
   {\left(\vec{F}_{\text{dri}}^{B_3} \right)}_z &=\;\;\; a \cdot \text{sin} \left(  \frac{2 \pi (t + \varphi)}{T}\right), \\
\label{eq:FB1}
   {\left(\vec{F}_{\text{dri}}^{B_1}\right)}_z &= - \left(  {\left(\vec{F}_{\text{dri}}^{B_2}\right)}_z  + {\left(\vec{F}_{\text{dri}}^{B_3}\right)}_z\right).
\end{align}
Here $a$ is the amplitude, $\omega$ is the driving frequency, $T$ is the oscillation period, and $\varphi$ is the phase shift, kept constant at a value of $\varphi = T/4$.
In equation~(\ref{eq:FB1}), we apply the negative sum of the forces of the two outer bodies $B_2$ and $B_3$ on the middle body $B_1$ in order to ensure that the net driving force acting on the system at each instant of time is zero. Figure~\ref{fig:forces} illustrates these forces.

\begin{figure}[htb]
\centering
\subfigure[Theoretical positions prescribed by the cycling strategy.]{
  \includegraphics[width=0.4 \textwidth]{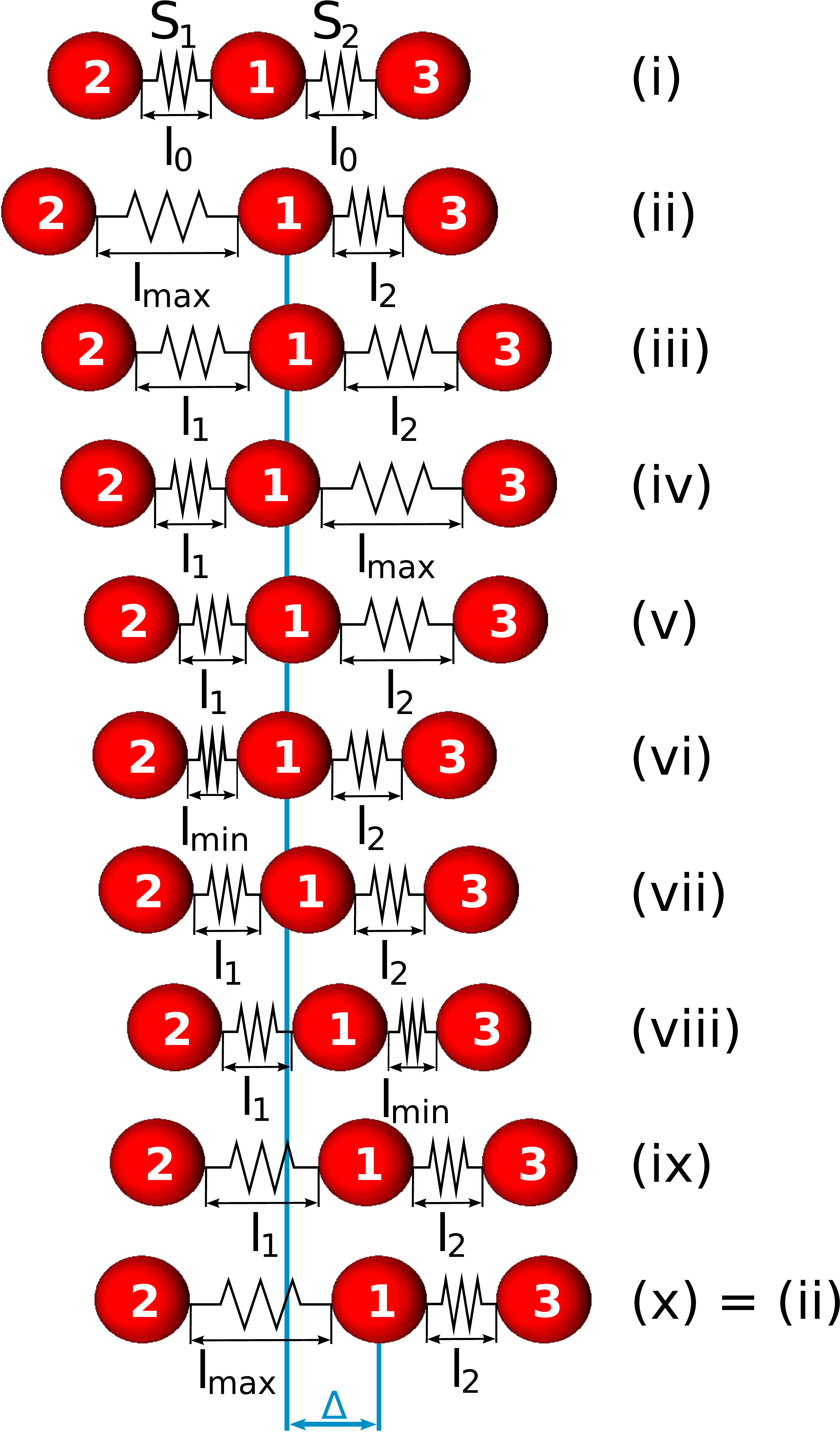}
  \label{fig:swimmingCycleTheo} 
}
\hfill
\subfigure[Position plot from the simulation. The z-positions are given in lattice cells.]{
  \includegraphics[width=0.52 \textwidth]{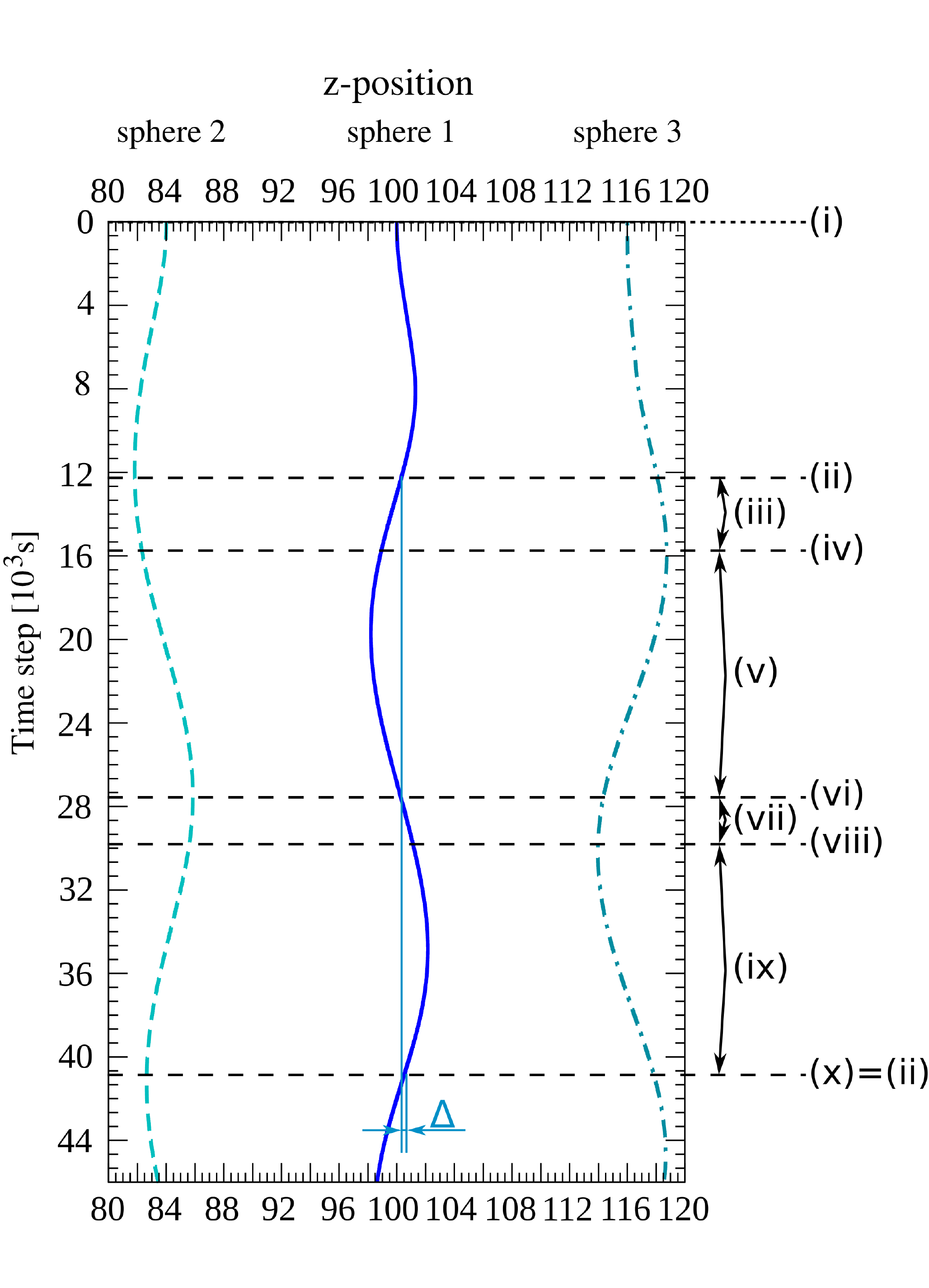}
  \label{fig:swimmingCycleSim} 
}
\caption{\label{fig:swimmingCycle}Cycling strategy of the non-time-reversible motion, exemplified on the three-sphere swimmer. For both damped harmonic potentials, $l_\text{min}$ is the minimum length of one arm,  $l_0$ is the rest length, and $l_{\text{max}}$ is the maximal extended armlength.
$\Delta$ is the distance covered by the swimmer after one cycle.}
\end{figure}

Applying this force protocol to our swimmer, we end up with our cycling strategy illustrated in Figure~\ref{fig:swimmingCycle}. 
Starting from the rest position, we have to delay the force on body $B_3$ by a fourth of the pulse length.
Step (i) depicts the initial position of our swimming device. The two \textquotedblleft arms \textquotedblright, i.e.~the connecting harmonic oscillators $S_1$ and $S_2$, are at their rest length $l_0$. In the following, $l_1$ and $l_2$ denote the current lengths of $S_1$ and $S_2$, respectively.
From step (i) to step (ii), we first apply the driving force $\vec{F}_{\text{dri}}^{B_2}$ on body $B_2$ and $\vec{F}_{\text{dri}}^{B_1}$ on body $B_1$. As shown in Figure~\ref{fig:forces}, and also apparent from equation~(\ref{eq:FB2}), $\vec{F}_{\text{dri}}^{B_2}$ starts in the negative direction until it reaches its negative maximum at which point the oscillator $S_1$ obtains its maximum length $l_\text{max}$. $S_2$ is also influenced by the force $\vec{F}_{\text{dri}}^{B_2}$ and, therefore, also gets extended to the length $l_0 < l_1 < l_\text{max}$.
From step (ii) onwards, $\vec{F}_{\text{dri}}^{B_2}$ increases and we start to exert the positive driving force $\vec{F}_{\text{dri}}^{B_3}$ on body $B_3$.
We reach the intermediate step (iii), where both oscillators $S_1$ and $S_2$ have the length $l_0 < l_1 = l_2 < l_\text{max}$.
In the transition from step (iii) to step (iv), $\vec{F}_{\text{dri}}^{B_2}$ goes to 0, i.e.~$S_1$ has decreasing length $l_0 < l_1 < l_\text{max}$. Moreover, $\vec{F}_{\text{dri}}^{B_3}$ reaches its positive maximum, and thus $S_2$ obtains its maximum length $l_\text{max}$.
With $\vec{F}_{\text{dri}}^{B_2}$ still increasing and $\vec{F}_{\text{dri}}^{B_3}$ starting to decrease, we attain another intermediate state in step (v). Here, the length of $S_1$ is $ l_\text{min} < l_1 < l_0$,  and the length of $S_2$ is $l_0 < l_2 < l_\text{max}$.
In step (vi), $\vec{F}_{\text{dri}}^{B_2}$ reaches its positive maximum. As a result, $S_1$ takes its minimum length $l_\text{min}$. Furthermore, $\vec{F}_{\text{dri}}^{B_3}$ goes to 0. Accordingly, $S_1$ relaxes to the length $l_\text{min} < l_1 < l_0$.
Now, we start to decrease $\vec{F}_{\text{dri}}^{B_2}$ and also decrease $\vec{F}_{\text{dri}}^{B_3}$ further. Passing by the temporary step (vii), where both oscillators 
$S_1$ and $S_2$ are at the length $l_0 < l_1= l_2 < l_\text{min}$, we pass into step (viii).
Here, $\vec{F}_{\text{dri}}^{B_3}$ reaches its negative maximum, resulting in $S_2$ obtaining its minimum length $l_\text{min}$. Additionally, 
$\vec{F}_{\text{dri}}^{B_2}$ goes to 0 and therefore $S_1$ relaxes to the length $l_\text{min} < l_1 < l_0$.
While $\vec{F}_{\text{dri}}^{B_2}$ continues to grow in the negative direction and $\vec{F}_{\text{dri}}^{B_3}$ decreases, our swimmer moves forward to another 
transitional state (ix), for which the conditions $l_0 < l_1 < l_\text{max}$ and $ l_\text{min} < l_2 < l_0$  hold for the oscillators' lengths.
Finally, $\vec{F}_{\text{dri}}^{B_2}$ reaches its negative maximum again, and $\vec{F}_{\text{dri}}^{B_3}$ becomes 0, resulting in a length of $l_0 < l_2 < l_\text{max}$ for the oscillator 
$S_2$. On reaching the state (x) (which is equal to state (ii)), one swimming cycle is completed.
Subsequently, our swimmer can begin another cycle of motion.

%% file: vacuum_validation.tex
\section{Validation of the simulation of swimmers with the \pe}
\label{sec:validationVacuum}

In order to quantitatively validate the application of springs in the \pe{} physics engine, we perform simulations of the motion of the three-sphere swimmer (Figure~\ref{fig:designA}) in vacuum, and compare these results with the analogous analytic model.
The basis of that model is the Lagrangian of a non-dissipative assembly
\begin{align}
\begin{split}
L = \, m\, \tfrac{\dot{\vec{x}}_1^2 + \dot{\vec{x}}_2^2 + \dot{\vec{x}}_3^2}{2} &- k\, \tfrac{(\vec{x}_1-\vec{x}_2-\vec{l}_0)^2 + (\vec{x}_3-\vec{x}_1-\vec{l}_0)^2}{2} +\\
 & \;\;\;\;\;\;\;\;\;\;\;+  \vec{x}_1 \cdot \vec{F}_{\text{dri}}^{B_1} + \vec{x}_2 \cdot \vec{F}_{\text{dri}}^{B_2} + \vec{x}_3 \cdot \vec{F}_{\text{dri}}^{B_3}.
\end{split}
\end{align}
Here, $\dot{\vec{x}}_i$ ($i=1,2,3$) refers to the derivative of the positions of the three spheres with respect to time and $\vec{l}_0$ denotes the rest 
length of the springs.
The first term in the Lagrangian
consists of the kinetic energy of the three spheres, the second gives the energy stored in the two springs due to their deformation, and the remaining terms account
for the driving forces (equations (\ref{eq:FB2})--(\ref{eq:FB1})).

As discussed previously (equation (\ref{eq:damp_i})), the damping forces in the simulation are proportional to the velocity of the spring deformation, with the factor of proportionality being $\gamma$.
Taking the damping into account, the equation of motion for each sphere becomes
\begin{equation}
 \frac{d}{dt}\;\; \frac{\partial L}{\partial \dot{\vec{x}}_i} = \frac{\partial L}{\partial \vec{x}_i} + \vec{F}_\text{damp}^{B_i} \;\;\;\;\;\; \text{for } i =1,2,3.
\end{equation}
This set of three differential equations can be solved analytically assuming appropriate initial conditions.
\begin{figure}[htb]
\begin{center}
\includegraphics[width=13.2cm]{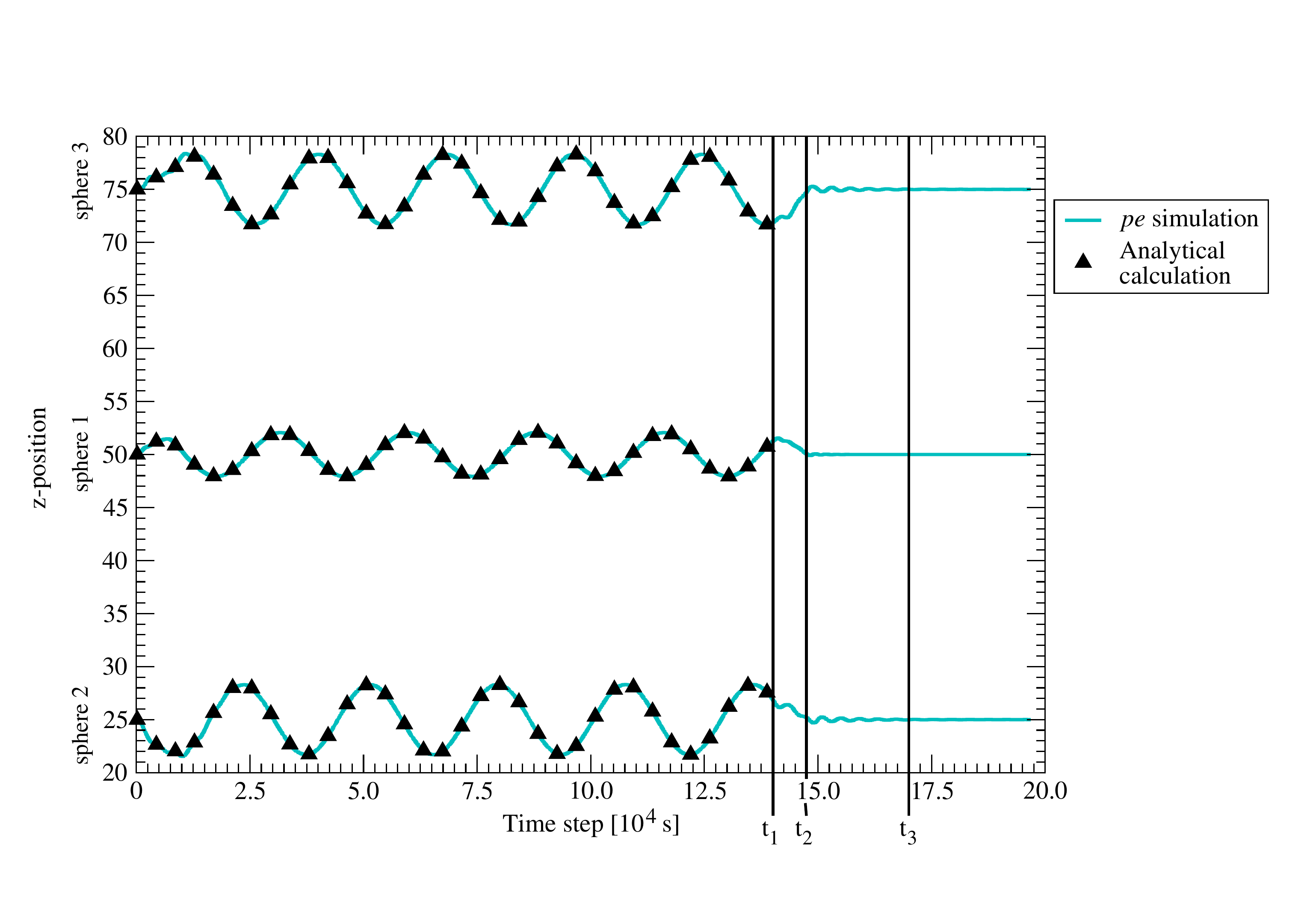}
\end{center}
\caption{Plot of the z-positions of the three spheres vs.~time (weak damping) of one of our simulations. The values of the parameters used are: $m=5.44 \times 10^{-13}$~kg, $k=1.72965$ kg/s$^2$, $(\vec{l}_0)_{z}=25$ lattice cells (with discretization $\Delta x = 10^{-6}$~m), $\gamma=1.57237 \times 10^{-7}$~kg/s, $\omega=296057.86703$~1/s, $a=1.0 \times 10^{-5}$~kg$\,$m/s$^2$, $\left(\vec{x}_2(0)\right)_{z}=25$ lattice cells, $\left(\vec{x}_1(0)\right)_{z}=50$ lattice cells, $\left(\vec{x}_3(0)\right)_{z}=75$ lattice cells, $\dot{\vec{x}}_2(0)=\vec{0}$~m/s, $\dot{\vec{x}}_1(0)=\vec{0}$~m/s, $\dot{\vec{x}}_3(0)=\vec{0}$~m/s. The simulation is run for 196812 time steps, while the driving force on the left body is switched off at $t_1=140580$ time steps, and on the right and hence also on the middle body at $t_2=147609$ time steps. The springs obtain their starting configuration at $t_3\approx179000$ time steps.}
\label{fig:positionPlot}
\end{figure}

Figure~\ref{fig:positionPlot} shows both the simulation and the analytically 
calculated results for a typical variation of the sphere positions with time in vacuum, in the case of weak damping. The used parameter values are stated in the caption. This particular setup of the damped harmonic oscillators and the driving forces is equal to the setup for the fluid simulation in Section \ref{sec:results}. For this number of time steps the lattice Boltzmann algorithm is stable and we also end up in the Stokes regime.
As expected, each sphere performs an oscillatory motion in the steady state with its frequency being the same as the driving frequency. Moreover, Figure~\ref{fig:positionPlot} indicates the 
force neutrality of our total cycle. After we switch off the driving force on the left body at $t_1=140580$ time steps and on the right body and hence also on the middle body at $t_2=147609$
time steps (which equal 3/4 of the total time steps), the springs in our swimmer start to revert to 
their rest lengths, and the swimmer soon obtains its starting configuration at $t_3\approx179000$ time steps, causing the flat ends of the position curves.
\begin{figure}[htb]
\centering
    \includegraphics[width=11cm]{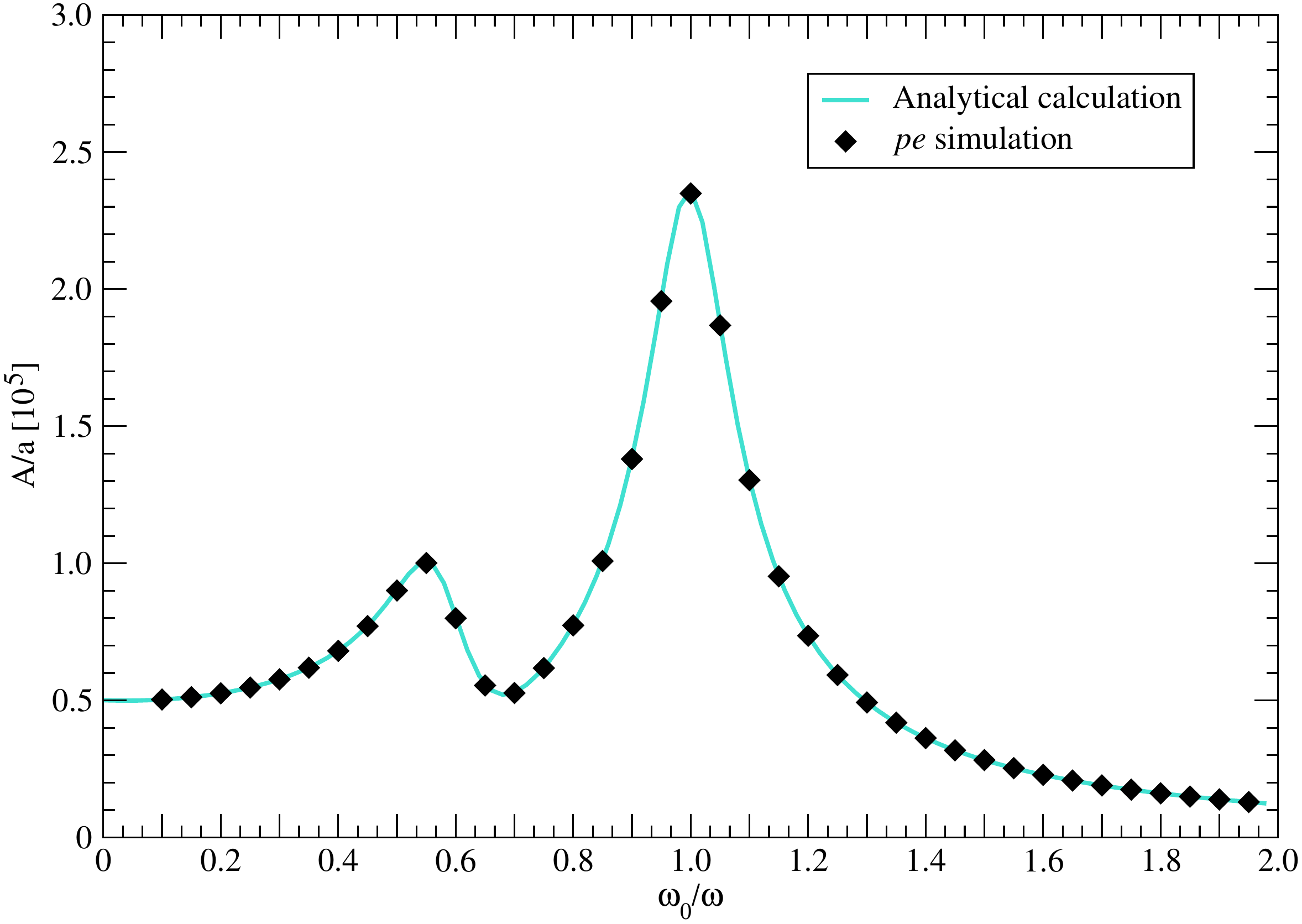}
\caption{Amplitude of oscillation $A$ scaled by amplitude of the driving forces $a$ (both given on the lattice) on the example of sphere 2 vs.~$\omega_{0}/\omega$, with $\omega$ fixed at a value of $296057.86703$~1/s, for very weak damping ($D=0.07$).}
\label{fig:vacuumVal2}
\end{figure}

Figure~\ref{fig:vacuumVal2} shows the graph of the amplitude of the oscillation $A$ of the sphere 2 (scaled by the amplitude of the driving forces $a$) as a function of the natural spring 
frequency $\omega_0=\sqrt{k/m}$ (scaled by constant $\omega$), with the fixed characteristic damping ratio $D=0.07$ (equation (\ref{eq:dampingFactor})).

For this graph, the mass of each sphere was fixed at $m=5.44 \times 10^{-13}$~kg,
while the stiffness $k$ of the spring was varied between the values of $4.76813 \times 10^{-3}$~kg/s$^2$ and $0.181309$~kg/s$^2$.
Accordingly, the damping $\gamma$ lies within the bounds of $2.27017 \times 10^{-9}$~kg/s and $4.414 \times 10^{-8}$~kg/s.
Both ranges yield a damping ratio of $D=0.07$ in each run.
In order to guarantee the achievement of a steady state in all of the simulations, we perform $10^{6}$ time steps.

Both Figure~\ref{fig:positionPlot} and \ref{fig:vacuumVal2} demonstrate excellent agreement between the simulation results and the analytically obtained results.
In Figure~\ref{fig:vacuumVal2}, the 38 selected simulation scenarios match the analytically obtained curve with the two characteristic maxima at $\omega_0 \sim \omega$ and $\omega_0 \sim \omega/\sqrt{3}$ for not less than 10 digits.

%% file: results.tex
\section{Simulation of swimmers in a fluid}
\label{sec:results}
In the following we perform simulations of swimmers in a highly viscous fluid. Apart from exploring the parameter space in order to achieve simulations in a physically-stable regime, we also 
show that the shape of a swimmer has a considerable effect on its swimming velocity.

\subsection{Parameters of simulations}
\label{subsec:res:design}
The setup of the domain for the design parameter study is a cuboidal channel with $(x \times y \times z ) = (100\times100\times200)$ lattice cells, each of which is a cube with a side length of $10^{-6}$~m. The z-axis corresponds to the axis of movement of the device. The highly viscous fluid has a kinematic viscosity of $\nu = 7.36 \times10^{-5}$~m$^2$/s and a density of $\rho = 1.36$~kg/m$^3$, these being typical values at room temperature for honey.
In order to make the different designs comparable, the mass of all spheres and capsules has been set to $m_\text{sph} = m_\text{cap} =5.44 \times 10^{-13}$~kg. The spheres in the designs (a)-(e), (g)-(j) and (l)-(o) all have the radius $r_\text{sph}=4 \times 10^{-6}$~m. The capsules used in the designs (b)-(p) all have the radius $r_\text{cap}=4 \times 10^{-6}$~m and the length $l_\text{cap}=8 \times 10^{-6}$~m. In order to observe the influence of the distances to the walls of the simulation box, we also consider a three-sphere swimmer (q) 
with larger radii $r_\text{sphbig}=8 \times 10^{-6}$~m. 

At $t = 0$, the swimmer is centered in the channel, with the center of the middle body being at the center of the channel with respect to all three dimensions. As the time-steps increase, the z-positions of the different bodies of the swimmer change, while the x- and the y-positions do not. Also, at $t = 0$, the springs are at their rest length of 8 lattice cells ($ = 8 \times 10^{-6}$~m) in the designs (a)-(f) and (l)-(q). In order to make the respective armlengths of the corresponding designs (b)-(f) and (g)-(k) equal, the springs in the designs (g)-(k) have a rest length of 16 lattice cells ($=16 \times 10^{-6}$~m) between perpendicular capsules, and those between a sphere and a perpendicular capsule have a rest length of 12 lattice cells ($=12 \times 10^{-6}$~m). The other springs in those designs have a rest length of 8 lattice cells ($=8 \times 10^{-6}$~m).
All springs are characterized by a stiffness of $k=1.72965$~kg/s$^2$ and a damping constant of $\gamma=1.57237 \times 10^{-7}$~kg/s. This results in a damping ratio of $D=0.07 \ll 1$, i.e.~our system is in the weak damping regime.

The simulations run for $196812$ time steps, performing five swimming cycles with an oscillation period of $T=28116$ time steps. The sinusoidal driving forces on the left and the middle body are applied at $T = 0$, whereas the force on the right body is implemented with a delay of $\varphi= T/4 =7029$ time steps. The amplitude of the driving force on the right and the left body is $a=1.0 \times 10^{-5}$~kg$\,$m/s$^2$. The force on the middle body is at each instant the negative of the sum of the forces on the two other bodies, so that the swimmer has no net external driving force at any time step. This means that the amplitude of the driving force on the middle body is $a$ for $t \leq 7029$ and $t \geq 140580$ time steps, and $\sqrt{2} a$ otherwise. All simulations are terminated by switching off the driving at $t_1 = 140580$ time steps for the left body and $t_2= 147609$ time steps (which equals 3/4 of the total simulation time) for all other bodies and the system is allowed to relax on its own for the remaining time. It should be noted that the same driving forces are applied on the bodies in each simulation, and any differences in their motion are due to their different shapes.
\begin{table}
\centering
\begin{tabular}{|cc|c|c|c|c|c|c|}
\hline
\multicolumn{2}{|c|}{\multirow{2}{*}{Design}} & {Rest}  & {Total} & \multirow{2}{*}{$Re_\text{swim}[10^{-3}]$}   & \multirow{2}{*}{$Re^{B_2}$} & 
\multirow{2}{*} {$Re^{B_1}$} & \multirow{2}{*}{$Re^{B_3}$} \\ 
&                                                            & {Length} & {Width} &                                              &                             &                                                 
             &                             \\  
\hline
{(a)}&\parbox[2cm][1cm][c]{2cm}{\centering{\includegraphics[height=0.8cm]{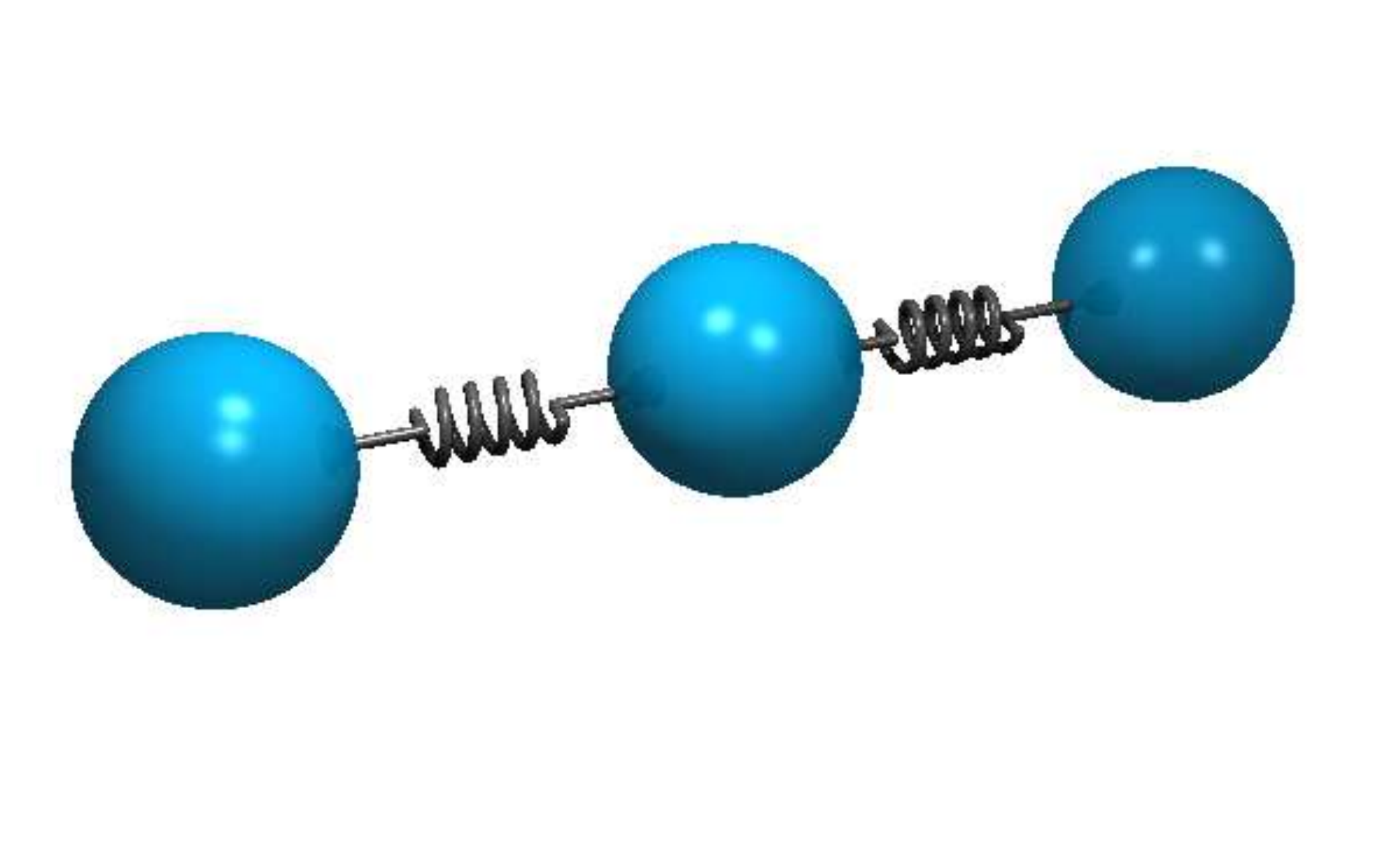}}}  & 40 &  8 & 6.51 & 0.066 & 0.059 & 0.078\\
\hline
{(b)}&\parbox[2cm][1cm][c]{2cm}{\centering{\includegraphics[height=0.8cm]{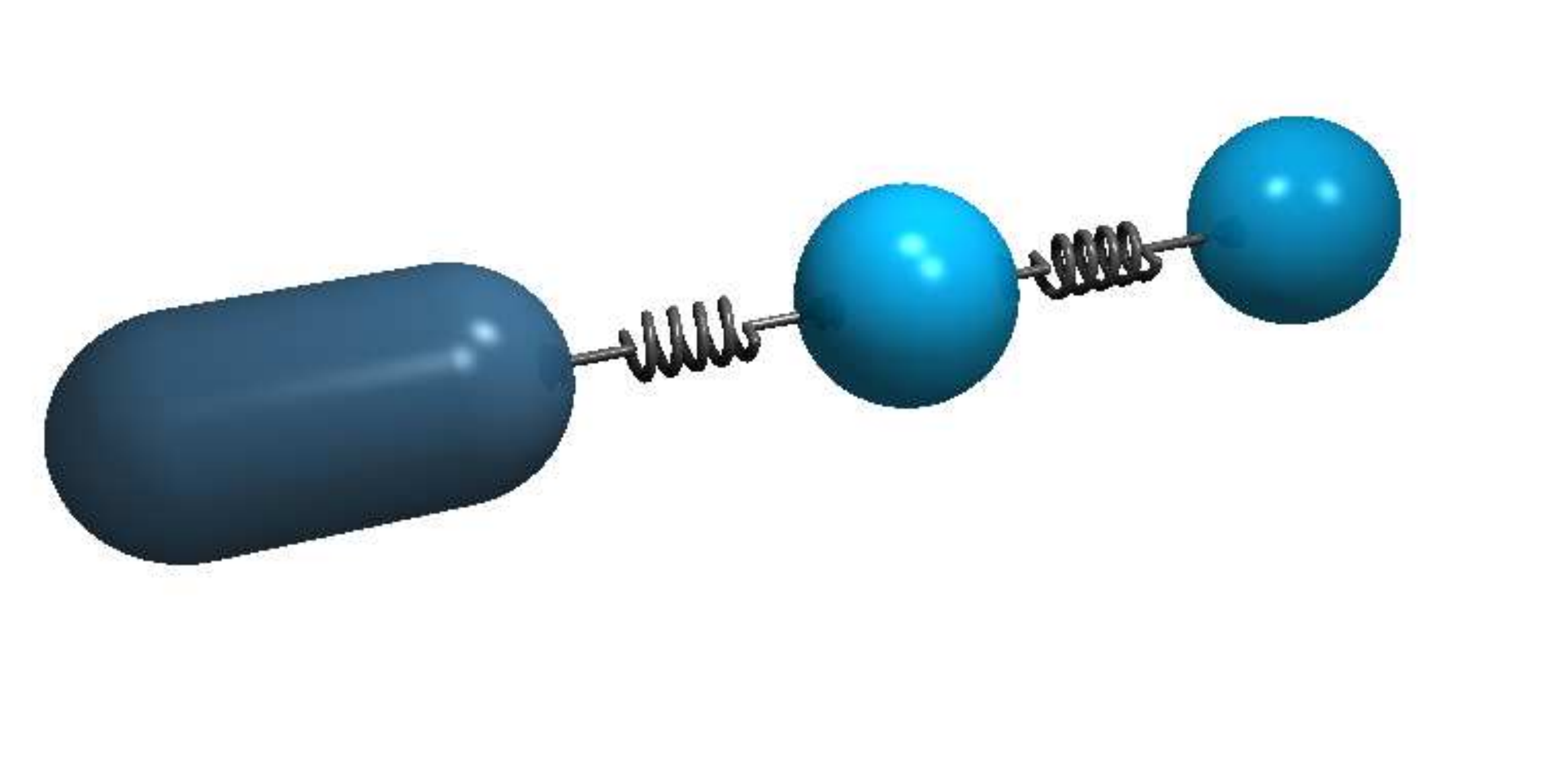}}}  & 48 &  8 & 6.32 & 0.128 & 0.047 & 0.130\\
\hline
{(c)}&\parbox[2cm][1cm][c]{2cm}{\centering{\includegraphics[height=0.8cm]{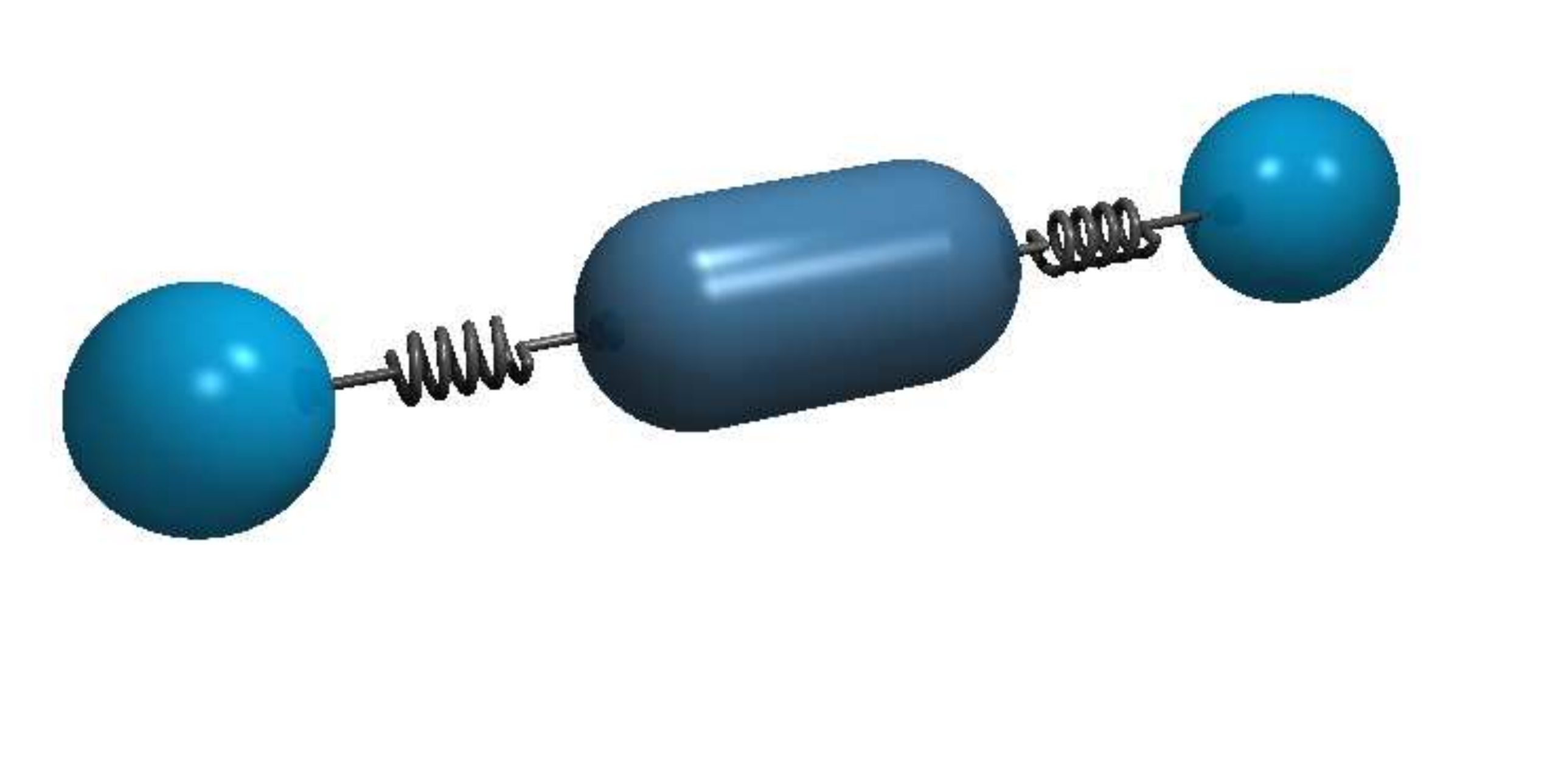}}}  & 48 &  8 & 5.75 & 0.053 & 0.123 & 0.081\\
\hline
{(d)}&\parbox[2cm][1cm][c]{2cm}{\centering{\includegraphics[height=0.8cm]{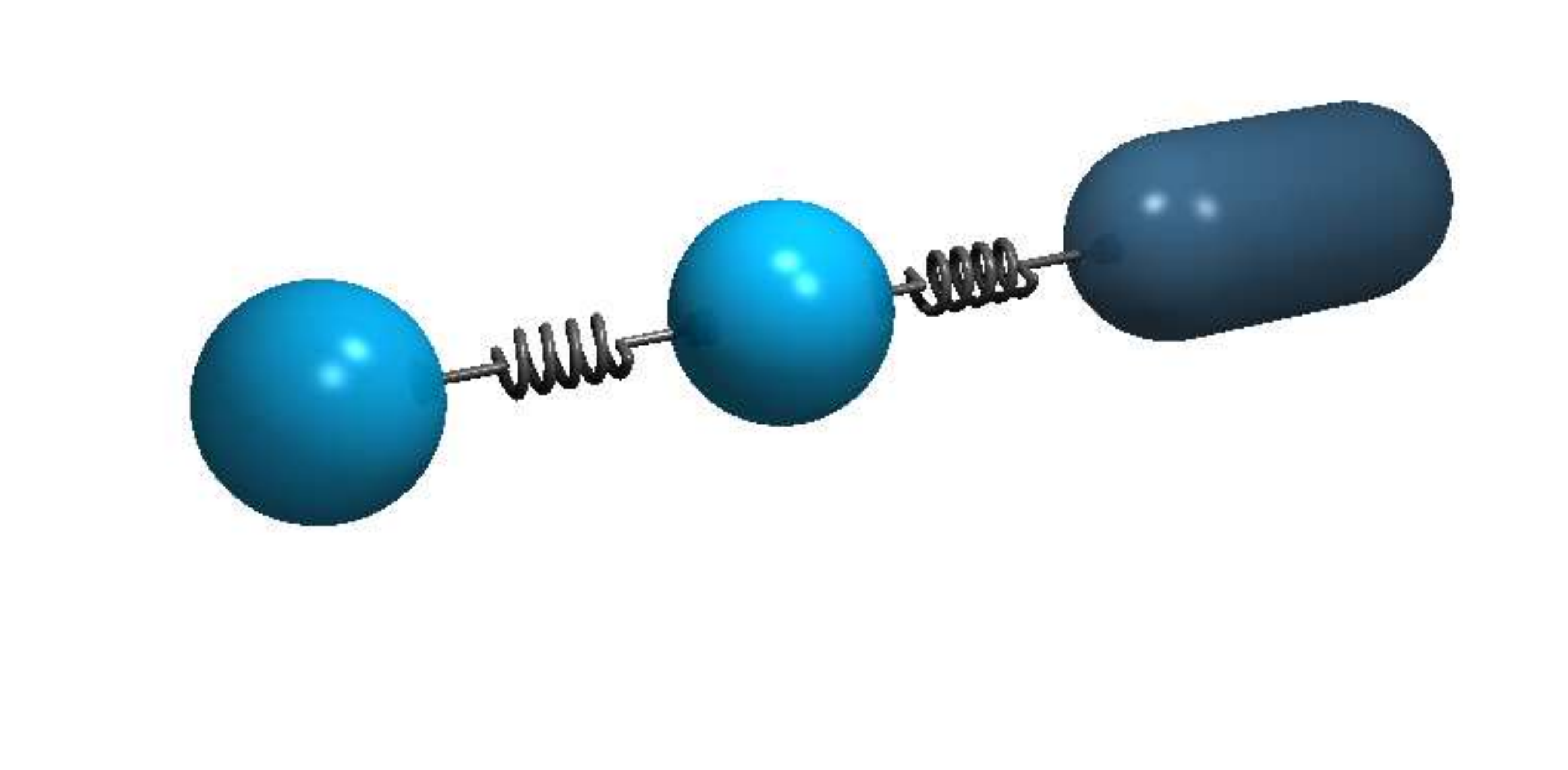}}}  & 48 &  8 & 6.14 & 0.074 & 0.060 & 0.121\\
\hline
{(e)}&\parbox[2cm][1cm][c]{2cm}{\centering{\includegraphics[height=0.8cm]{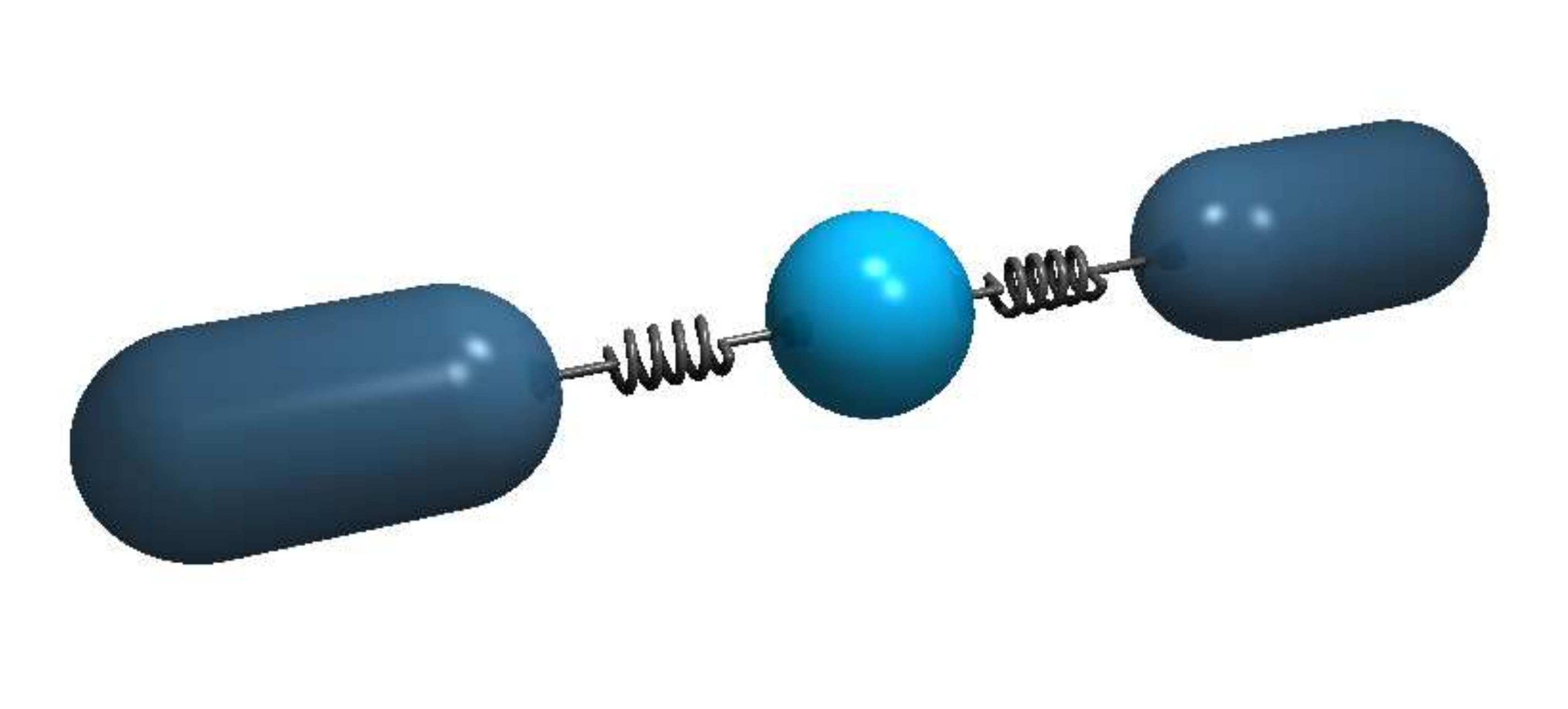}}}  & 56 &  8 & 5.72 & 0.143 & 0.047 & 0.130\\
\hline
{(f)}&\parbox[2cm][1cm][c]{2cm}{\centering{\includegraphics[height=0.8cm]{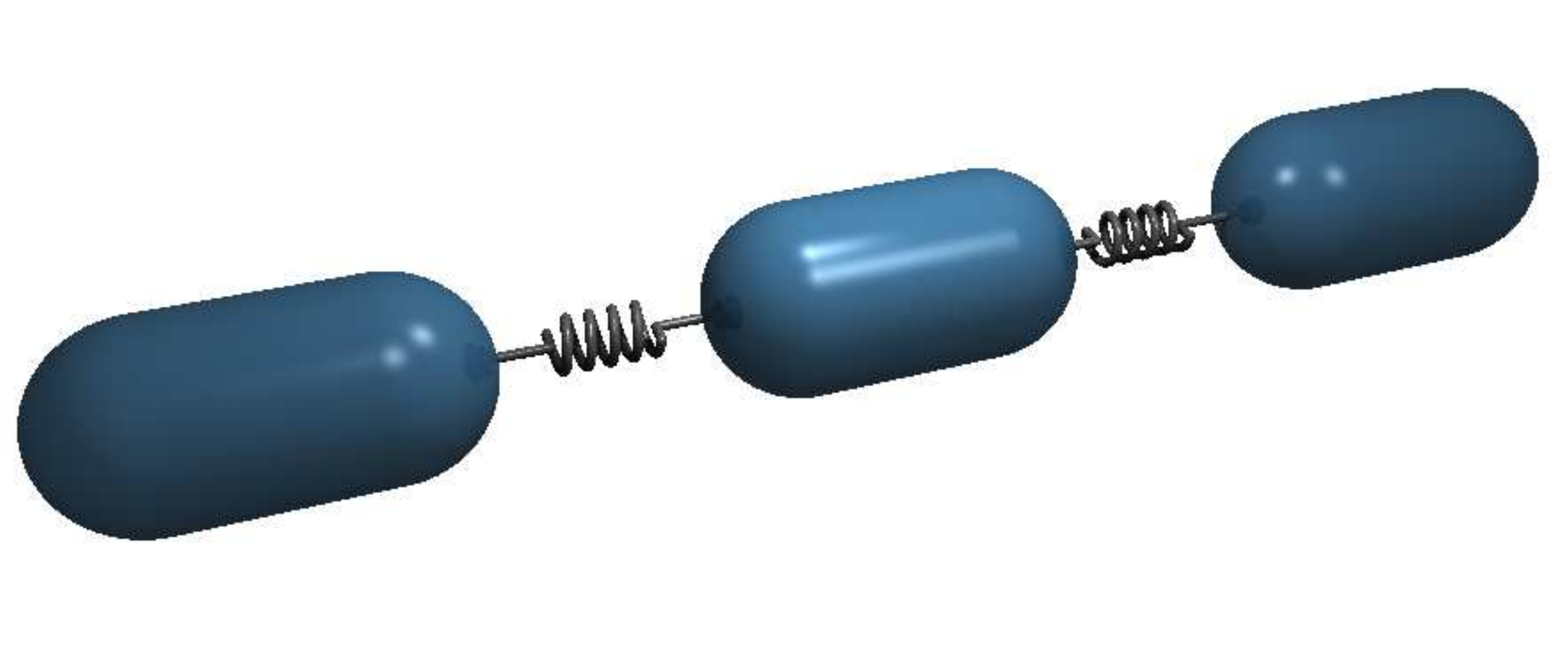}}}  & 64 &  8 & 5.51 & 0.119 & 0.096 & 0.136\\
\hline
{(g)}&\parbox[2cm][1cm][c]{2cm}{\centering{\includegraphics[height=0.8cm]{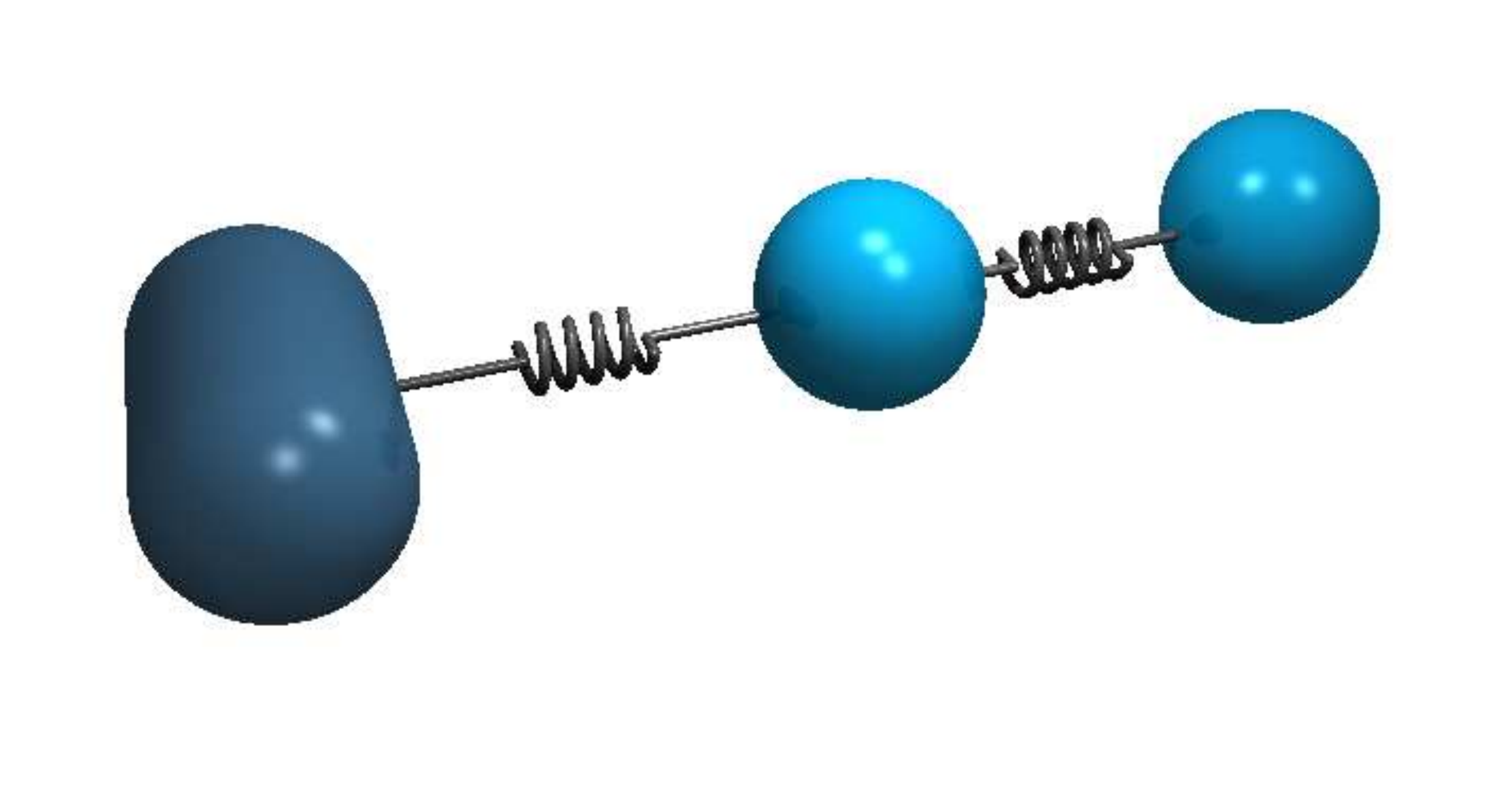}}}  & 44 & 16 & 5.07 & 0.064 & 0.041 & 0.086\\
\hline
{(h)}&\parbox[2cm][1cm][c]{2cm}{\centering{\includegraphics[height=0.8cm]{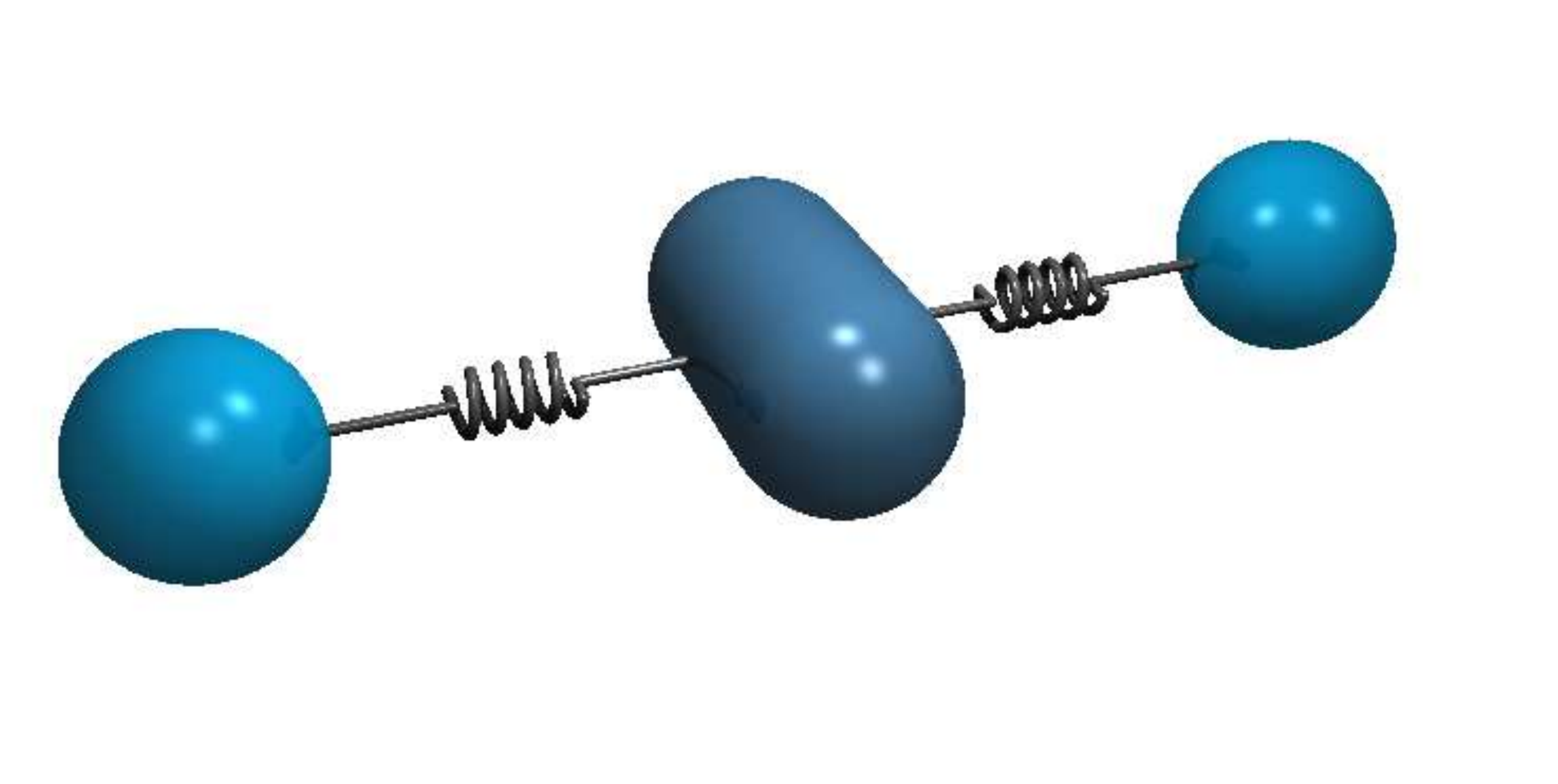}}}  & 48 & 16 & 4.93 & 0.047 & 0.065 & 0.083\\
\hline
{(i)}&\parbox[2cm][1cm][c]{2cm}{\centering{\includegraphics[height=0.8cm]{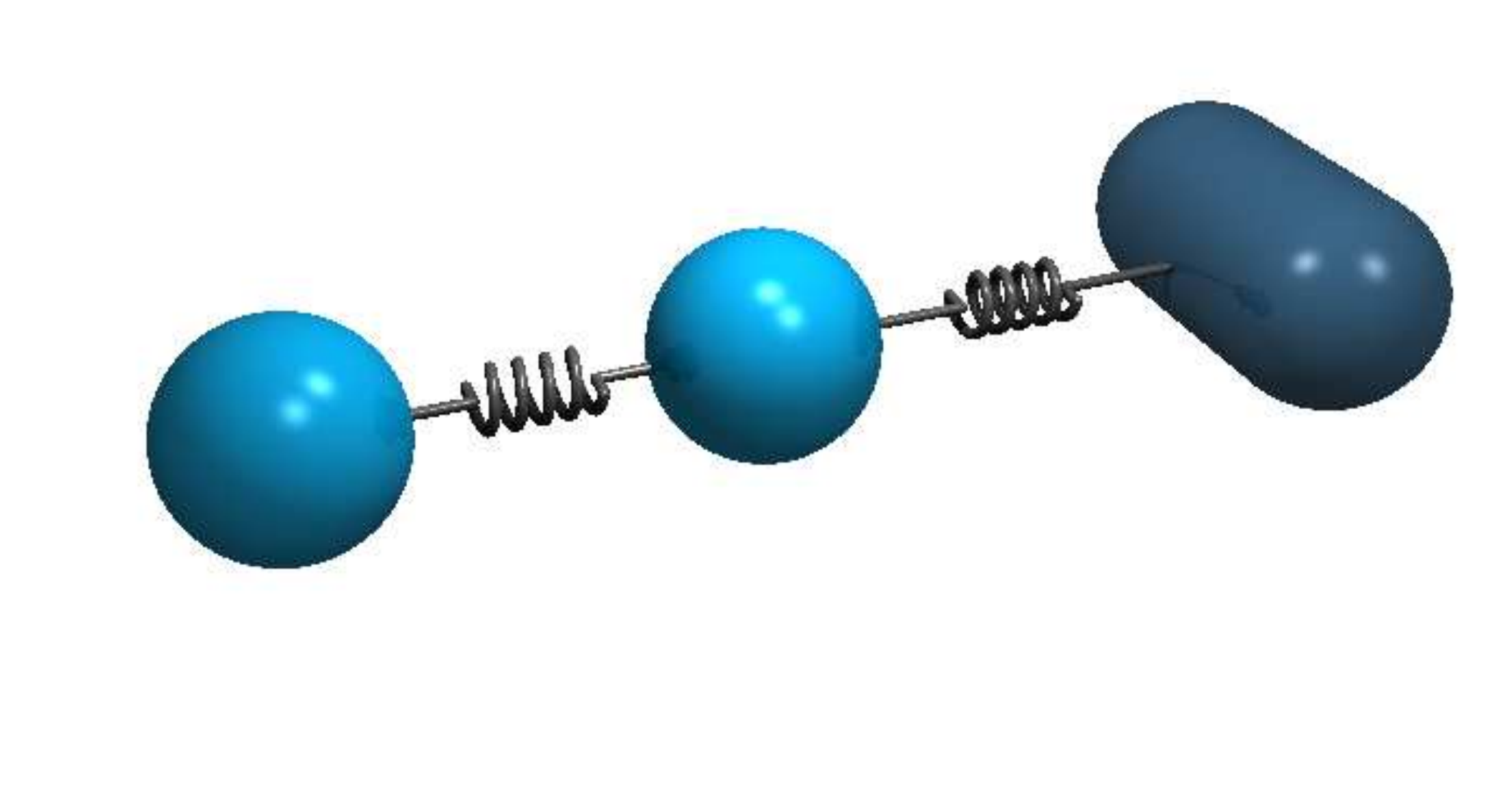}}}  & 44 & 16 & 4.93 & 0.077 & 0.060 & 0.053 \\
\hline
{(j)}&\parbox[2cm][1cm][c]{2cm}{\centering{\includegraphics[height=0.8cm]{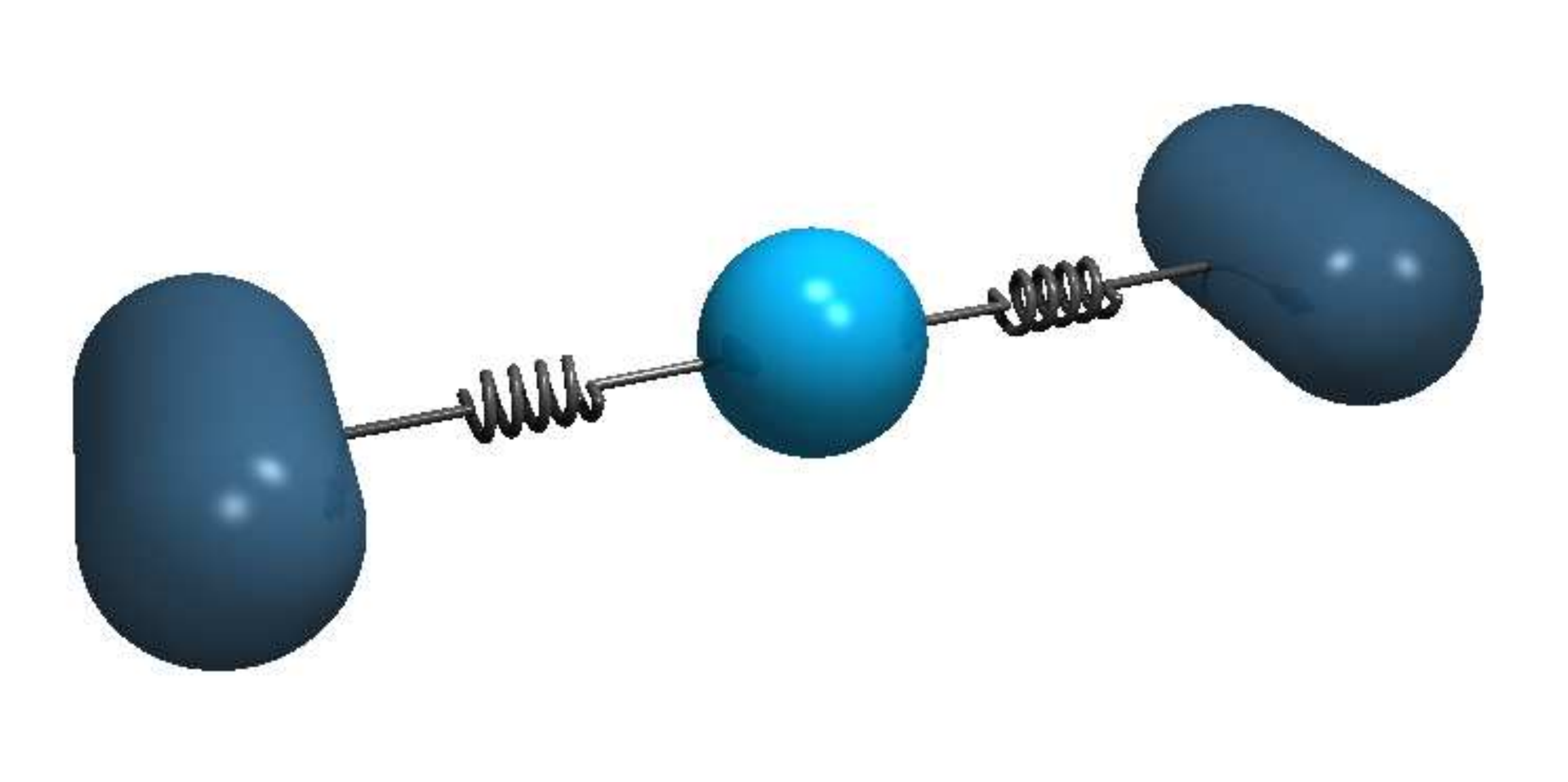}}}  & 48 & 16 & 3.64 & 0.074 & 0.042 & 0.059\\
\hline
{(k)}&\parbox[2cm][1cm][c]{2cm}{\centering{\includegraphics[height=0.8cm]{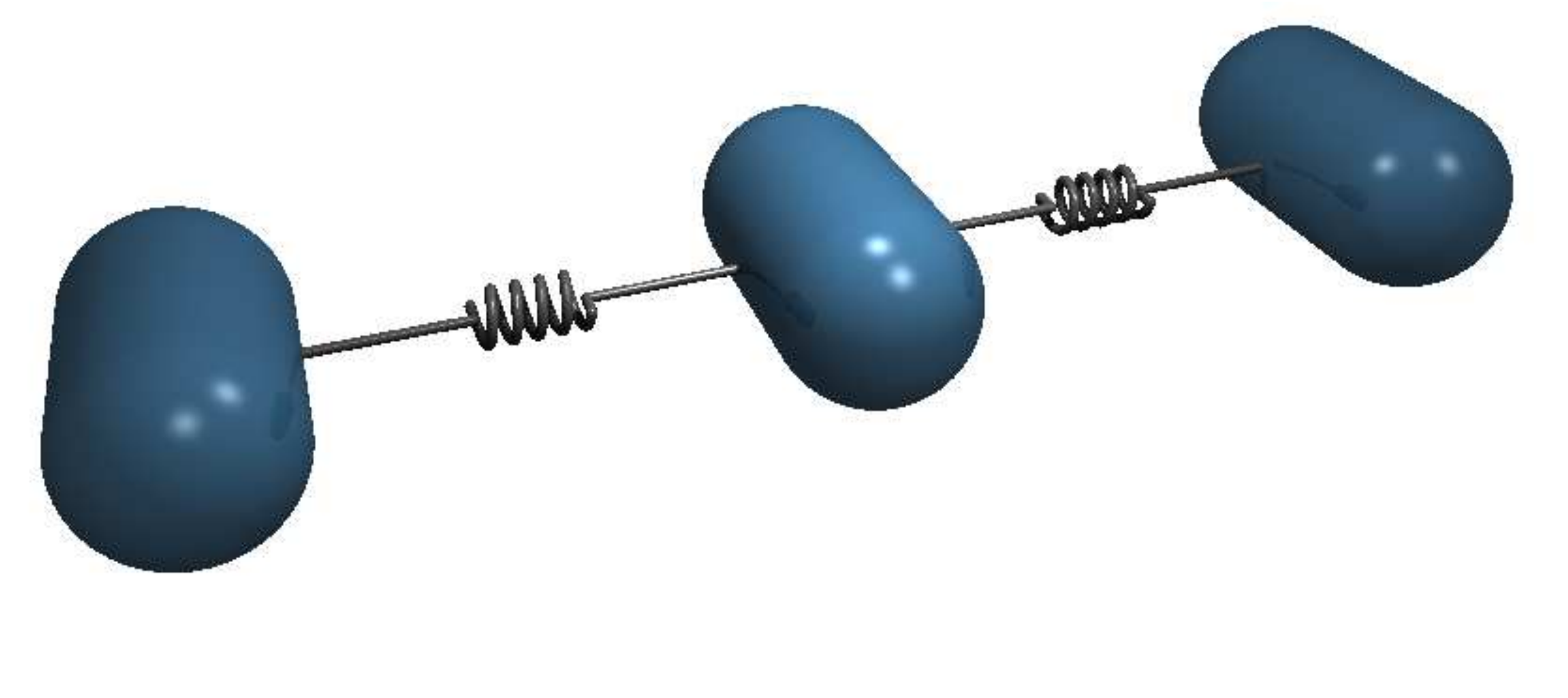}}}  & 56 & 16 & 3.72 & 0.056 & 0.044 & 0.063\\
\hline
{(l)}&\parbox[2cm][1cm][c]{2cm}{\centering{\includegraphics[height=0.8cm]{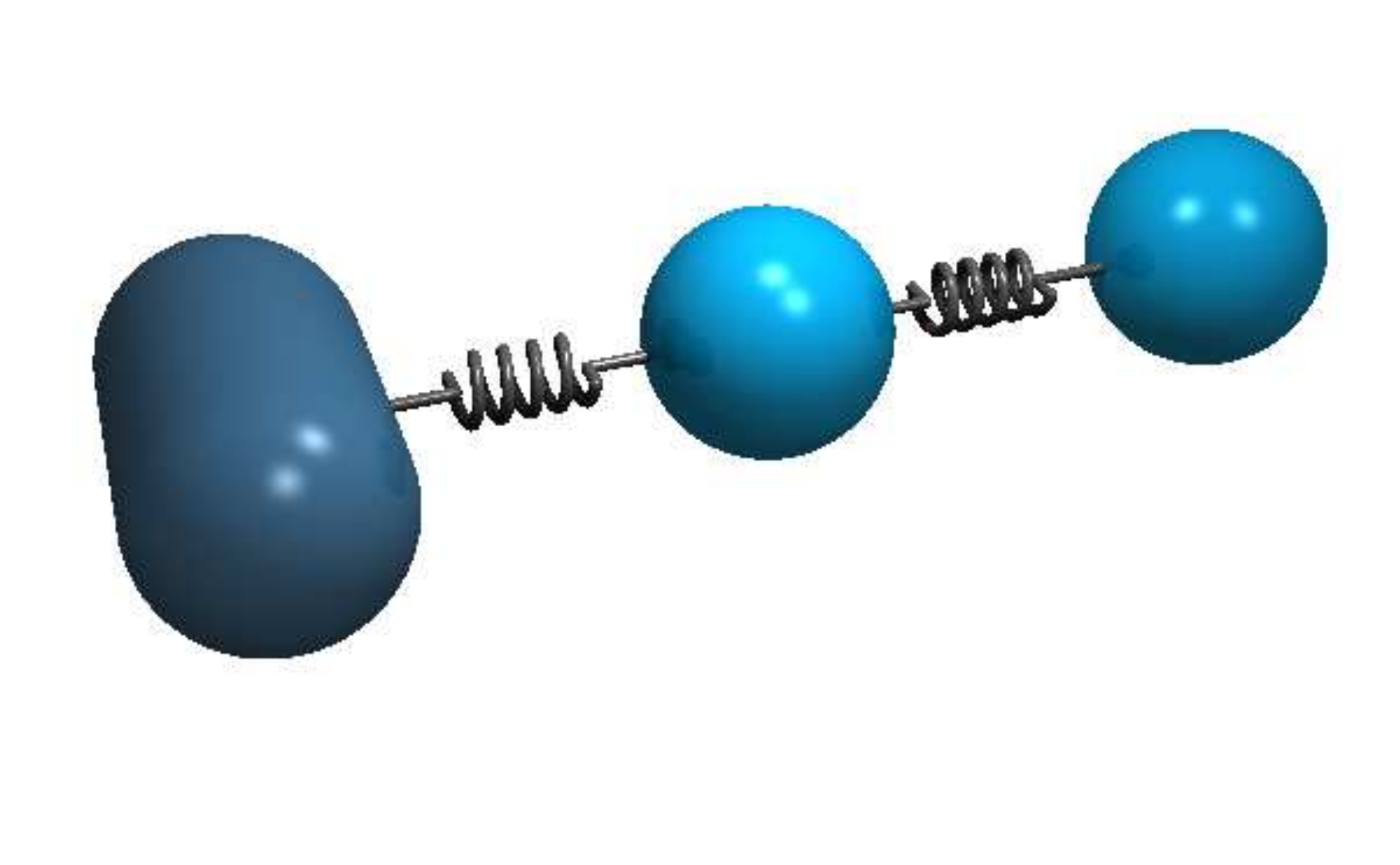}}}  & 40 & 16 & 4.86 & 0.062 & 0.041 & 0.085\\
\hline
{(m)}&\parbox[2cm][1cm][c]{2cm}{\centering{\includegraphics[height=0.8cm]{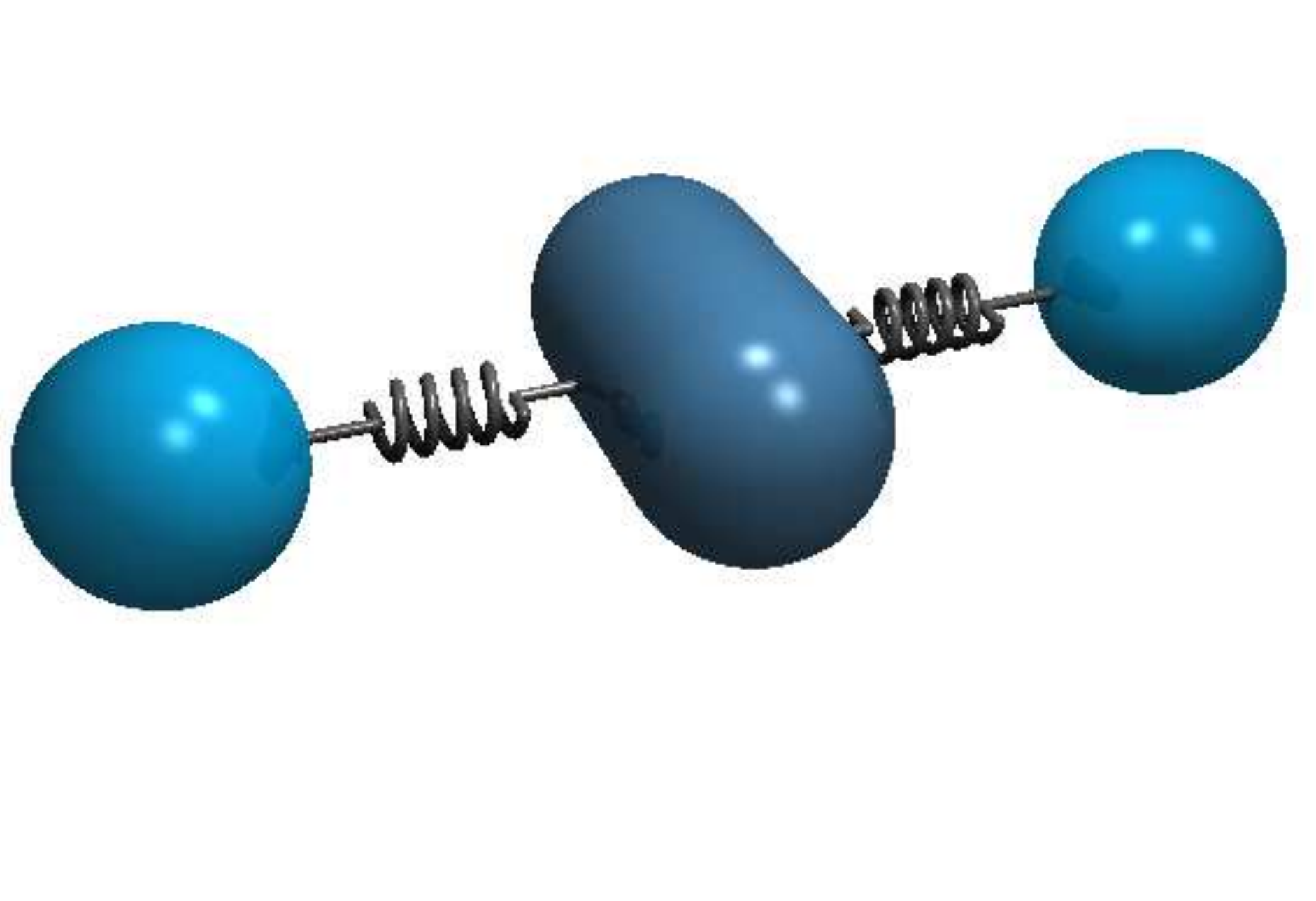}}}  & 40 & 16 & 4.54 & 0.045 & 0.063 & 0.078\\
\hline
{(n)}&\parbox[2cm][1cm][c]{2cm}{\centering{\includegraphics[height=0.8cm]{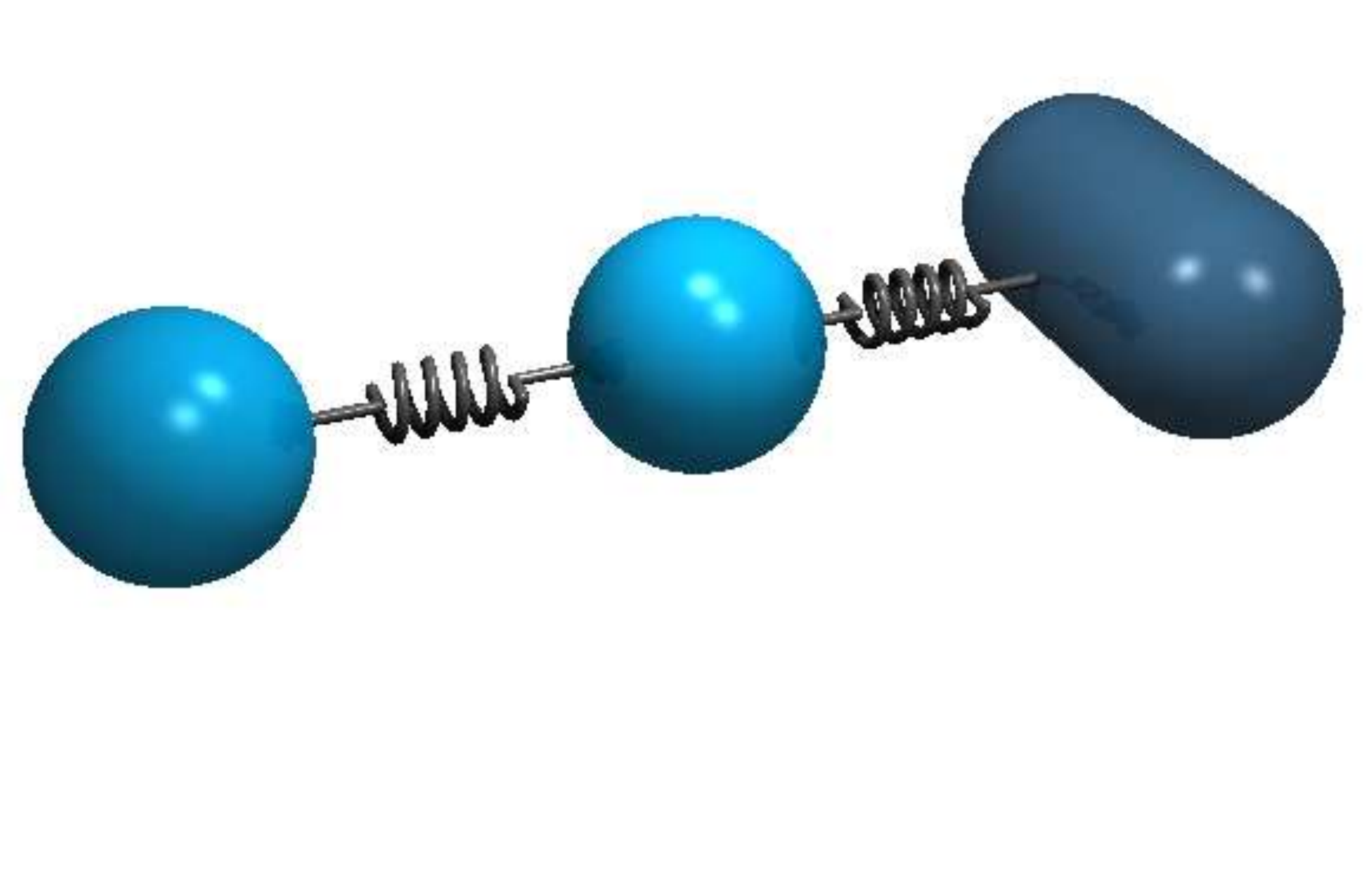}}}  & 40 & 16 & 4.64 & 0.078 & 0.058 & 0.050 \\
\hline
{(o)}&\parbox[2cm][1cm][c]{2cm}{\centering{\includegraphics[height=0.8cm]{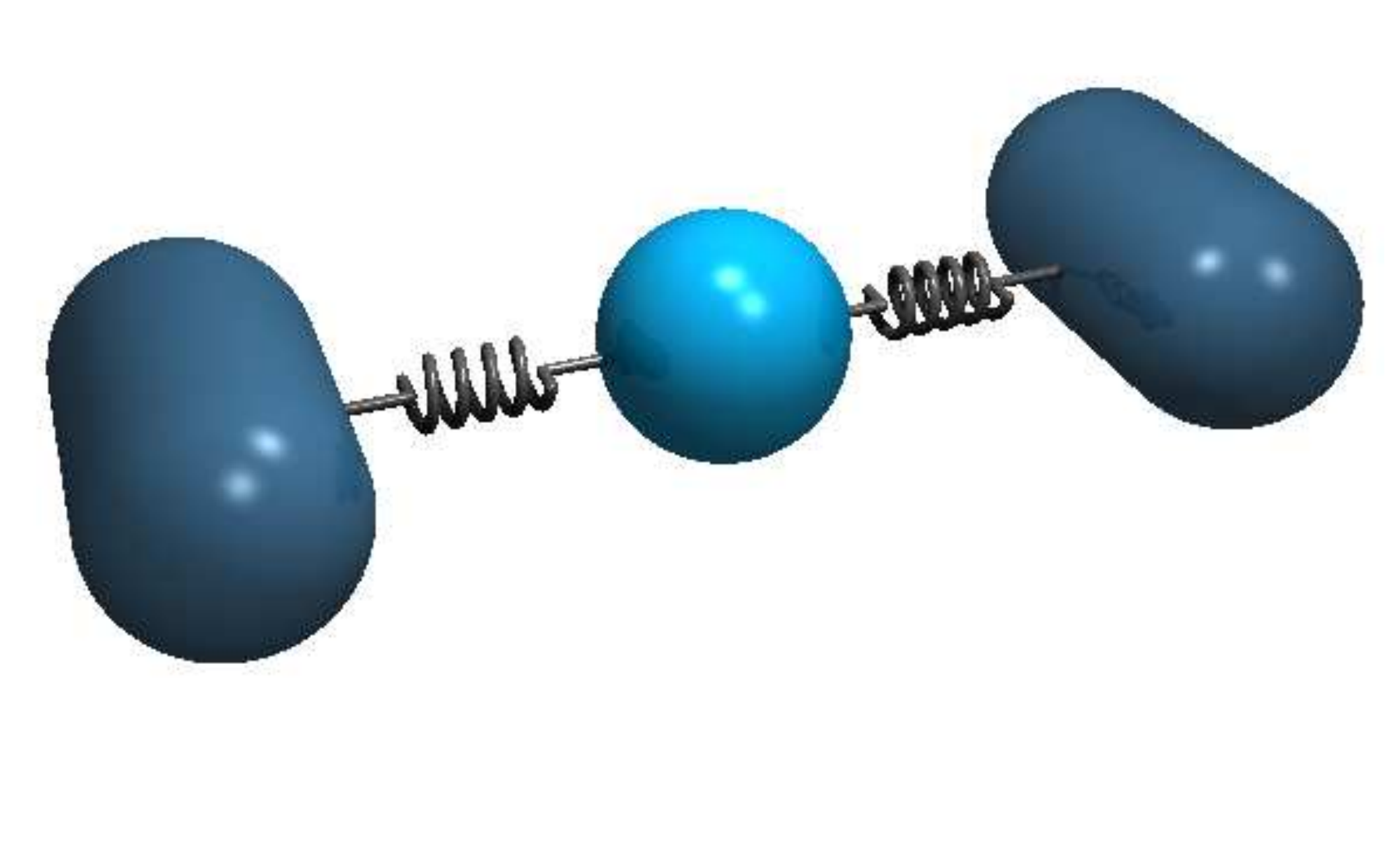}}}  & 40 & 16 & 3.64 & 0.074 & 0.041 & 0.054\\
\hline
{(p)}&\parbox[2cm][1cm][c]{2cm}{\centering{\includegraphics[height=0.8cm]{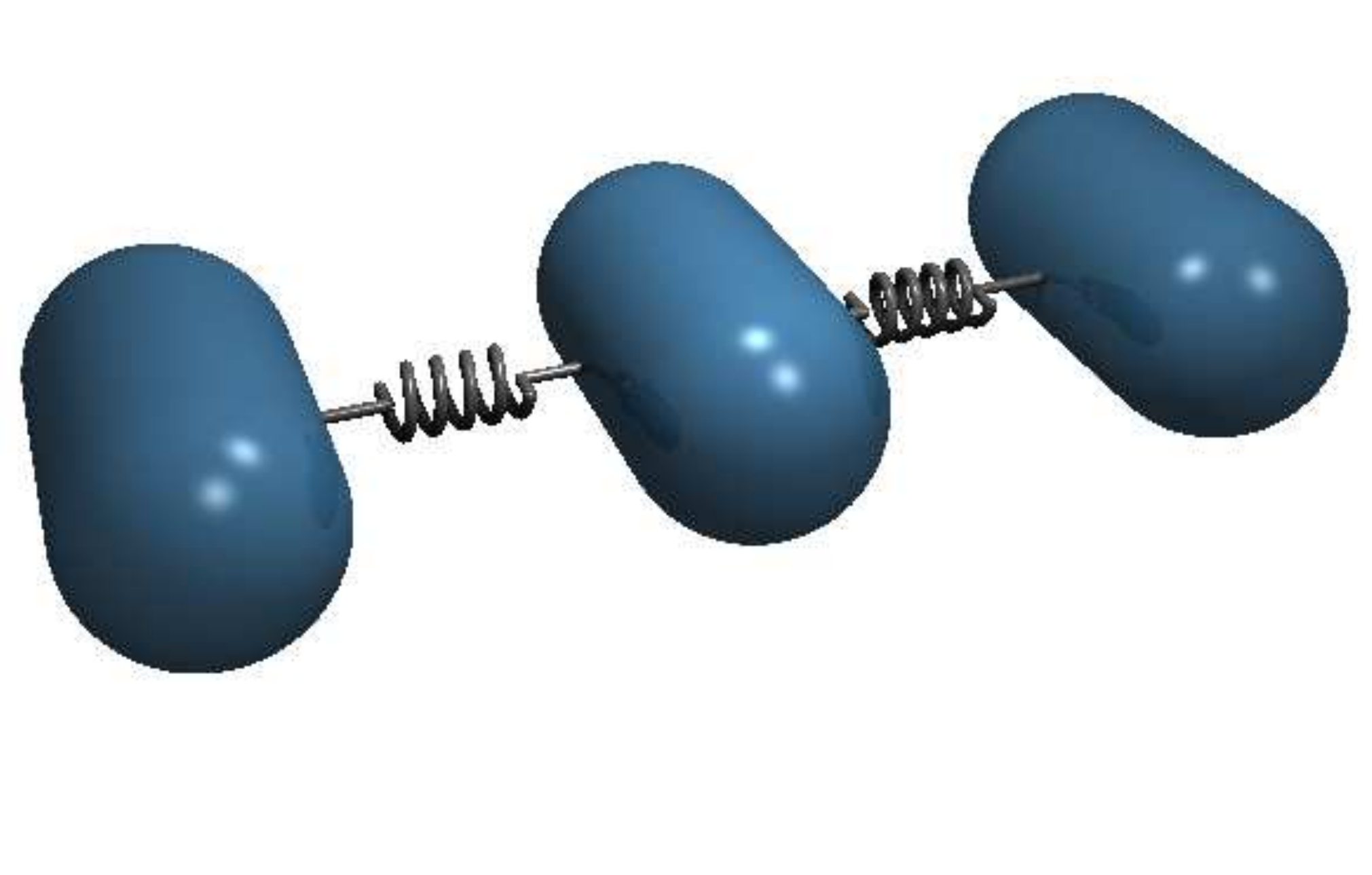}}}  & 40 & 16 & 3.53 & 0.051 & 0.041 & 0.053\\
\hline
{(q)}&\parbox[2cm][1cm][c]{2cm}{\centering{\includegraphics[height=0.8cm]{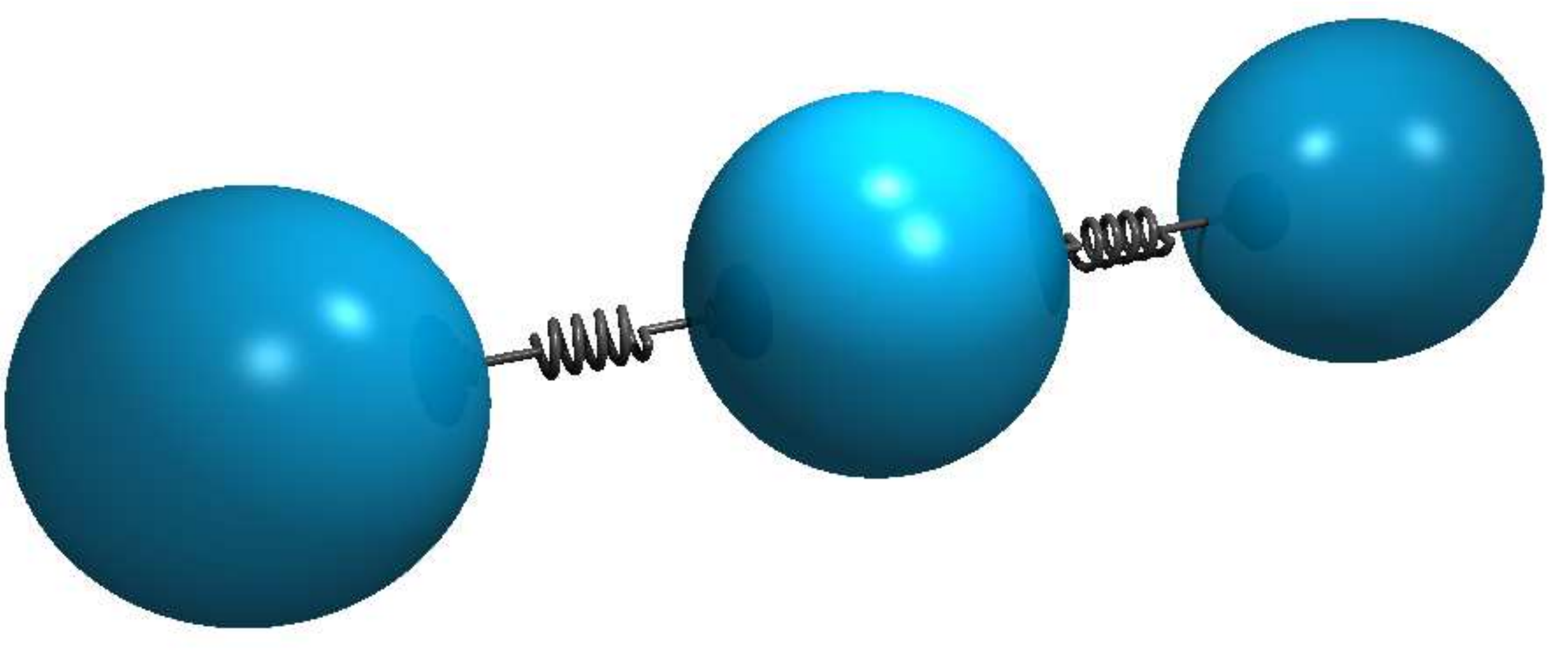}}}  & 64 & 16 & 2.33 & 0.080 & 0.061 & 0.083\\
\hline
\end{tabular}
\caption{{Parameters of the different swimmers. $Re^{B_i}$ denotes the Reynolds number of the $i$th body in each swimmer, while $Re_{\text{swim}}$ indicates the Reynolds number of the whole swimmer.}}
\label{tab:designs}
\end{table}

\subsection{Stability}
\label{subsec:stability}
We show that our choice of parameters results in our swimmers being in the regime of low Reynolds numbers \cite{Purcell:1977:LALRN}.
The latter is given by
\begin{equation}\label{eq:Re}
   Re = \frac { u \, l}{\nu} \ll 1,
\end{equation}
with the kinematic viscosity of the fluid $\nu$, the characteristic velocity $u$ and the characteristic length-scale $l$.

In order to simulate our swimmer in a stable regime, we define two different Reynolds numbers. The first is the Reynolds number of the whole swimming device (Table~\ref{tab:designs}, fourth column), defined as
\begin{equation}\label{eq:ReSwimmer}
Re_\text{swim} = \frac{u_\text{swim} \, l_\text{swim}}{\nu}\ll 1
\end{equation}
with respect to the characteristic velocity of the swimmer $u_\text{swim}$. The latter is calculated from the movement $\Delta$ of the central body in the swimmer during one swimming cycle (here chosen to be the last full cycle), divided by the time-period $T$, i.e.~$u_\text{swim} = \Delta / T$. The characteristic length-scale $l_\text{swim}$ of the different swimming devices is the rest length in the direction of motion for all the different swimmer designs (Table~\ref{tab:designs}, second column).

Additionally, we verify the Reynolds number not only for the whole swimmer but also for each of the bodies (Table~\ref{tab:designs}, third, fourth and fifth column), which may move with considerably higher velocities than the swimming device.
Here, the Reynolds number of the body $B_i$ is
\begin{equation}\label{eq:ReBody}
Re^{B_i} = \frac{u^{B_i} \, l^{B_i}}{\nu} \ll 1,
\end{equation}
and $u^{B_i}$ is the maximum velocity of the body $B_i$ and $l^{B_i}$ is its length in the direction of motion. As Table~\ref{tab:designs} confirms, all the Reynolds numbers satisfy the $Re \ll 1$ condition.

\subsection{Swimming velocities}

In the following, we investigate the performance of each swimmer as a function of its design geometry, as also detailed in Pickl et al.~\cite{Pickl:2011:ESES}. Specifically, we study the dependence of the amplitudes of the oscillations of the bodies in each swimmer on their shapes, and relate this to the overall swimming velocity.

\subsubsection{The three-sphere swimmers}
\label{sec:sphere_swimmers}

\begin{figure}[htb]
\centering
    \includegraphics[width=14cm]{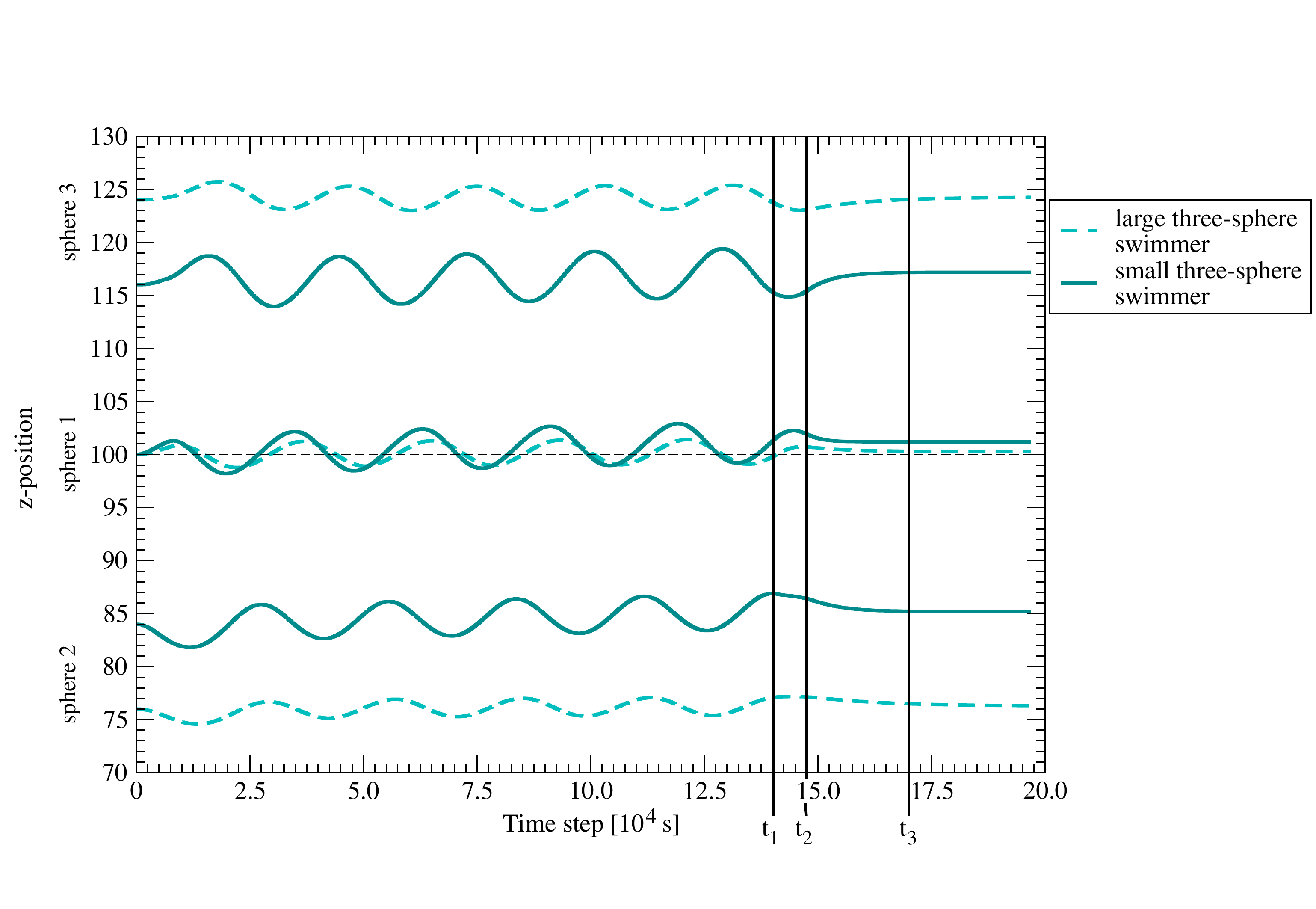}
\caption{Trajectories of each sphere in the two three-sphere swimmers. The dashed curves and the solid curves show the trajectories of the spheres in the large and the small swimmer, respectively. The simulation is run for 196812 time steps, while the driving force on the left body is switched off at $t_1=140580$ time steps, and on the right and hence also on the middle body at $t_2=147609$ time steps. The springs obtain their starting configuration at $t_3\approx179000$ time steps.}
\label{fig:traj}
\end{figure}

First we look at the two swimmers which consist solely of spheres. Table~\ref{tab:spheres} shows the velocity of each swimmer obtained from the simulations ($u_\text{swim}$). We find that the swimmer consisting of smaller spheres moves significantly faster than the one with larger spheres. Figure~\ref{fig:traj} shows the plot of the simulation trajectories of each body in the two swimmers.

\begin{table}
\centering
\begin{tabular}{|cc|c|c|c|c|}
\hline
\multicolumn{2}{|c|}{Design} & \parbox[0cm][0.6cm][c]{0cm}{}$u_\text{swim} [ 10^{-6} ] $ & $u_\text{GA} [ 10^{-6}] $ &  Error & Efficiency $[ 10^{-4} ]$ \\  
\hline
{(q)}&\parbox[2cm][1cm][c]{2cm}{\centering{\includegraphics[height=0.8cm]{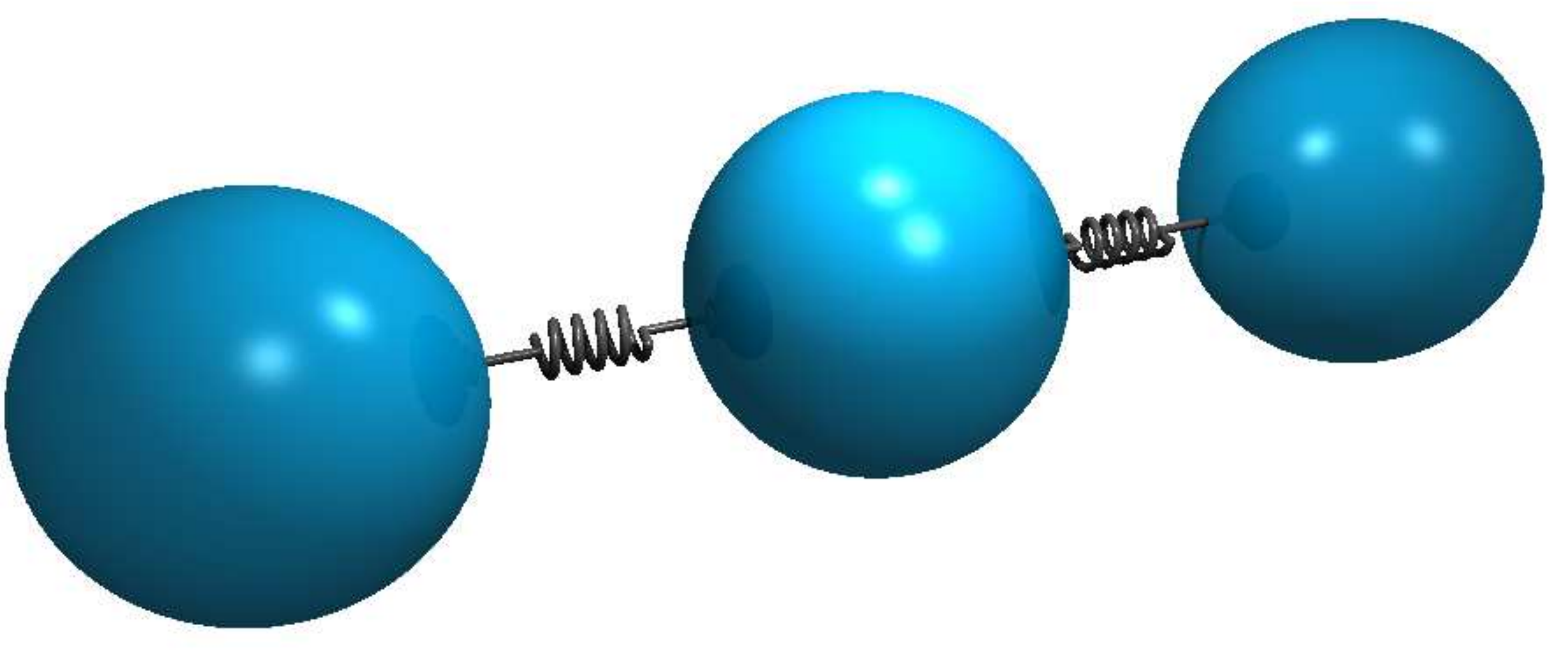}}} & 2.02 & 2.45 & $< 18 \%$ & 0.88\\
\hline
{(a)}&\parbox[2cm][1cm][c]{2cm}{\centering{\includegraphics[height=0.8cm]{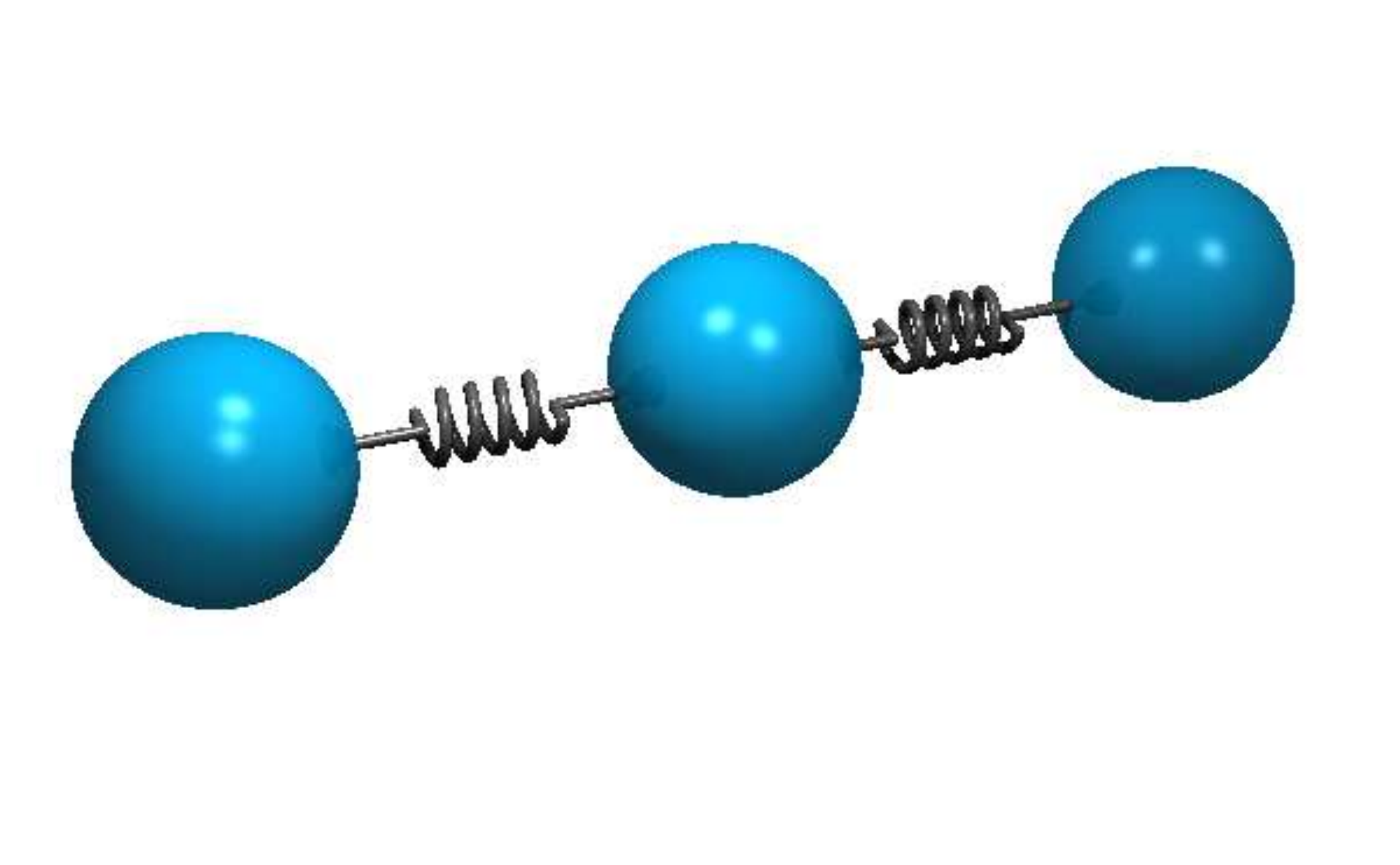}}} & 9.05 & 8.98 & $<  1 \%$ & 10.44\\
\hline
\end{tabular}
\caption{{Comparison of swimming velocities from the simulations ($u_\text{swim}$) and from the prediction of Golestanian and Ajdari ($u_\text{GA}$), in the case of spheres. All the quantities are given in terms of their values on the lattice.}}
\label{tab:spheres}
\end{table}

In previous work, this design has been analytically modeled by Golestanian and Ajdari \cite{Golestanian:2008:ARS}, who predict the velocity of a swimmer (here called $u_\text{GA}$, Table~\ref{tab:spheres}, third column) if the amplitude of oscillation of its two arms is known. Here, an arm is defined as the distance between the centers of two bodies connected by a spring.  In their model, Golestanian and Ajdari assume that the deformation of each arm is given by
\begin{equation}\label{eq:arms}
l_i(t) = \bar{l}_i + d_i \cos(\omega t + \varphi_i), \,\,\,\,\,i = 1, 2
\end{equation}
Here $\bar{l}_i$ is the average length and $l_i(t)$ the instantaneous length of arm $i$, and $d_i$, $\omega$ and $\varphi_i$ are the amplitude, the frequency and the phase shift of the oscillation of arm $i$. Under these conditions, the formula for the average velocity of a three-sphere swimmer is
\begin{equation}\label{eq:V_GA}
u_\text{GA} = K\, d_1\, d_2\, \omega\, \sin(\varphi_1-\varphi_2)
\end{equation}
where $K$ is a geometrical factor given by
\begin{equation}\label{eq:K}
K = \frac{3 r_1 r_2 r_3}{2 (r_1+r_2+r_3)^2}\left [\frac{1}{\bar{l}_1^2}+\frac{1}{\bar{l}_2^2}-\frac{1}{(\bar{l}_1+\bar{l}_2)^2}\right ].
\end{equation}
Here, $r_1, r_2$ and $r_3$ are the radii of the three spheres. It is also assumed that $r_i \ll \bar{l}_j$ and $d_i \ll \bar{l}_j$, for all $i$ and $j$.

Since this model assumes a velocity protocol in contrast to our force protocol (i.e., we start off with known forces which are applied on the bodies whereas the model of Golestanian and Ajdari assumes known deformations of the arms, regardless of the forces that cause them), we first extract $\{d_i\}$ and $\{\varphi_i\}$ from the simulation trajectories (Table~\ref{tab:sphere_params}), and use them with the known values of $\{\bar{l}_i\}$, $\{r_i\}$ and $\omega$ to find $u_\text{GA}$ from equation (\ref{eq:V_GA}). To extract $\{d_i\}$ and $\{\varphi_i\}$, we fit the armlength trajectories for the last complete period of motion (from $t = 4 T$ to $t = 5 T$) with equation (\ref{eq:arms}).

\begin{table}
\centering
\begin{tabular}{|cc|c|c|c|c|}
\hline
\multicolumn{2}{|c|}{Design}  & $K$ & $d_1$ & $d_2$ & $\sin(\varphi_1-\varphi_2)$ \\  
\hline
{(q)}&\parbox[2cm][1cm][c]{2cm}{\centering{\includegraphics[height=0.8cm]{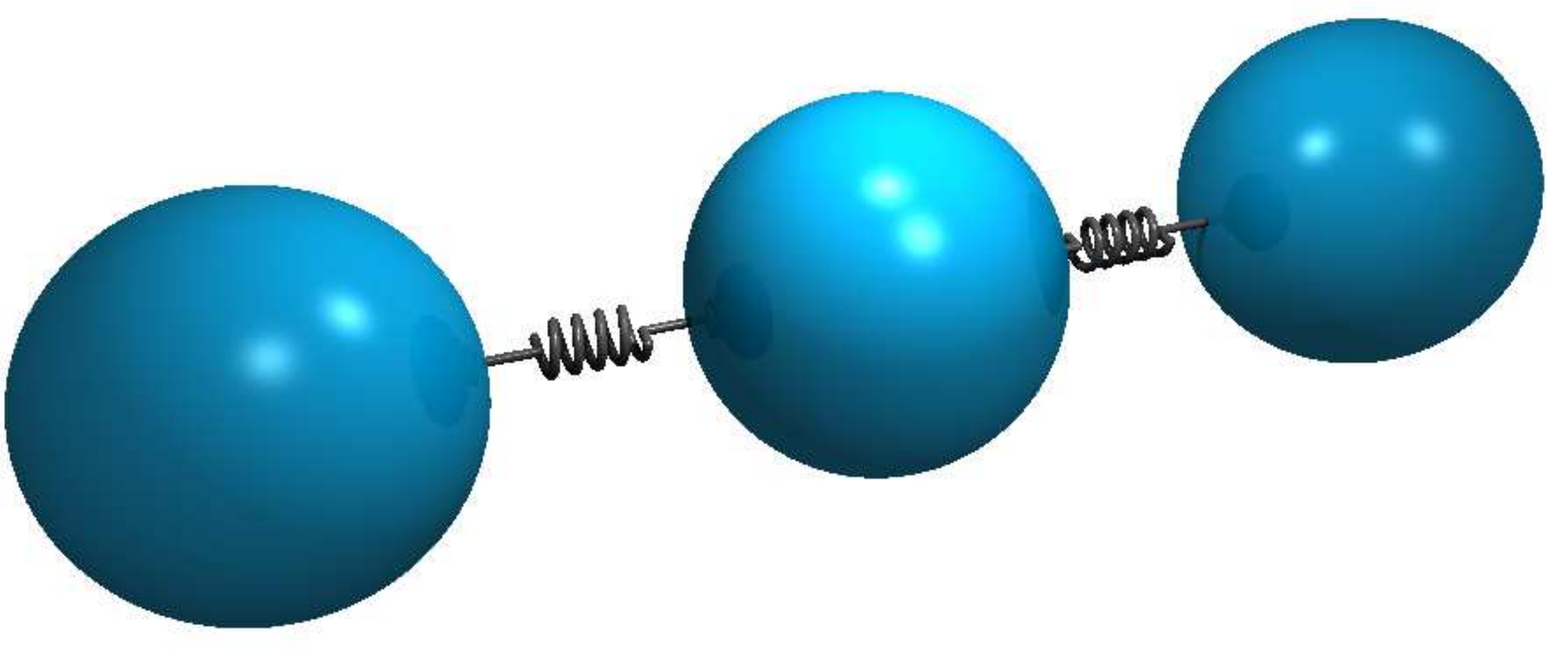}}} & 0.0040 & 1.59 & 2.07 & 0.83\\
\hline
{(a)}&\parbox[2cm][1cm][c]{2cm}{\centering{\includegraphics[height=0.8cm]{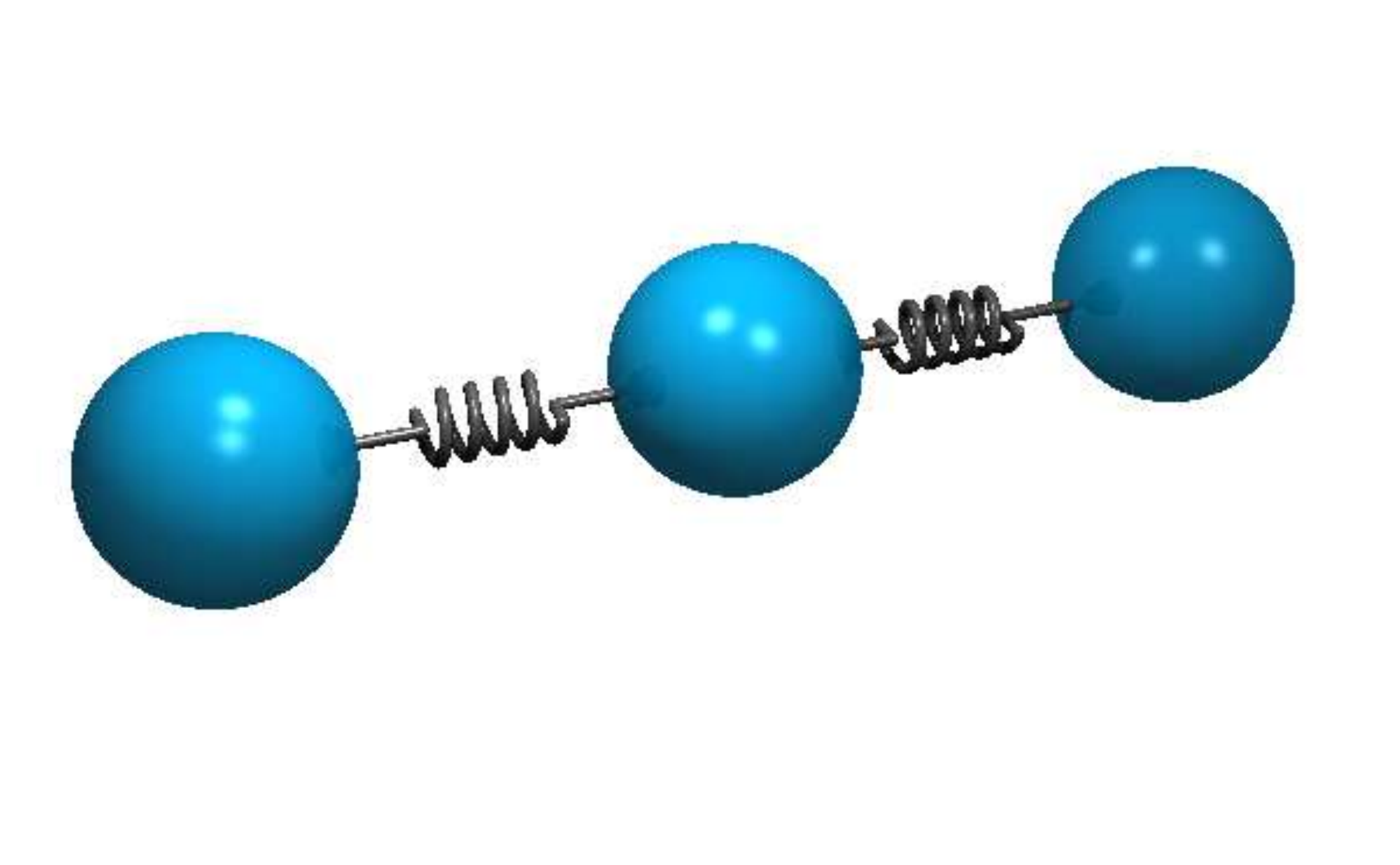}}} & 0.0045 & 2.53 & 3.80 & 0.94 \\
\hline
\end{tabular}
\caption{{Parameters for the three-sphere swimmers. $\omega$ equals 0.00022. All the quantities are given in terms of their values on the lattice.}}
\label{tab:sphere_params}
\end{table}

We define the error in $u_\text{swim}$, when compared with the theoretically-expected $u_\text{GA}$ value, as
\begin{equation}
\text{Error} = \left |\dfrac{u_\text{GA} - u_\text{swim}}{u_\text{GA}}\right | \times 100.
\end{equation}
For the swimmer with large spheres, this error comes out to be 18\%. This is possibly explained by the effect of the boundaries of the simulation box on the spheres of the larger swimmer. The theoretical result assumes infinite fluid, whereas the simulations are conducted in a finite box of the dimensions stated above. Another possible reason for the disagreement in the two velocities for the large-spheres swimmer is that the condition $r_i \ll \bar{l}_j$ is not satisfied very well. For the swimmer with small spheres, the effect of the finiteness of the swimming domain is smaller, and the $r_i \ll \bar{l}_j$ condition is also satisfied to a greater extent. Consequently, we find that in this case there is excellent agreement between $u_\text{GA}$ and $u_\text{swim}$, to an error of less than 1\% (Table~\ref{tab:spheres}). We also observe that for the swimmer with small spheres, the amplitude of oscillation of each arm is larger than that of the corresponding arm for the larger swimmer (Table~\ref{tab:sphere_params}, third column)), and that for each swimmer, the amplitude of oscillation of the leading arm ($d_2$) is greater than that of the trailing arm ($d_1$).

Table~\ref{tab:spheres} also shows the swimming efficiency of the two swimmers, as defined by Lighthill~\cite{Lighthill:1952:CPA},
\begin{equation}\label{eq:eff}
\text{Efficiency} = \frac{6\, \pi\, \eta\, r_\text{eff}\, u_\text{swim}^2}{\frac{1}{T}\int_{4T}^{5T} \sum_i \vec{F}_{\text{dri}}^{B_i}(t) \cdot \vec{u}^{B_i}(t)\,dt}.
\end{equation}
Here $r_\text{eff}$ is an effective hydrodynamic radius of the swimmer, and is approximated by the sum of the radii of the three spheres~\cite{Alouges:2009:EPJE}. $\vec{u}^{B_i}(t)$ is found from the simulation curves, using the last full swimming cycle (i.e., from $t = 4T$ to $t = 5T$). The integration is also done over this period, in order to work with the steady-state values. Table~\ref{tab:spheres} shows that the efficiency of the swimmer with small spheres is greater, as expected due to its significantly greater swimming velocity.

\subsubsection{Swimmers with capsules}
\label{sec:caps_swimmers}

The rest of the swimmers contain capsules, and can be separated into three distinct families. The first family consists of the swimmers labeled (b)-(f) in Table~\ref{tab:designs}. These contain capsules which are parallel to the swimming direction (henceforth referred to as parallel capsules), and have springs of a rest length of 8 lattice cells. The second family consists of the swimmers labeled (g)-(k) in Table~\ref{tab:designs}. These contain capsules which are perpendicular to the swimming direction (henceforth referred to as perpendicular capsules), and have springs of a rest length of 8, 12 or 16 lattice cells. The third family consists of the swimmers labeled (l)-(p) in Table~\ref{tab:designs}. These also contain perpendicular capsules, and their springs all have a rest length of 8, making these swimmers shorter than the swimmers in the second family.

Tables~\ref{tab:caps_family1}, \ref{tab:caps_family2} and \ref{tab:caps_family3} (second column) show the measured $u_\text{swim}$ for the swimmers with parallel capsules, with perpendicular capsules and longer springs, and with perpendicular capsules and shorter springs, respectively. In all the three tables, the swimmers are arranged in the increasing order of $u_\text{swim}$, from top-to-bottom. It can be observed that corresponding designs in each family occupy the same position in their respective tables. Here, two designs are said to be corresponding if one needs to replace the same spheres in a three-sphere swimmer to get both the designs, albeit by different kinds of capsules (and, possibly, with springs of different lengths). One also observes that the speed of swimming decreases with an increase in the number of capsules. This can be attributed to the fact that the small spheres have greater amplitudes of oscillation than the capsules, which means that the lengths of arms which contain spheres have greater amplitudes of oscillation than those that do not. A greater amplitude of oscillation of the arms results in a higher swimming velocity.

\begin{table}
\centering
\begin{tabular}{|cc|c|c|c|c|}
\hline
\multicolumn{2}{|c|}{Design} & \parbox[0cm][0.6cm][c]{0cm}{}$u_\text{swim} [ 10^{-6} ] $ & $u_\text{GA} [ 10^{-6} ] $ & {Error} & {Efficiency} $[ 10^{-4} ]$ \\  
\hline

{(f)}&\parbox[2cm][1cm][c]{2cm}{\centering{\includegraphics[height=0.8cm]{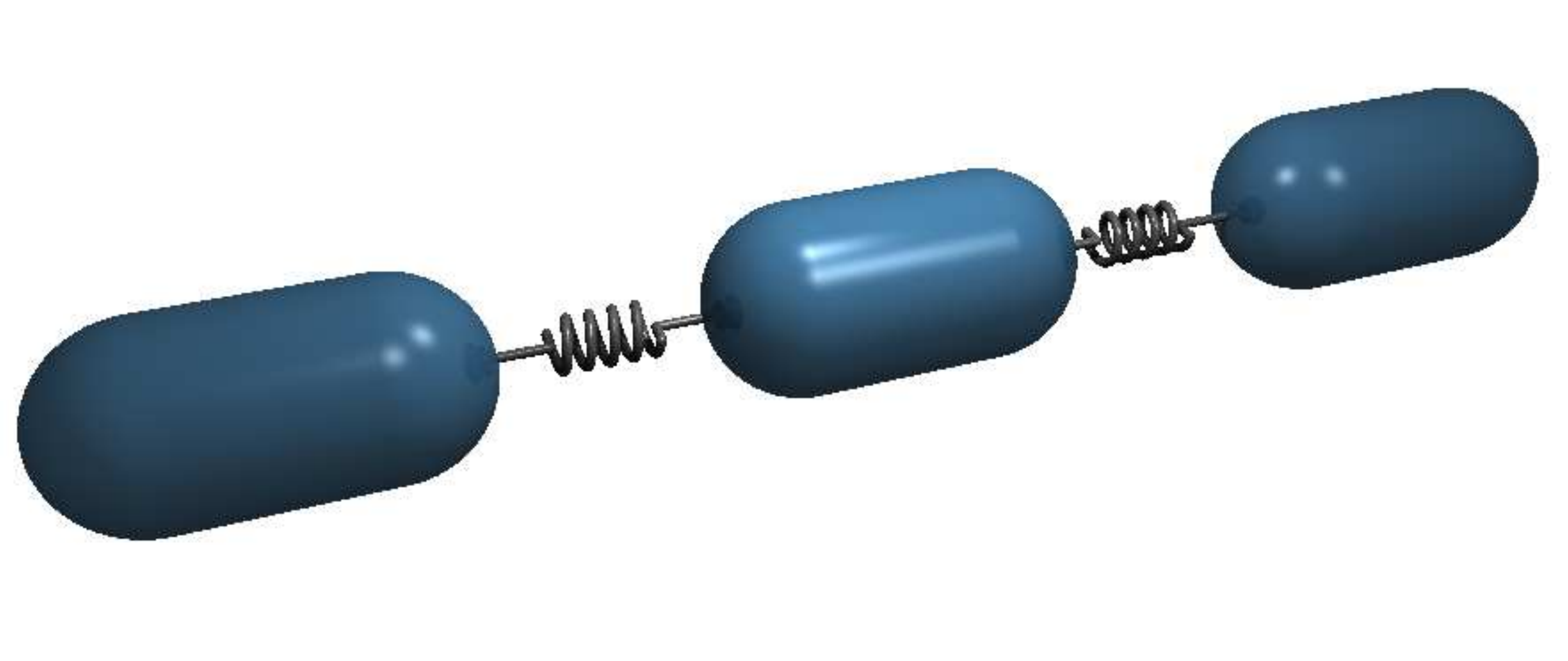}}}   & 4.78 & 4.50 & $< 7 \%$ & 3.69\\
\hline
{(e)}&\parbox[2cm][1cm][c]{2cm}{\centering{\includegraphics[height=0.8cm]{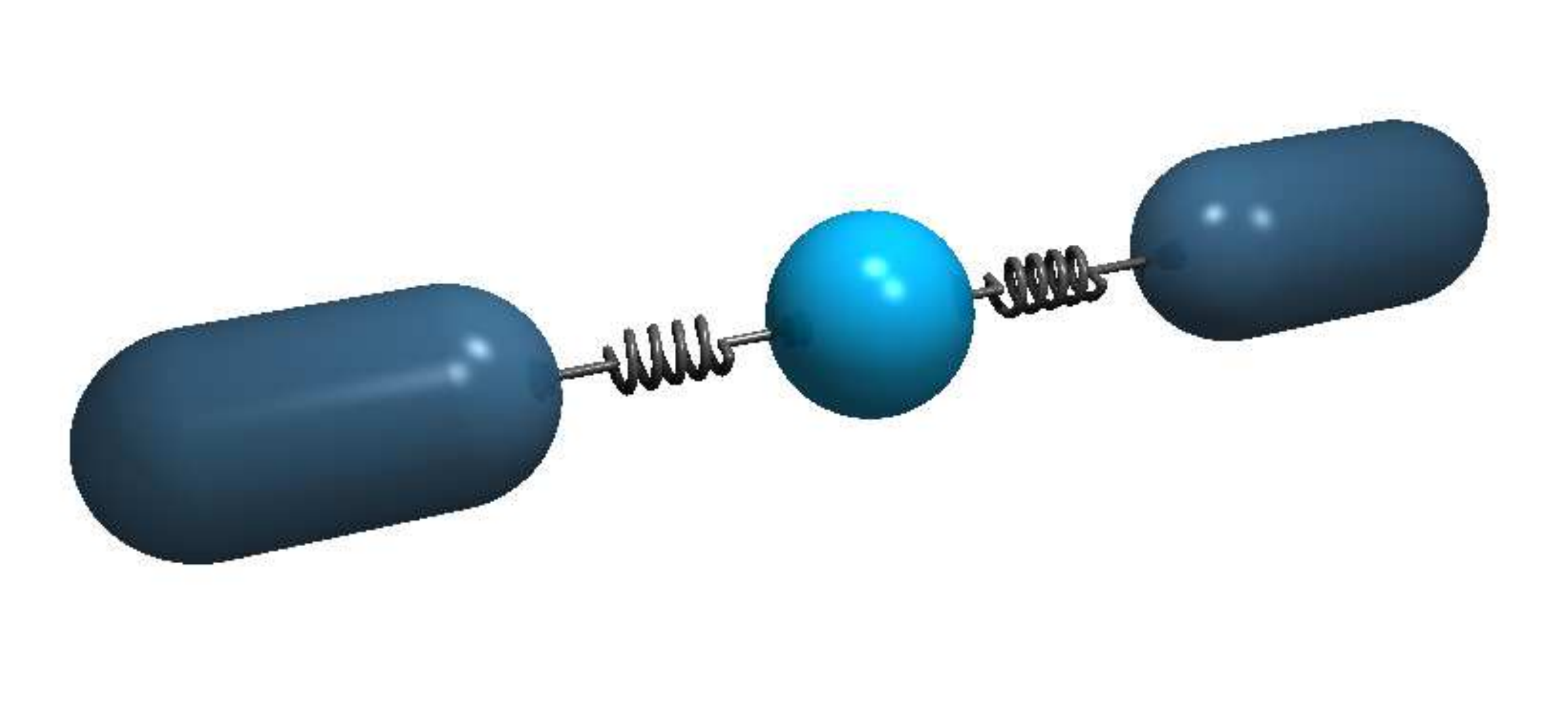}}}   & 5.68 & 5.48 & $< 4 \%$ & 5.31\\
\hline
{(c)}&\parbox[2cm][1cm][c]{2cm}{\centering{\includegraphics[height=0.8cm]{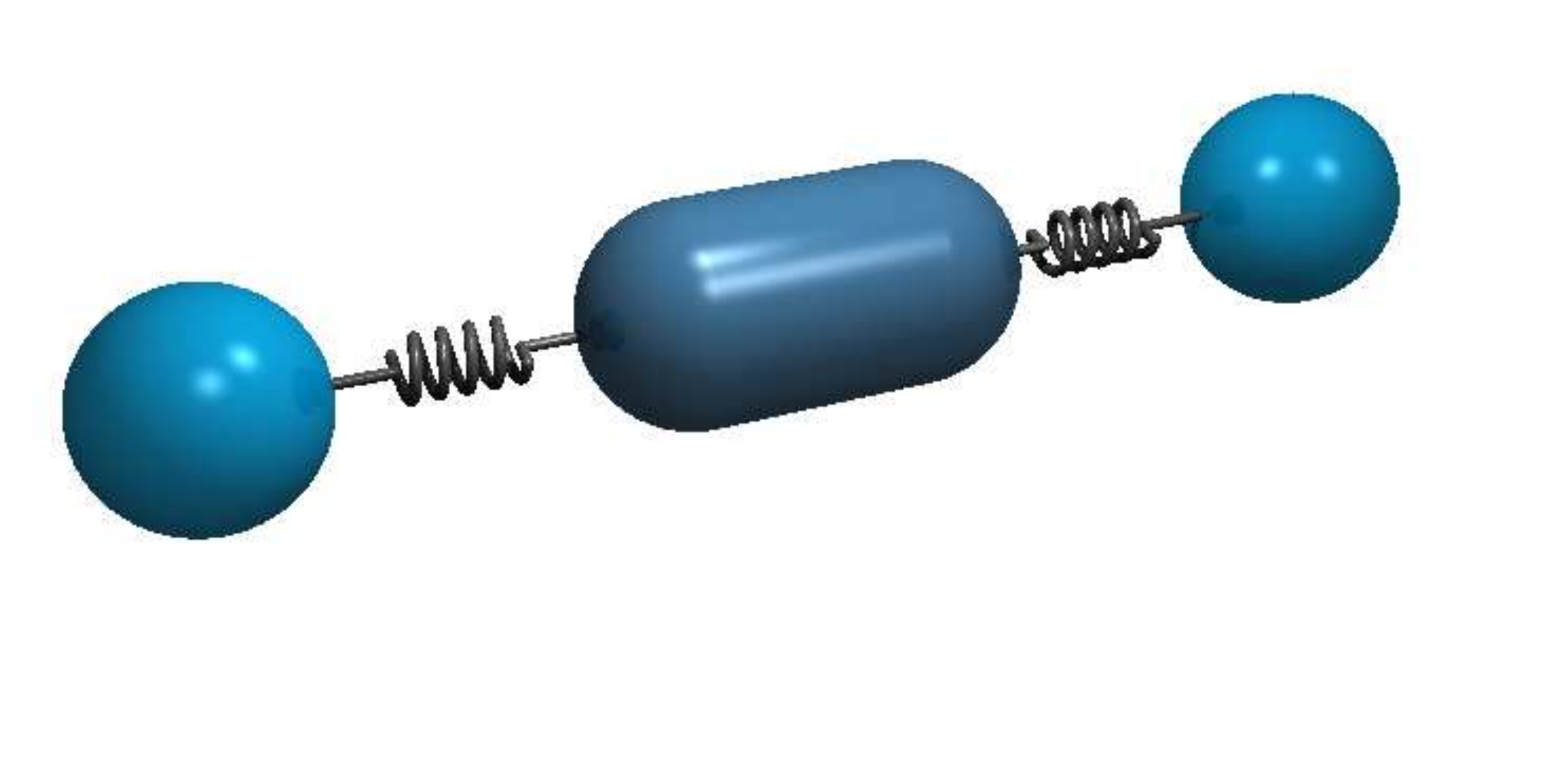}}}   & 6.66 & 6.31 & $< 6 \%$ & 5.86\\
\hline
{(d)}&\parbox[2cm][1cm][c]{2cm}{\centering{\includegraphics[height=0.8cm]{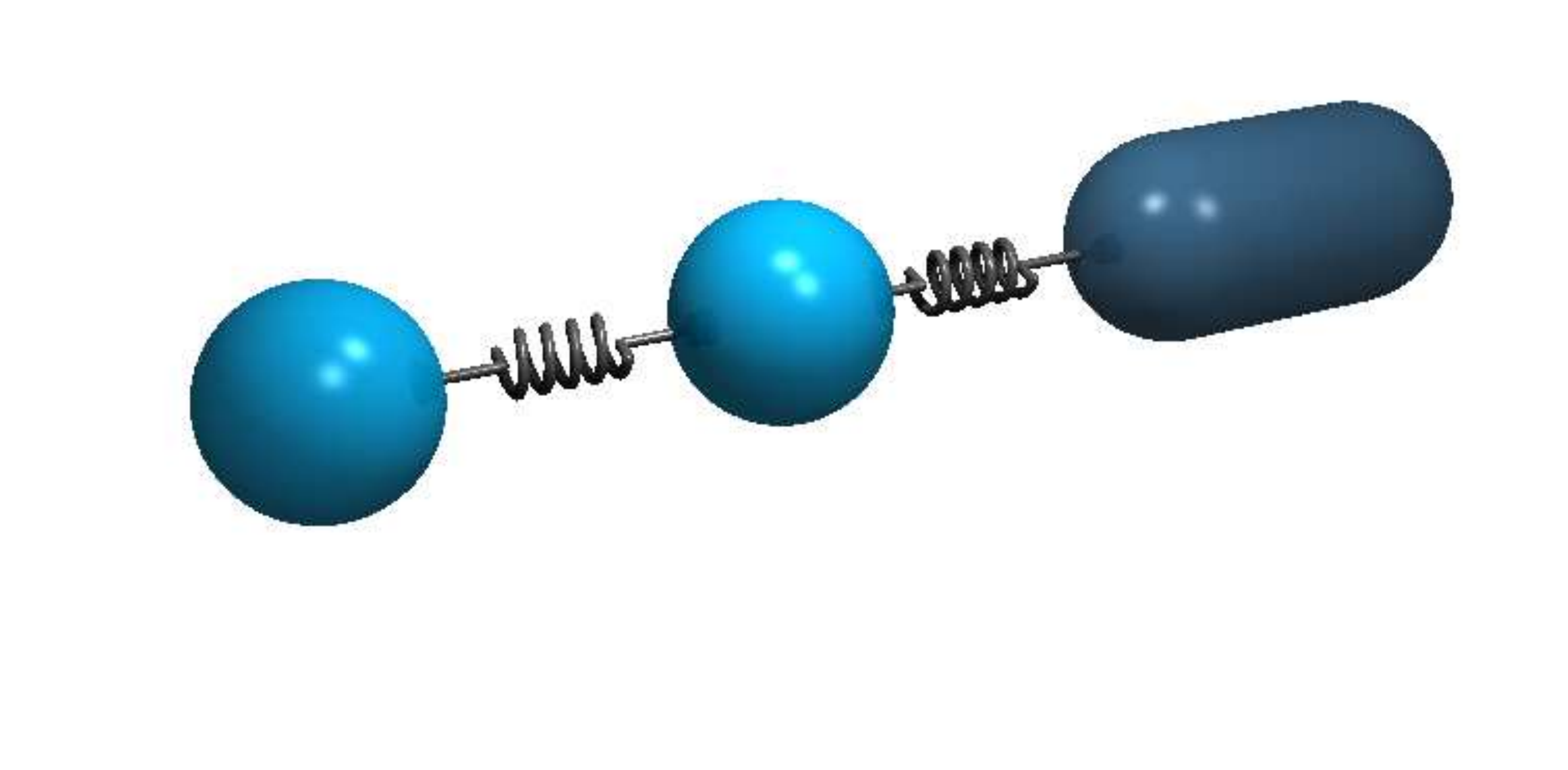}}} & 7.11 & 7.16 & $< 1 \%$ & 7.37\\
\hline
{(b)}&\parbox[2cm][1cm][c]{2cm}{\centering{\includegraphics[height=0.8cm]{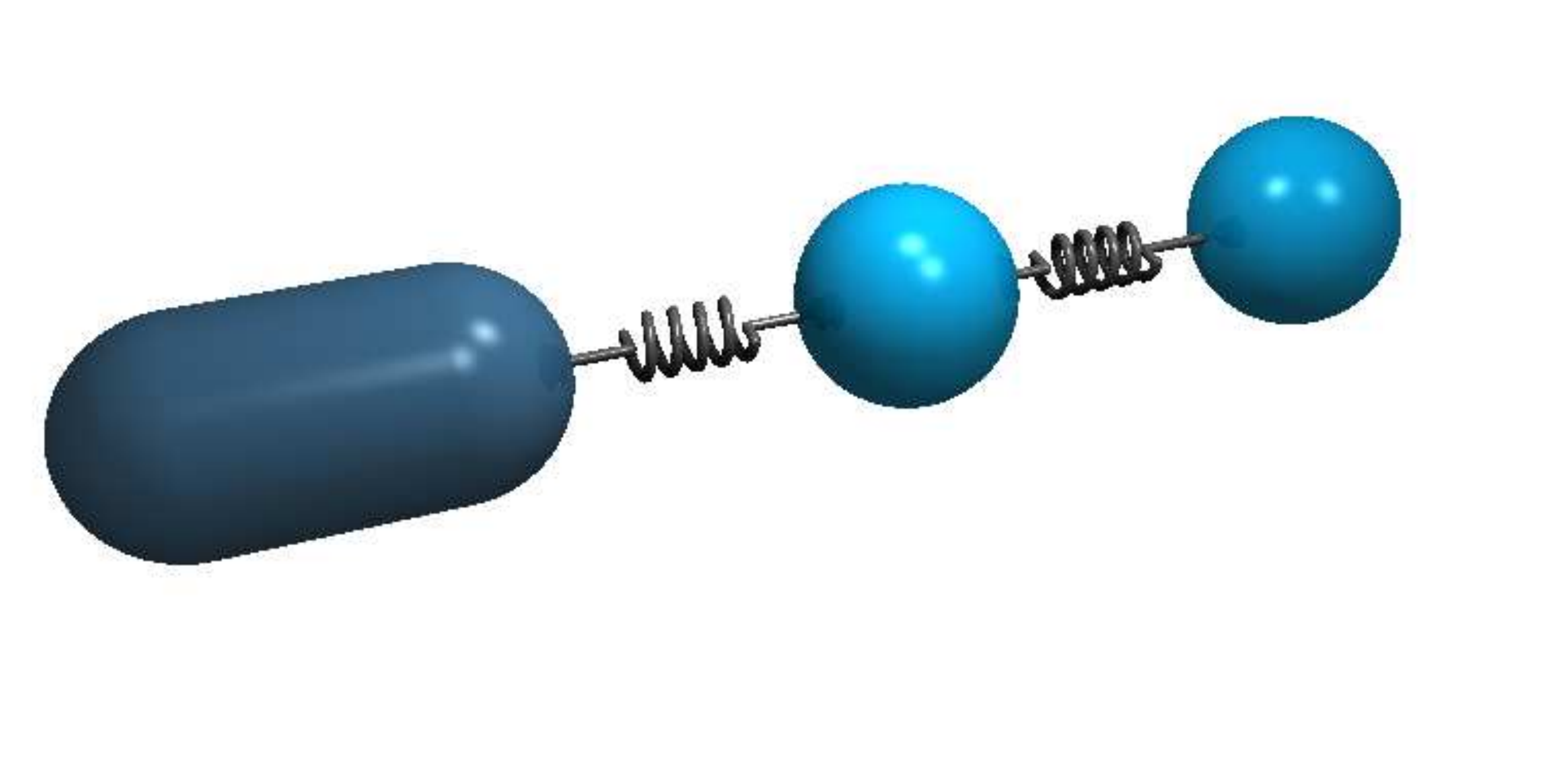}}} & 7.32 & 7.21 & $< 2 \%$ & 7.77\\
\hline
\end{tabular}
\caption{Comparison of swimming velocities from the simulations ($u_\text{swim}$) and from the prediction of Golestanian and Ajdari ($u_\text{GA}$), in the case of the swimmers with parallel capsules. All the quantities are given in terms of their values on the lattice.}
\label{tab:caps_family1}
\end{table}
\begin{table}
\centering
\begin{tabular}{|cc|c|c|c|c|}
\hline
\multicolumn{2}{|c|}{Design} & \parbox[0cm][0.6cm][c]{0cm}{}$u_\text{swim} [ 10^{-6} ] $ & $u_\text{GA} [ 10^{-6} ] $ & {Error} & {Efficiency} $[ 10^{-4} ]$ \\  
\hline
{(k)}&\parbox[2cm][1cm][c]{2cm}{\centering{\includegraphics[height=0.8cm]{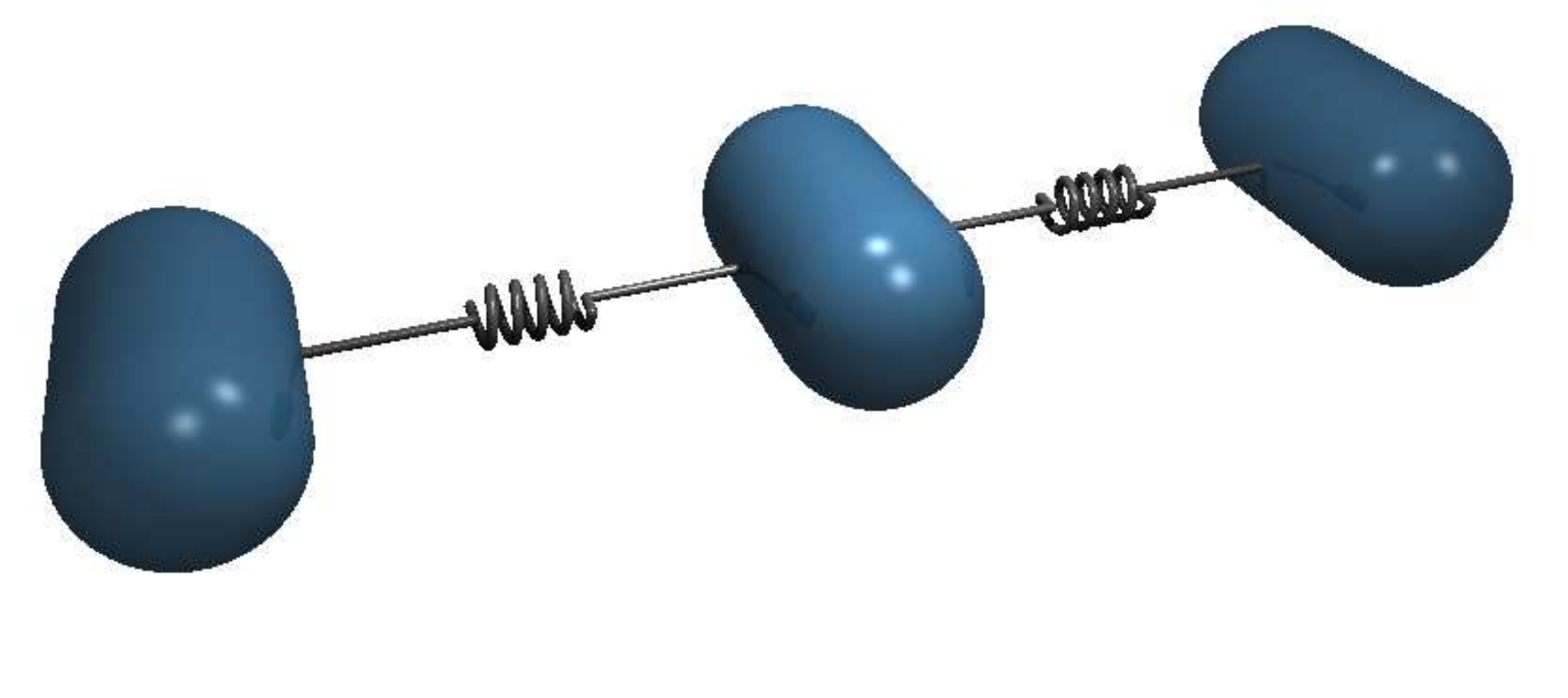}}} & 3.69 & 3.95 & $< 7 \%$ & 2.12\\
\hline
{(j)}&\parbox[2cm][1cm][c]{2cm}{\centering{\includegraphics[height=0.8cm]{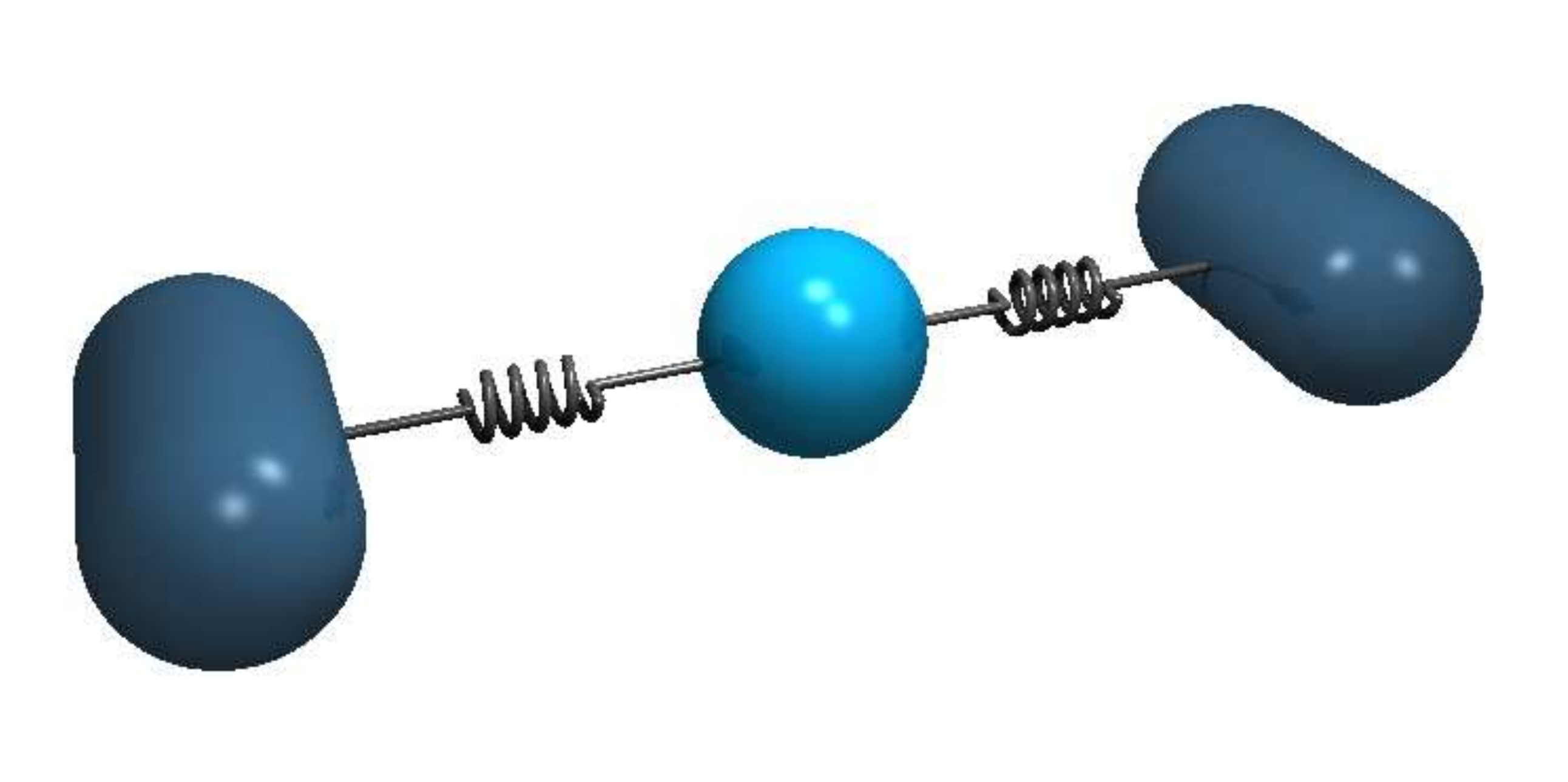}}} & 4.21 & 4.66 & $< 10 \%$ & 3.08\\
\hline
{(h)}&\parbox[2cm][1cm][c]{2cm}{\centering{\includegraphics[height=0.8cm]{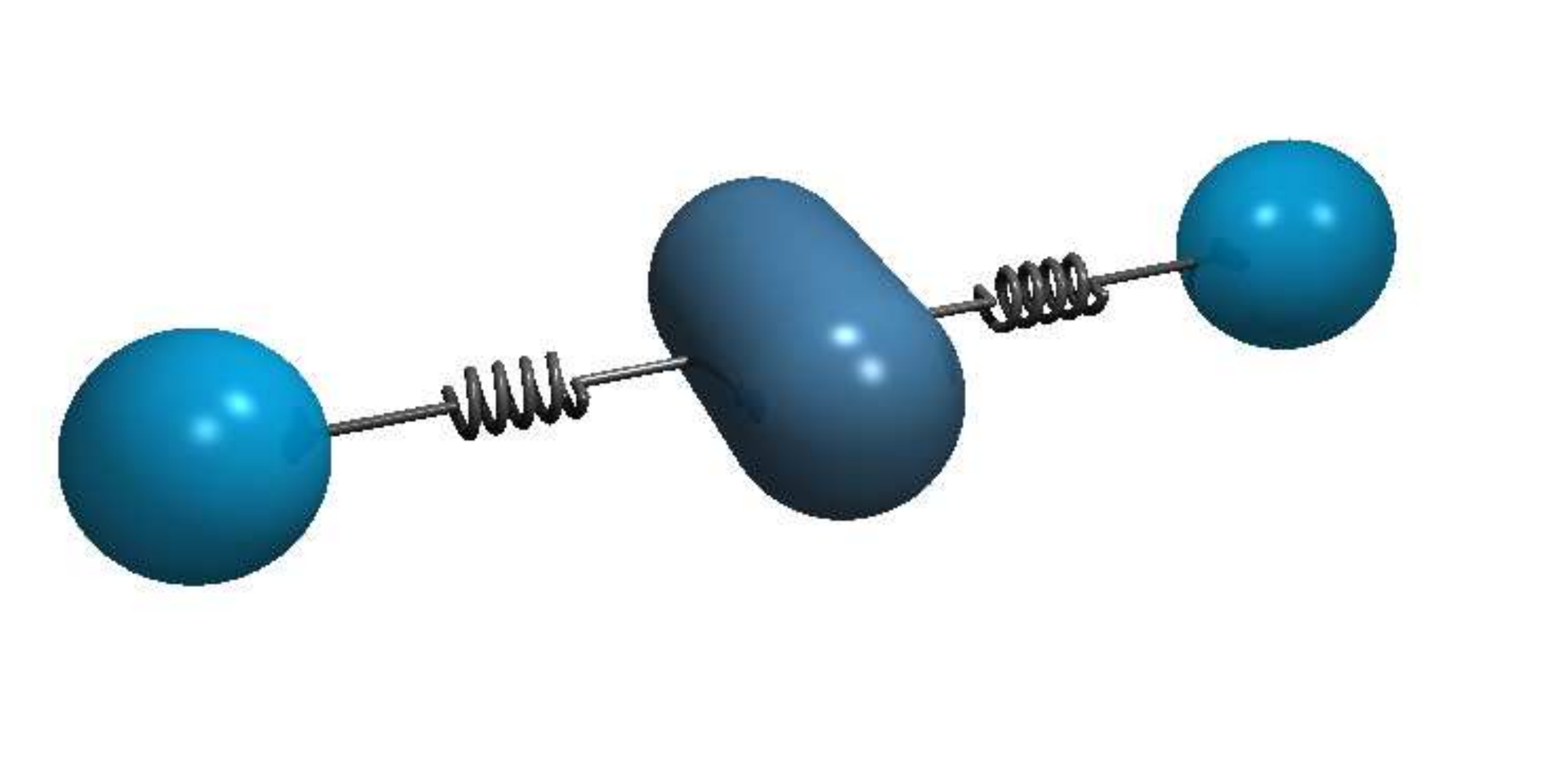}}} & 5.71 & 6.50 & $< 13 \%$ & 4.24\\
\hline
{(i)}&\parbox[2cm][1cm][c]{2cm}{\centering{\includegraphics[height=0.8cm]{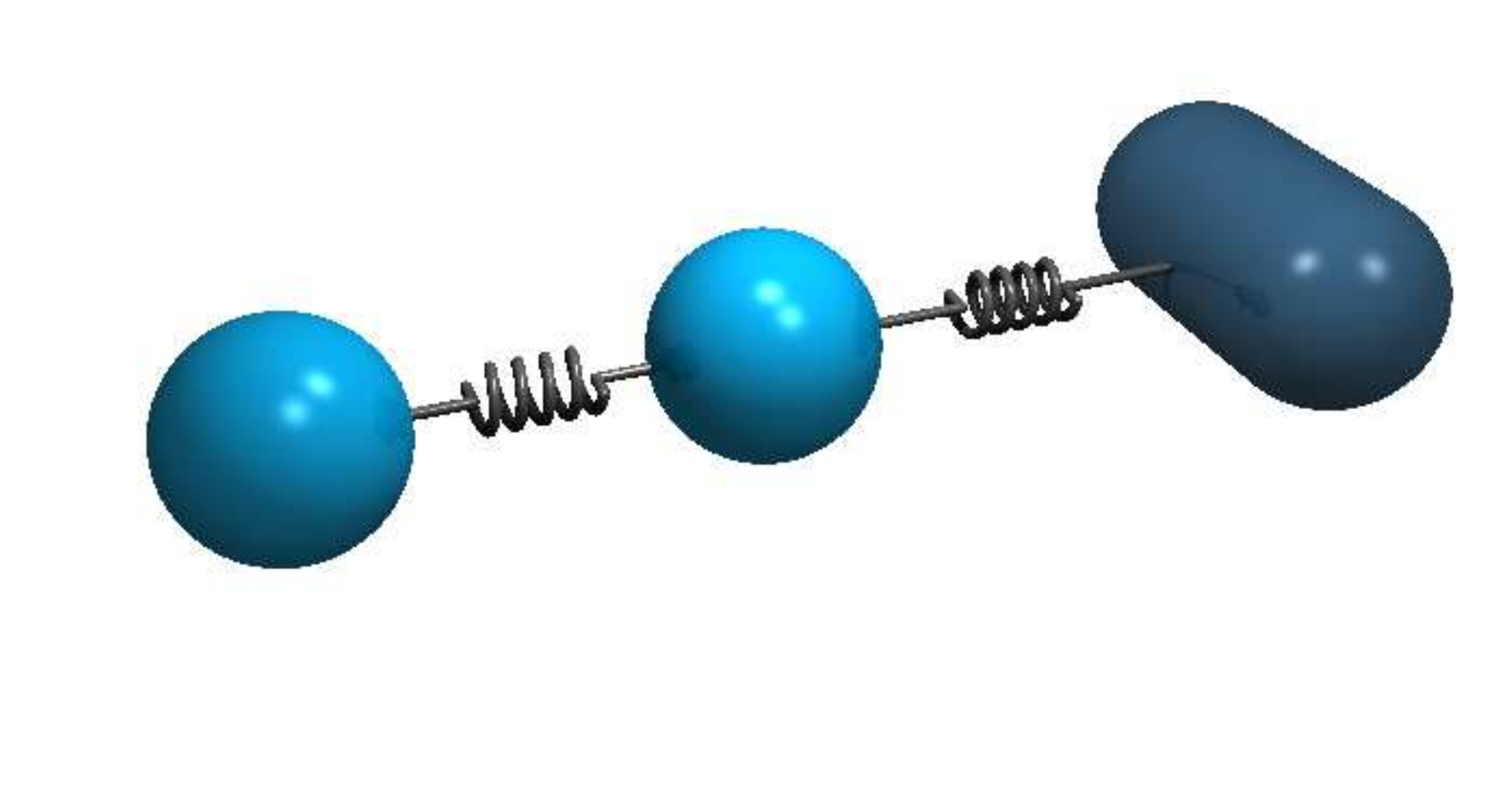}}} & 6.23 & 6.75 & $< 8 \%$ & 5.74\\
\hline
{(g)}&\parbox[2cm][1cm][c]{2cm}{\centering{\includegraphics[height=0.8cm]{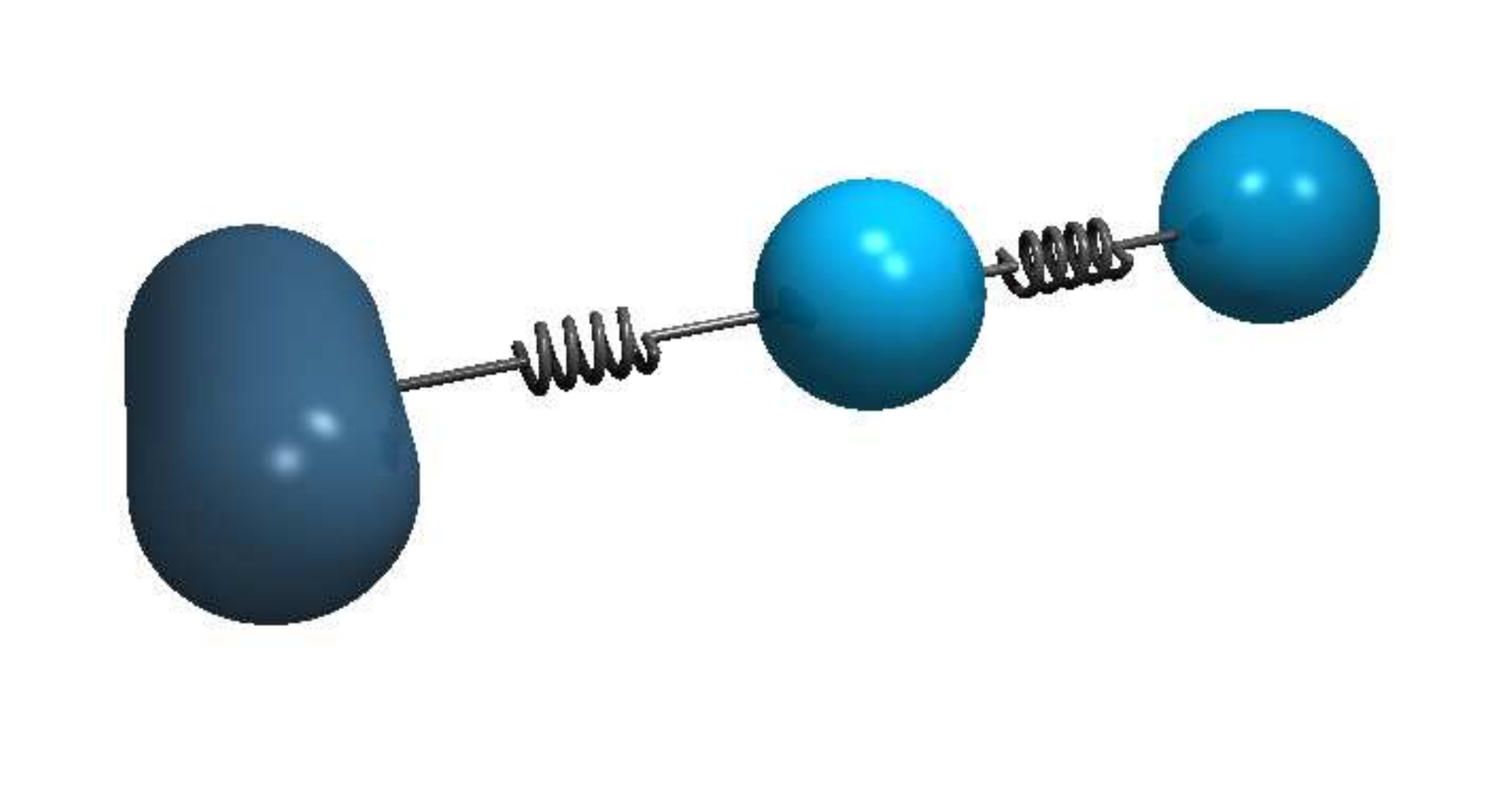}}} & 6.40 & 6.82 & $< 7 \%$ & 5.97\\
\hline
\end{tabular}
\caption{Comparison of swimming velocities from the simulations ($u_\text{swim}$) and from the prediction of Golestanian and Ajdari ($u_\text{GA}$), in the case of the second family of the swimmers with capsules. All the quantities are given in terms of their values on the lattice.}
\label{tab:caps_family2}
\end{table}

\begin{table}
\centering
\begin{tabular}{|cc|c|c|c|c|}
\hline
\multicolumn{2}{|c|}{Design} & \parbox[0cm][0.6cm][c]{0cm}{}$u_\text{swim} [ 10^{-6} ] $ & $u_\text{GA} [ 10^{-6} ] $ & {Error} & {Efficiency} $[ 10^{-4} ]$ \\  
\hline
{(p)}&\parbox[2cm][1cm][c]{2cm}{\centering{\includegraphics[height=0.8cm]{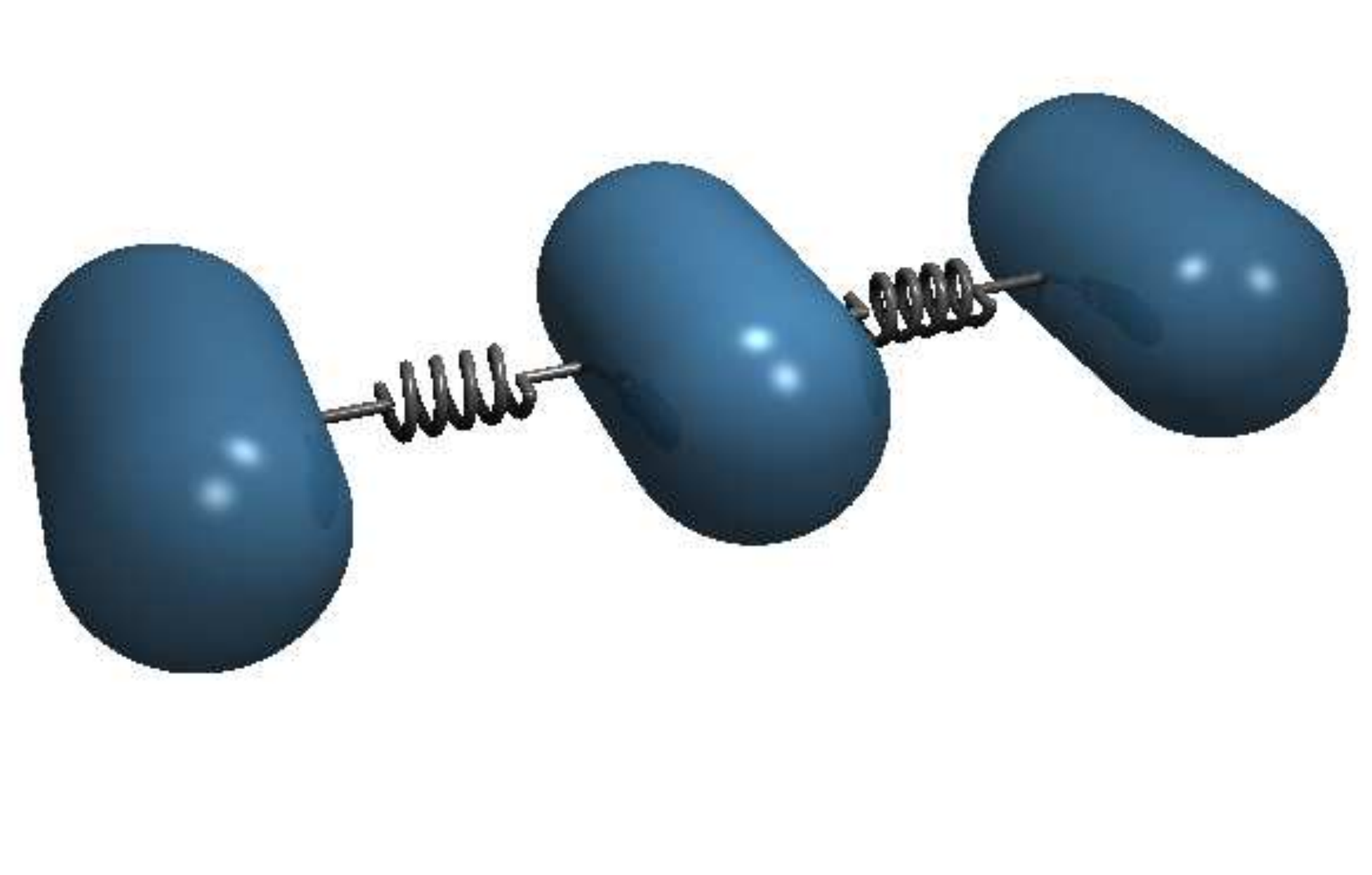}}} & 4.90 & 6.70 & $< 27 \%$ & 3.63\\
\hline
{(o)}&\parbox[2cm][1cm][c]{2cm}{\centering{\includegraphics[height=0.8cm]{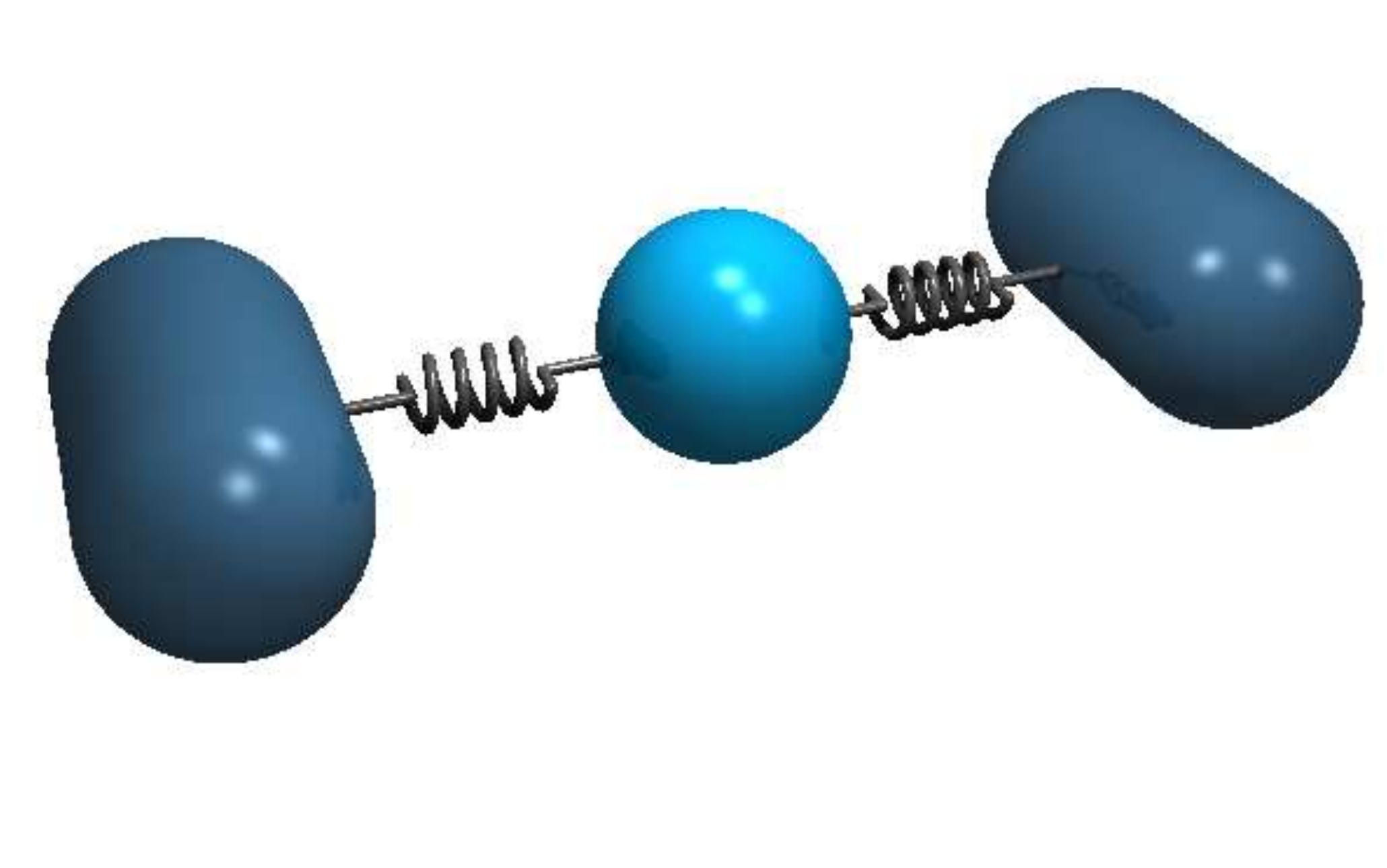}}} & 5.05 & 6.63 & $< 24 \%$ & 4.08\\
\hline
{(m)}&\parbox[2cm][1cm][c]{2cm}{\centering{\includegraphics[height=0.8cm]{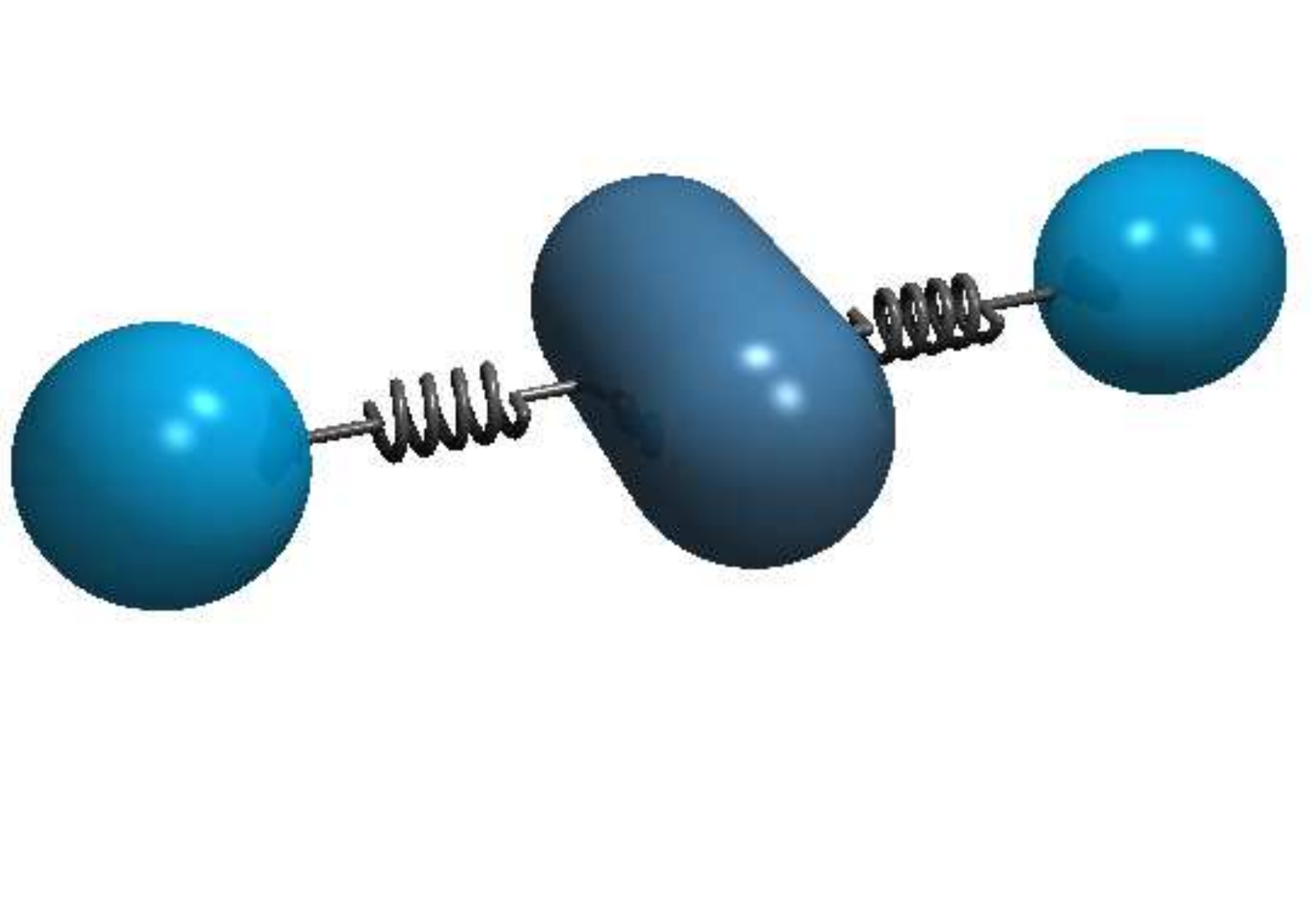}}} & 6.30 & 8.96 & $< 30 \%$ & 4.76\\
\hline
{(n)}&\parbox[2cm][1cm][c]{2cm}{\centering{\includegraphics[height=0.8cm]{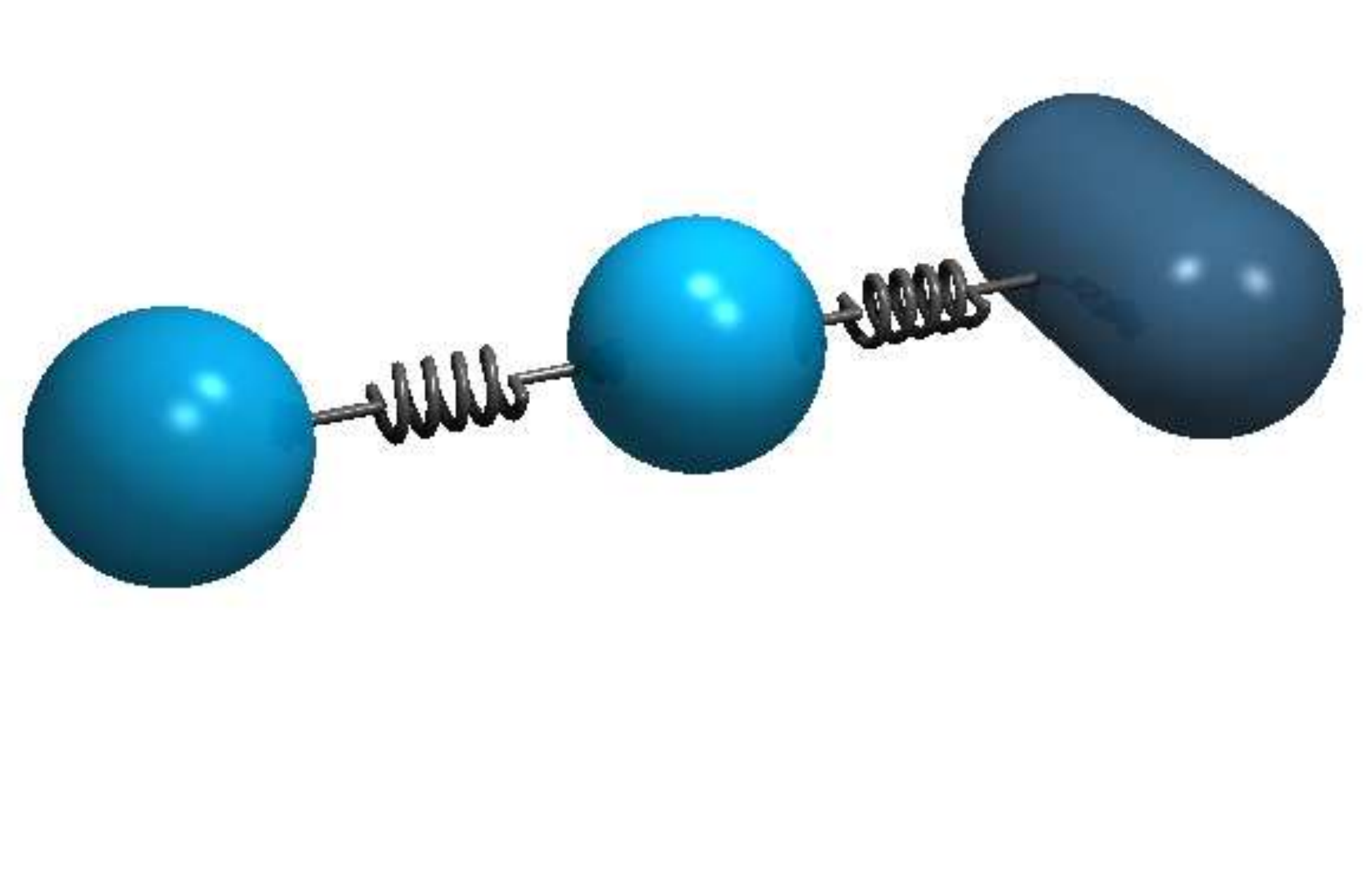}}} & 6.44 & 7.74 & $< 17 \%$ & 5.93\\
\hline
{(l)}&\parbox[2cm][1cm][c]{2cm}{\centering{\includegraphics[height=0.8cm]{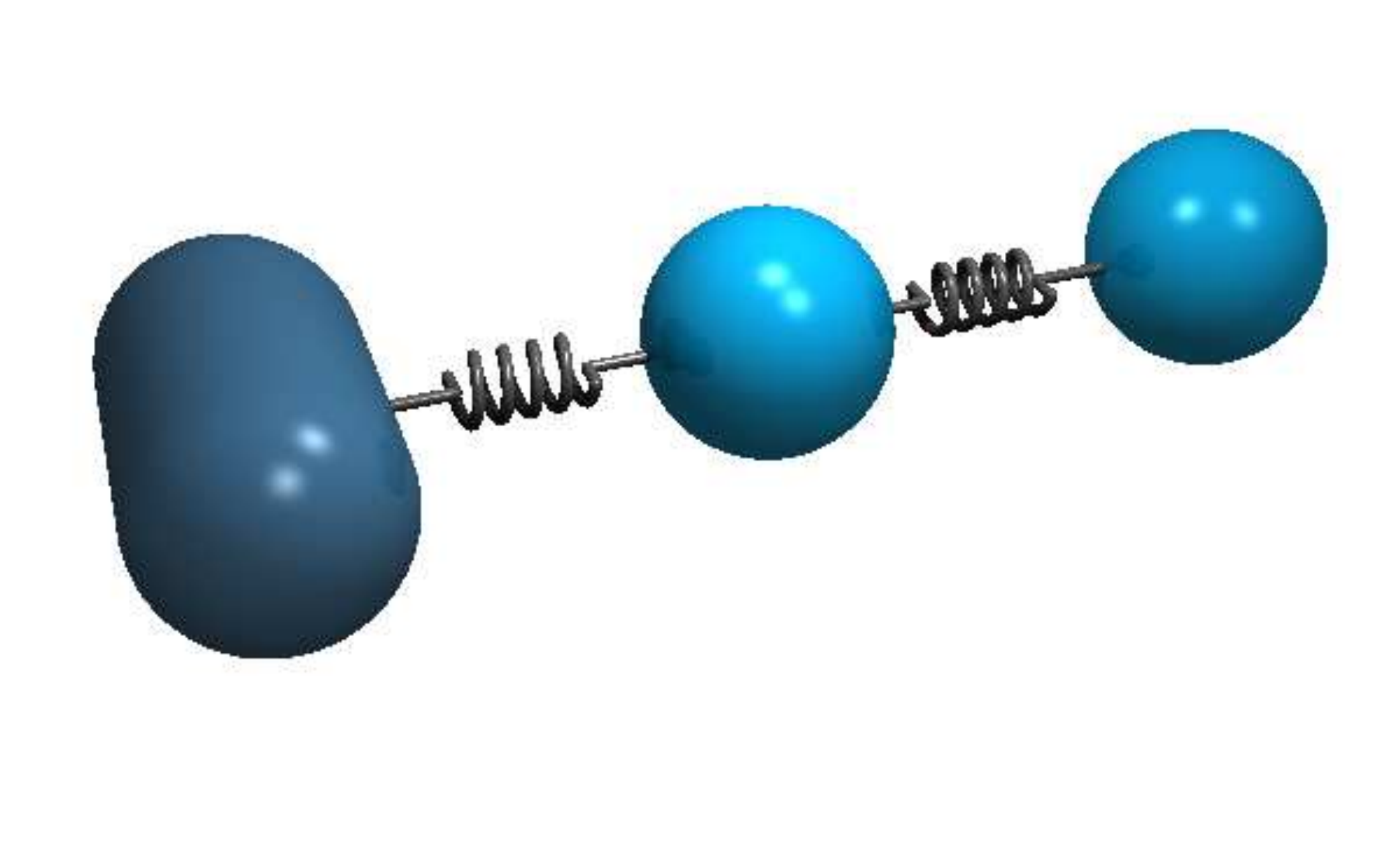}}} & 6.75 & 7.66 & $< 12 \%$ & 6.47\\
\hline
\end{tabular}
\caption{Comparison of swimming velocities from the simulations ($u_\text{swim}$) and from the prediction of Golestanian and Ajdari ($u_\text{GA}$), in the case of the third family of the swimmers with capsules. All the quantities are given in terms of their values on the lattice.}
\label{tab:caps_family3}
\end{table}

The simulation trajectories of all swimmers with capsules show that irrespective of the number of capsules in the swimmer, each body exhibits a smooth sinusoidal movement in the steady state, and one may attempt to use the formula of Golestanian and Ajdari (equation~(\ref{eq:V_GA})) for the analysis of the swimming motion. While this formula was developed assuming the bodies to be spherical, one may hypothesize that small deviations from the spherical shapes of the bodies will mostly affect the geometrical pre-factor $K$ in equation~(\ref{eq:V_GA}) and not the general expression. We test this hypothesis by extracting the amplitudes of the body oscillations and their phase shifts by fitting the trajectories with equation (\ref{eq:arms}). The geometric factor is obtained by approximating each capsule by a sphere of diameter 12, as that is the arithmetic mean of the dimensions of each capsule in the direction of movement and in a direction perpendicular to it. The mean armlengths are the same as in the original swimmers. For ease of reference, the values of all of these parameters are given in the appendix (tables~\ref{tab:caps_family1_params}, \ref{tab:caps_family2_params} and \ref{tab:caps_family3_params}). These tables show that as in the case of the three-sphere swimmers, the amplitude of oscillation of the leading arm ($d_2$ in the tables) is larger than that of the trailing arm ($d_1$ in the tables), for each swimmer.

The second and the third columns of the Tables~\ref{tab:caps_family1}, \ref{tab:caps_family2} and~\ref{tab:caps_family3} show the estimated $u_\text{GA}$ and the error in the measured $u_\text{swim}$ relative to $u_\text{GA}$ for each of the three families. It is interesting to note that the $u_\text{GA}$ approximation is still in good agreement with the results of the simulations. Especially in the first two families the error is about 10\% or less. For the third family, i.e.~for the swimmers with perpendicular capsules and shorter springs, the errors increase, to between 10\% and 30\%, probably because our approximation causes the spring lengths to become smaller than or equal to the sphere radii. Therefore the conditions necessary for the validity of Golestanian and Ajdari's formula are compromised. This is similar to the case of the three-sphere swimmer with large spheres, where the calculated and the observed velocities of the swimmer also disagreed due to the $r_i \ll \bar{l}_j$ condition not being satisfied well.

The tables also show the efficiencies, as defined in equation (\ref{eq:eff}), of the swimmers in each family. Here, the hydrodynamic radius $A$ has been calculated by assuming, as before, that each capsule is replaced by a sphere of diameter 12. The results show that in each family, the order of the efficiencies accords with the order of $u_\text{swim}$. This is as expected, since the driving forces for each swimmer in all our simulations are exactly the same, and so a more efficient swimmer would be expected to move faster than a less efficient one.

While this comparison has provided some useful insights, in order to fully understand our simulation results, the theory of Golestanian and Ajdari should be expanded to account for capsules, a task that we hope to undertake in the future.

%% file: conclusion.tex
\section{Conclusions and Future Work}
\label{sec:concAfw}

In this work we have demonstrated the successful integration of self-propelled micro-devices
into the coupled \pe{}-\Walberla{} framework.
Furthermore, we have shown the validity of this approach by comparing the results of simulations to analytical models.
We have taken advantage of the flexibility of our framework to simulate symmetric and asymmetric swimmers combining capsule and sphere geometries, which has not been possible so far.

The establishment of this framework is especially important for the study of systems that are inaccessible via experimental and analytical methods. In the future, we aim to harness its capabilities to address issues such as the problem of the three-body swimmer in regimes beyond those dictated by the approximations of Felderhof~\cite{Felderhof:2006:SwimAni} and Golestanian and Ajdari~\cite{Golestanian:2008:ARS}. Another interesting application would be to study the behavior of the swimmer in a narrow channel and explore the role of the boundary conditions at the wall.

\begin{figure}
\begin{center}
\subfigure[]{
  \includegraphics[width=13cm]{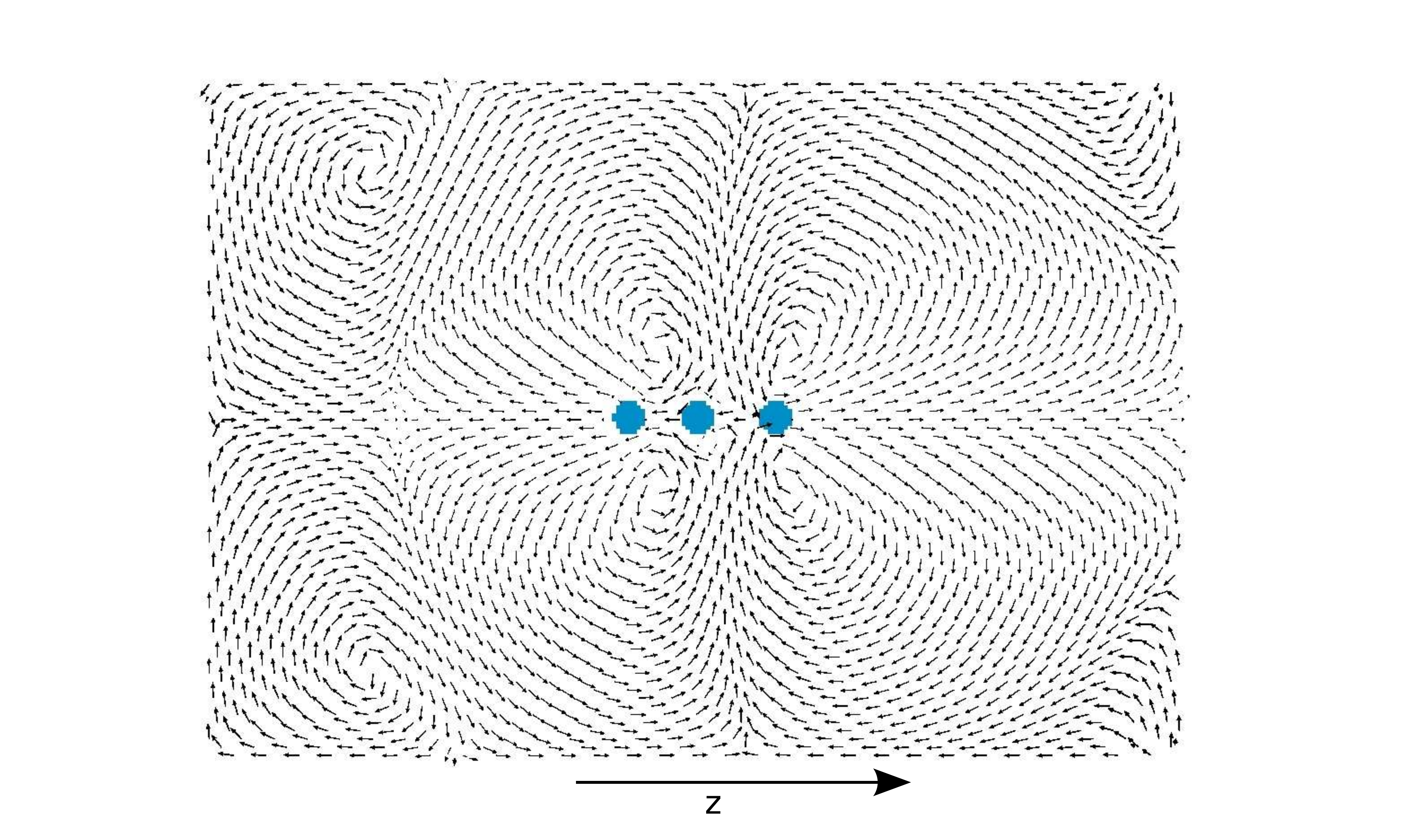}
  \label{fig:1swimmer}
}
\subfigure[]{
  \includegraphics[width=13cm]{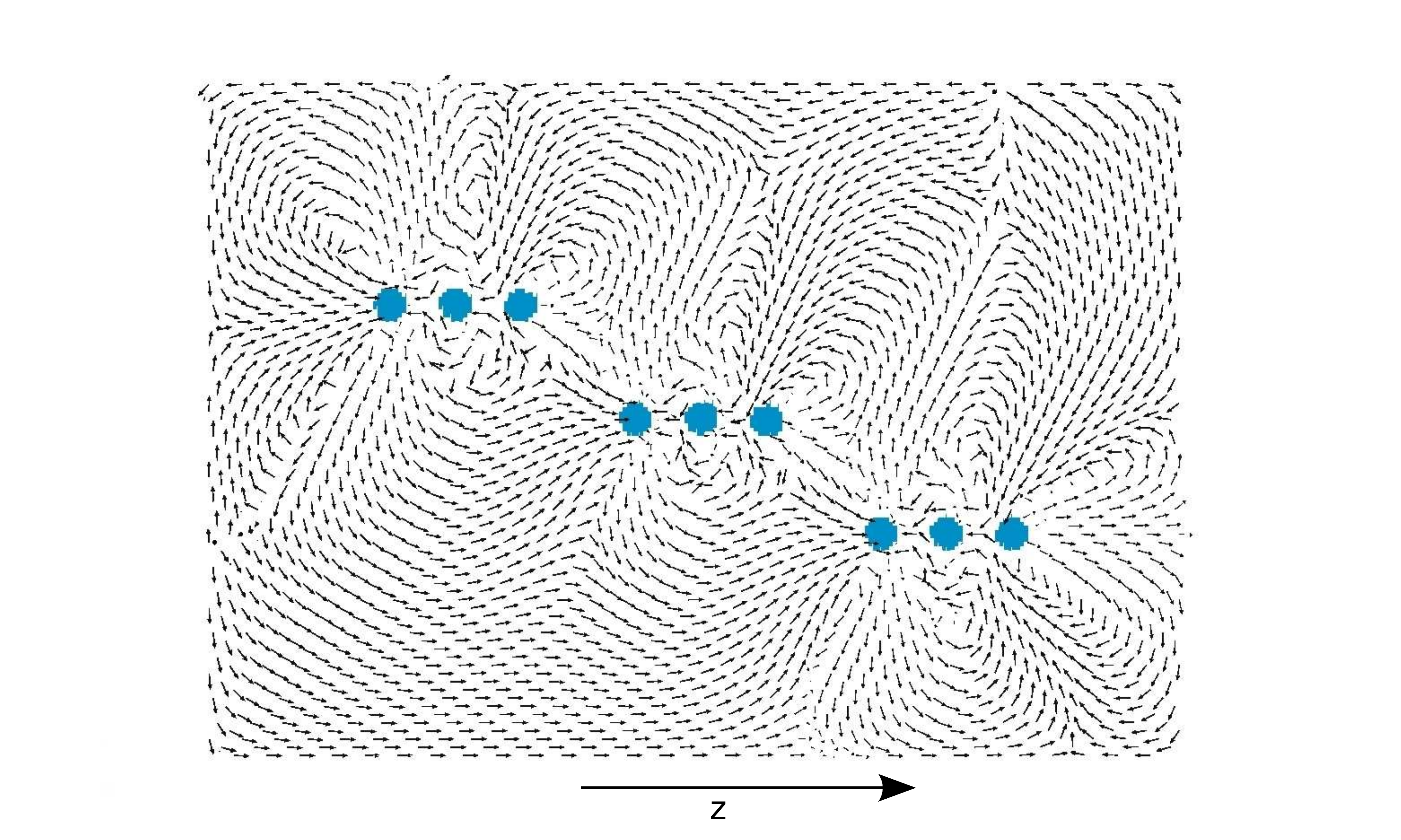}
  \label{fig:3swimmers}
}
\end{center}
\caption{\label{fig:flowfields13}Flow field of one and three three-sphere swimming devices, averaged over five swimming cycles. We used the same parameters for the simulation setup as in Section~\ref{subsec:res:design}, except for the channel, which now is $(x \times y \times z) = (164 \times 164 \times 240)$ lattice cells.}
\end{figure}

It would also be interesting to investigate the hydrodynamic interactions amongst many swimmers swimming together~\cite{Golestanian:2011:HSLRe, Kanevsky:2010:MSL}. In Figure~\ref{fig:flowfields13} we show the flow fields averaged over five cycles for the case of a single swimmer and also for three swimmers swimming together. The simulation of the three swimmers took 60.28 hours of total CPU-time on a single Intel Core i5-680 3.60GHz. To exclude the effects of the walls of the channel, one needs to set up even larger flow fields around the swimmers in this case than in the case of a single swimmer. Moreover, our work indicates that a steady state is reached after a significantly longer time in the case of many swimmers swimming together compared to the case of a single swimmer.

These factors suggest that in the current setting, the accurate simulation of many swimmers would take very long. To reduce this simulation time significantly, the large domains of simulation should be processed in a more efficient way by parallelizing the framework. Both \Walberla{} and \pe{} already feature massively parallel simulation algorithms for large scale particulate flow scenarios~\cite{Goetz:2010:SC10}, but the parallelization of the force generators (i.e.~springs) and velocity constraints is yet to be implemented. Once this is achieved, massive simulations with a large number of swimmers will be accessible, allowing us, finally, to address the problem of swarming.

%% file: appendix.tex
\label{sec:appendix}

\section{Values of different parameters of the swimmers with capsules}

Here we present the values of the different parameters of the swimmers with capsules, found from the simulation curves and used in equation (\ref{eq:V_GA}).
\begin{table}
\centering
\begin{tabular}{|cc|c|c|c|c|}
\hline
\multicolumn{2}{|c|}{Design} & $K$ & $d_1$ & $d_2$ & $\sin(\varphi_1-\varphi_2)$ \\  
\hline
{(f)}&\parbox[2cm][1cm][c]{2cm}{\centering{\includegraphics[height=0.8cm]{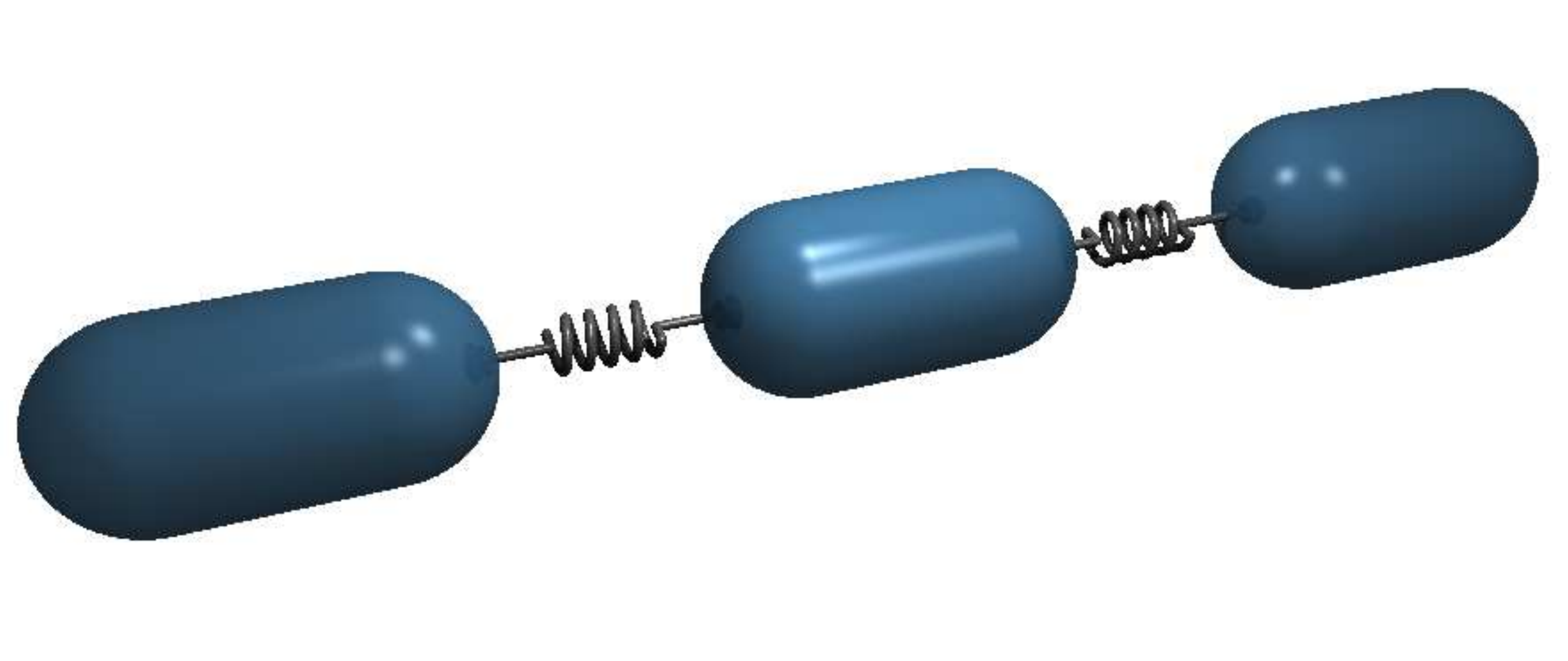}}} & 0.0030 & 2.24 & 3.39 & 0.89\\
\hline
{(e)}&\parbox[2cm][1cm][c]{2cm}{\centering{\includegraphics[height=0.8cm]{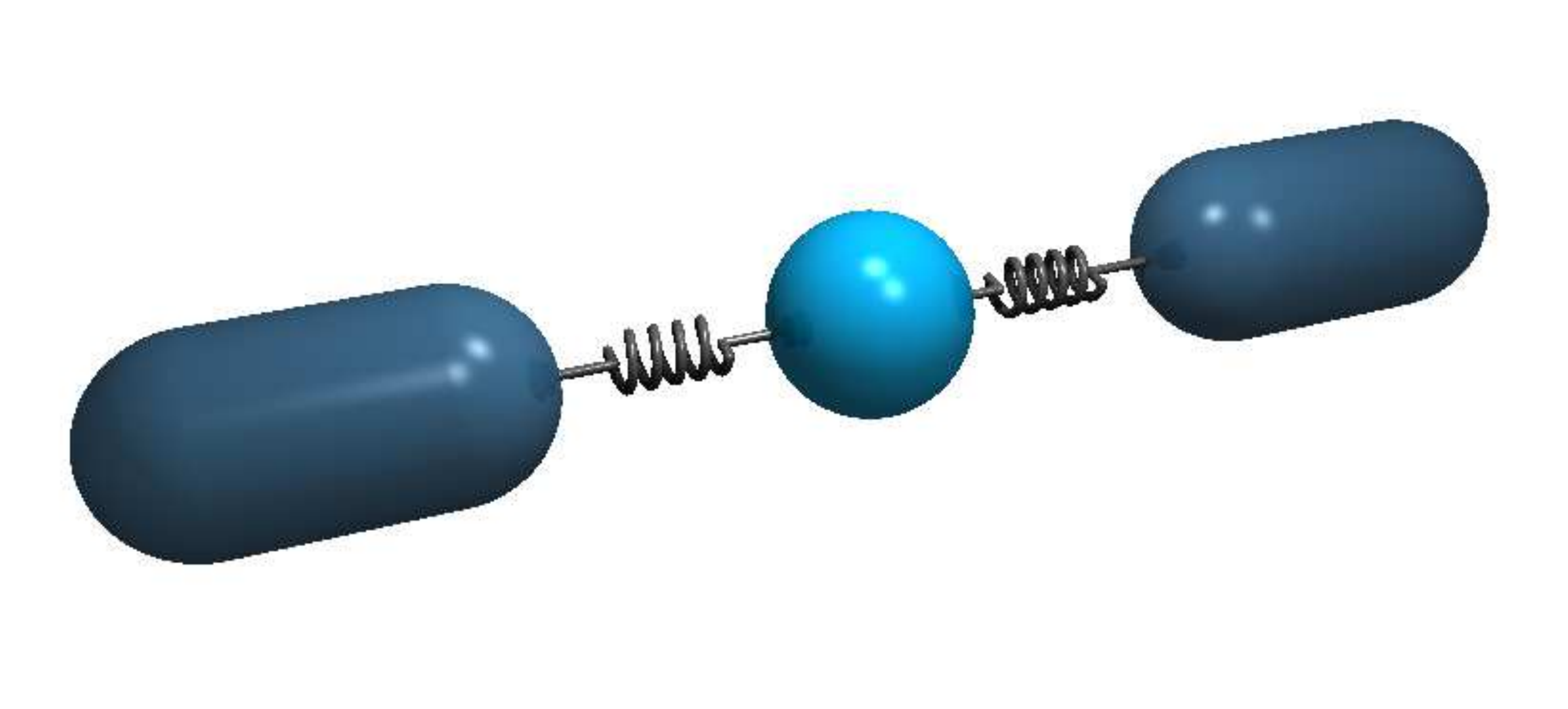}}} & 0.0037 & 2.21 & 3.63 & 0.84 \\
\hline
{(c)}&\parbox[2cm][1cm][c]{2cm}{\centering{\includegraphics[height=0.8cm]{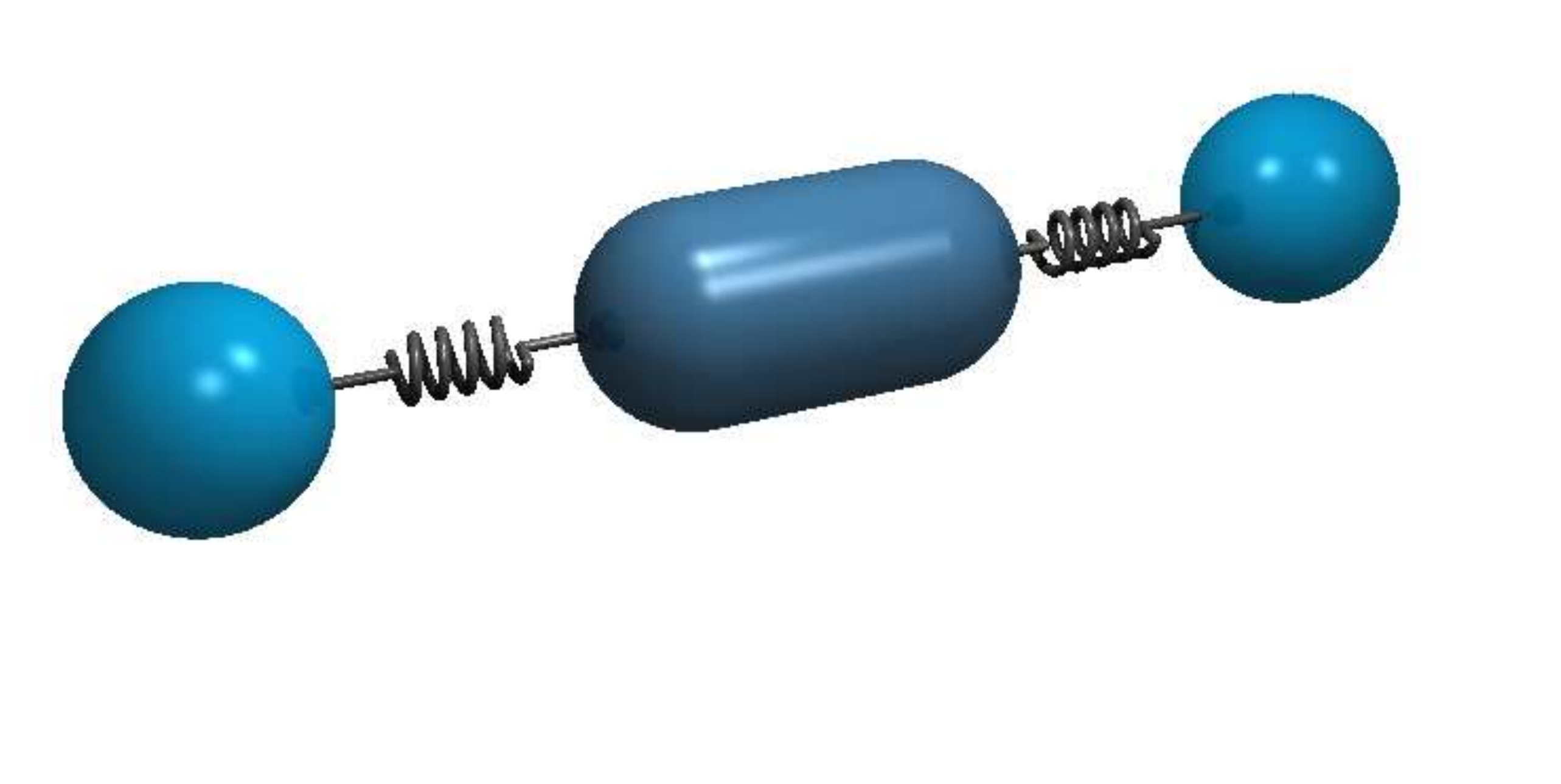}}} & 0.0032 & 2.59 & 3.59 & 0.96 \\
\hline
{(d)}&\parbox[2cm][1cm][c]{2cm}{\centering{\includegraphics[height=0.8cm]{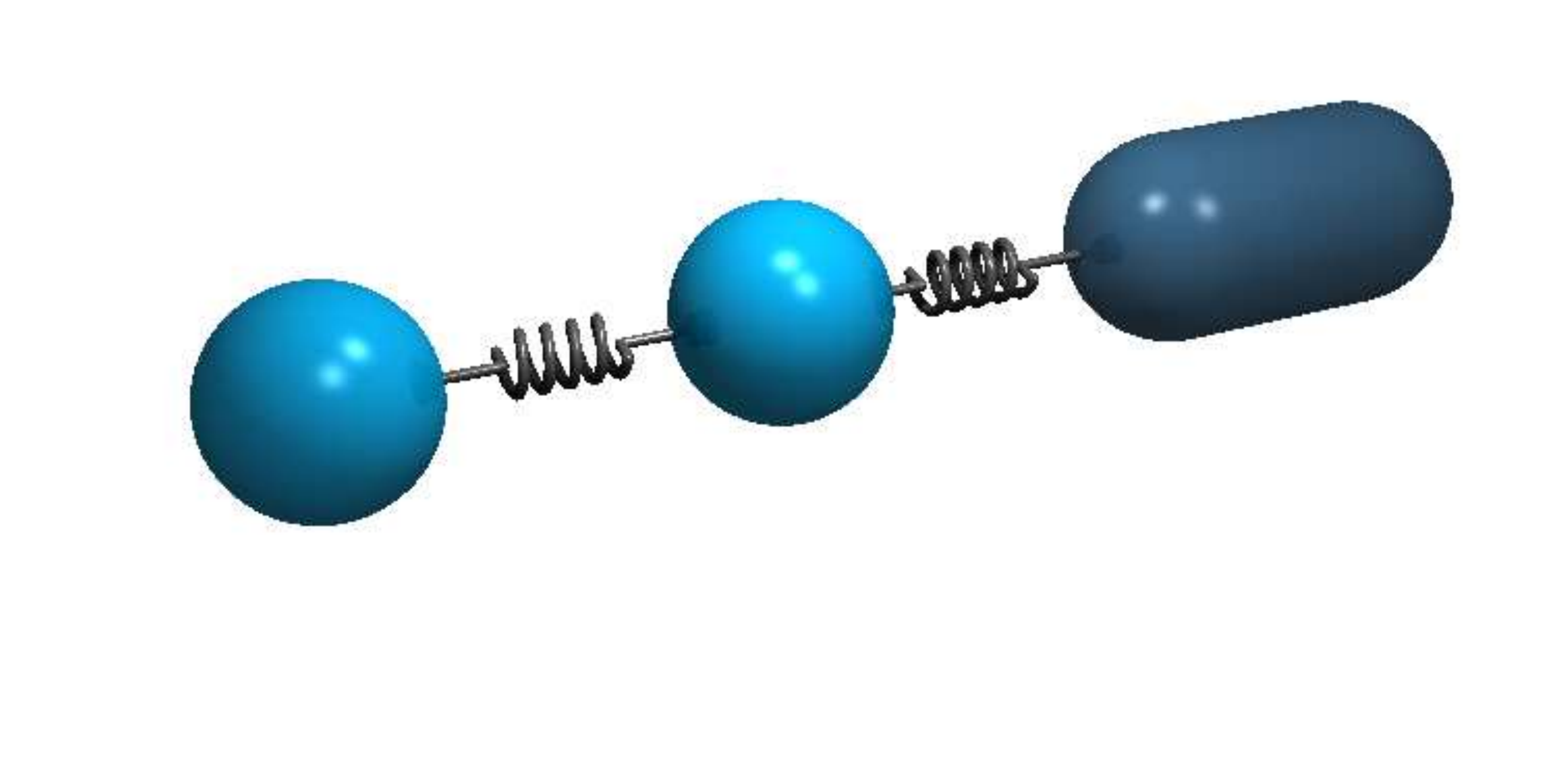}}} & 0.0041 & 2.47 & 3.65 & 0.87\\
\hline
{(b)}&\parbox[2cm][1cm][c]{2cm}{\centering{\includegraphics[height=0.8cm]{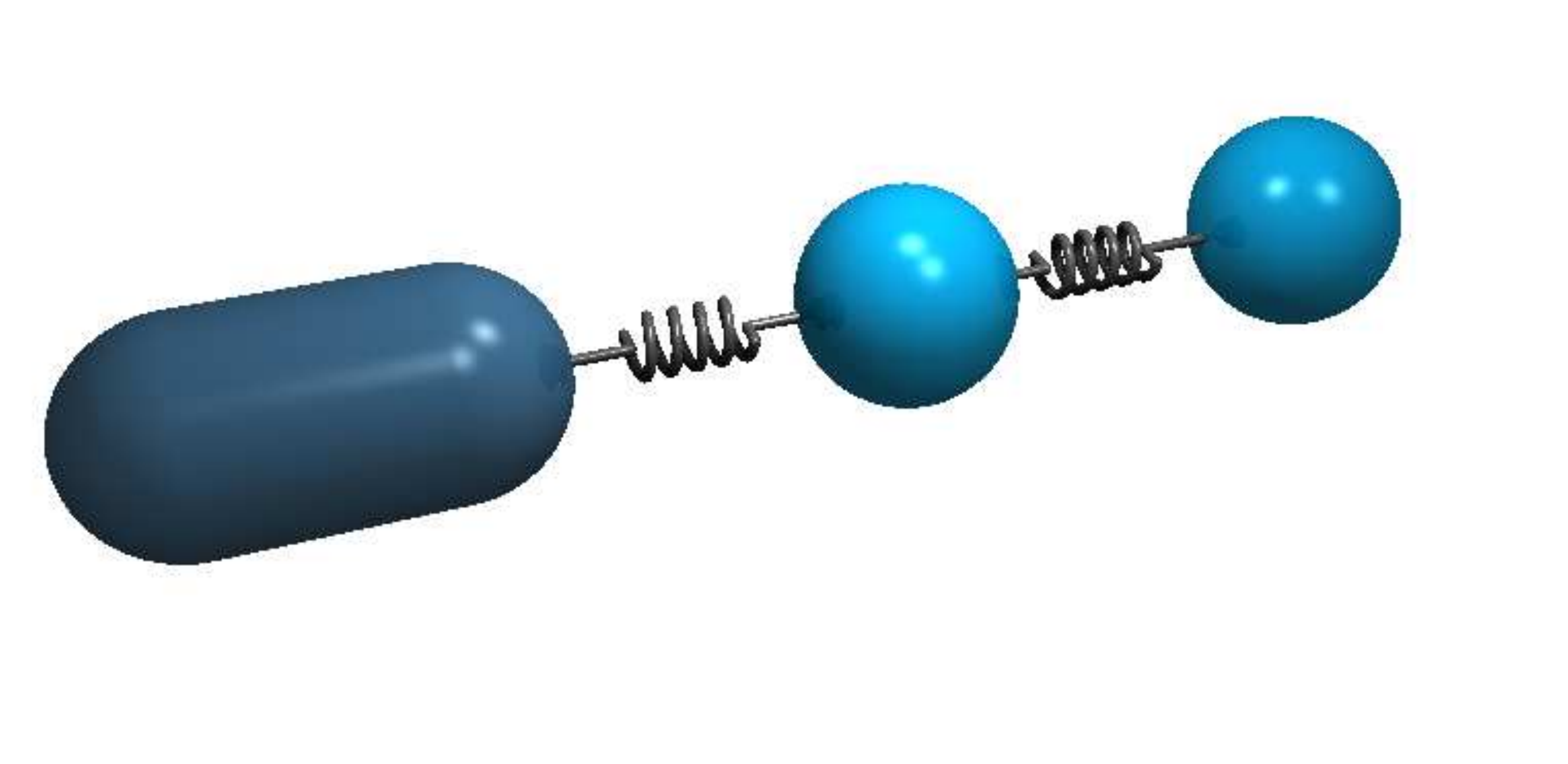}}} & 0.0041 & 2.26 & 3.79 & 0.92 \\
\hline
\end{tabular}
\caption{\footnotesize{The different parameters for the swimmers with parallel capsules. $\omega$ equals 0.00022. All the quantities are given in terms of their values on the lattice.}}
\label{tab:caps_family1_params}
\end{table}
\begin{table}
\centering
\begin{tabular}{|cc|c|c|c|c|}
\hline
\multicolumn{2}{|c|}{Design} & $K$ & $d_1$ & $d_2$ & $\sin(\varphi_1-\varphi_2)$ \\  
\hline
{(k)}&\parbox[2cm][1cm][c]{2cm}{\centering{\includegraphics[height=0.8cm]{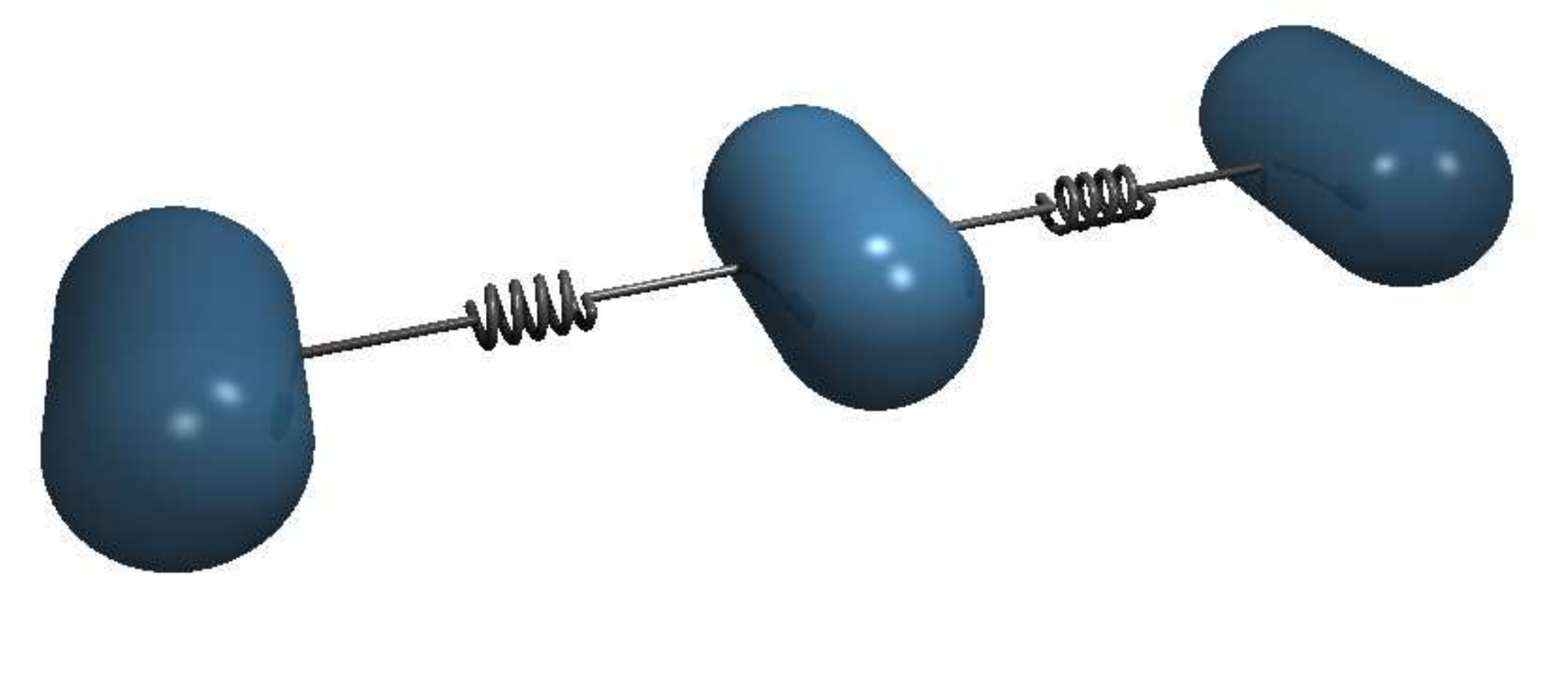}}} & 0.0030 & 2.14 & 3.26  & 0.85\\
\hline
{(j)}&\parbox[2cm][1cm][c]{2cm}{\centering{\includegraphics[height=0.8cm]{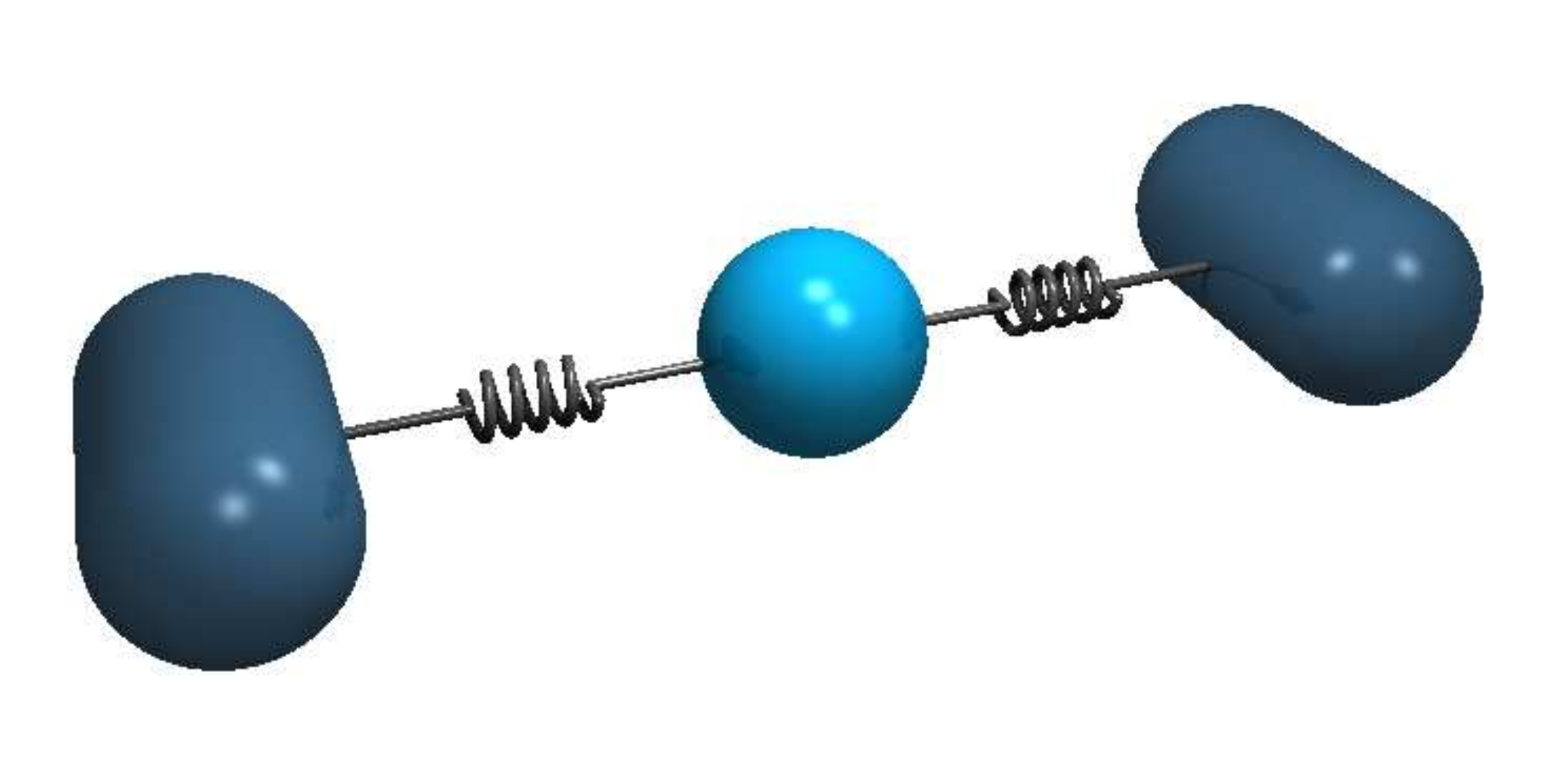}}} & 0.0037  & 2.11 & 3.62 & 0.75 \\
\hline
{(h)}&\parbox[2cm][1cm][c]{2cm}{\centering{\includegraphics[height=0.8cm]{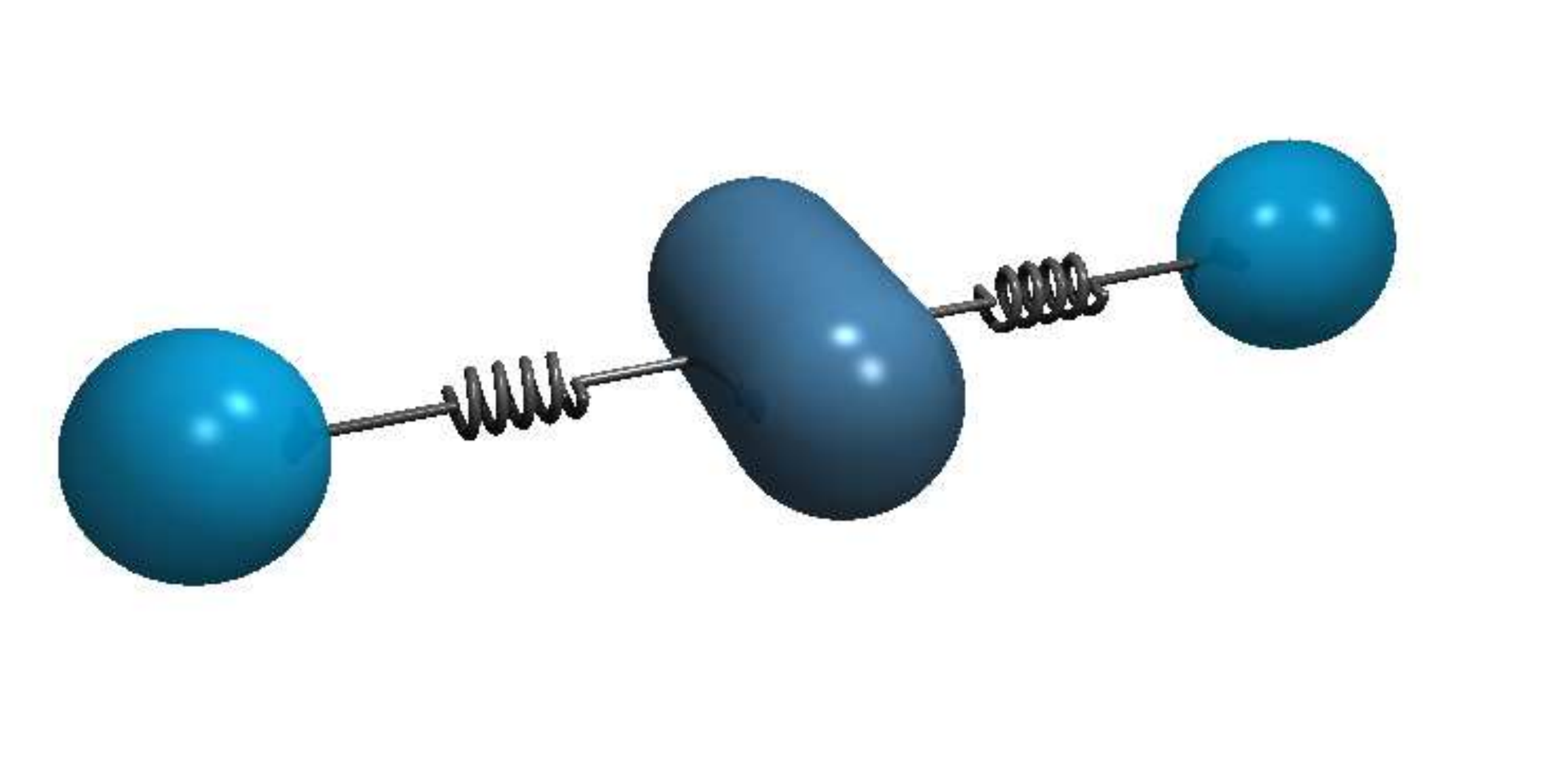}}}  & 0.0032 & 2.66 & 3.57 & 0.97 \\
\hline
{(i)}&\parbox[2cm][1cm][c]{2cm}{\centering{\includegraphics[height=0.8cm]{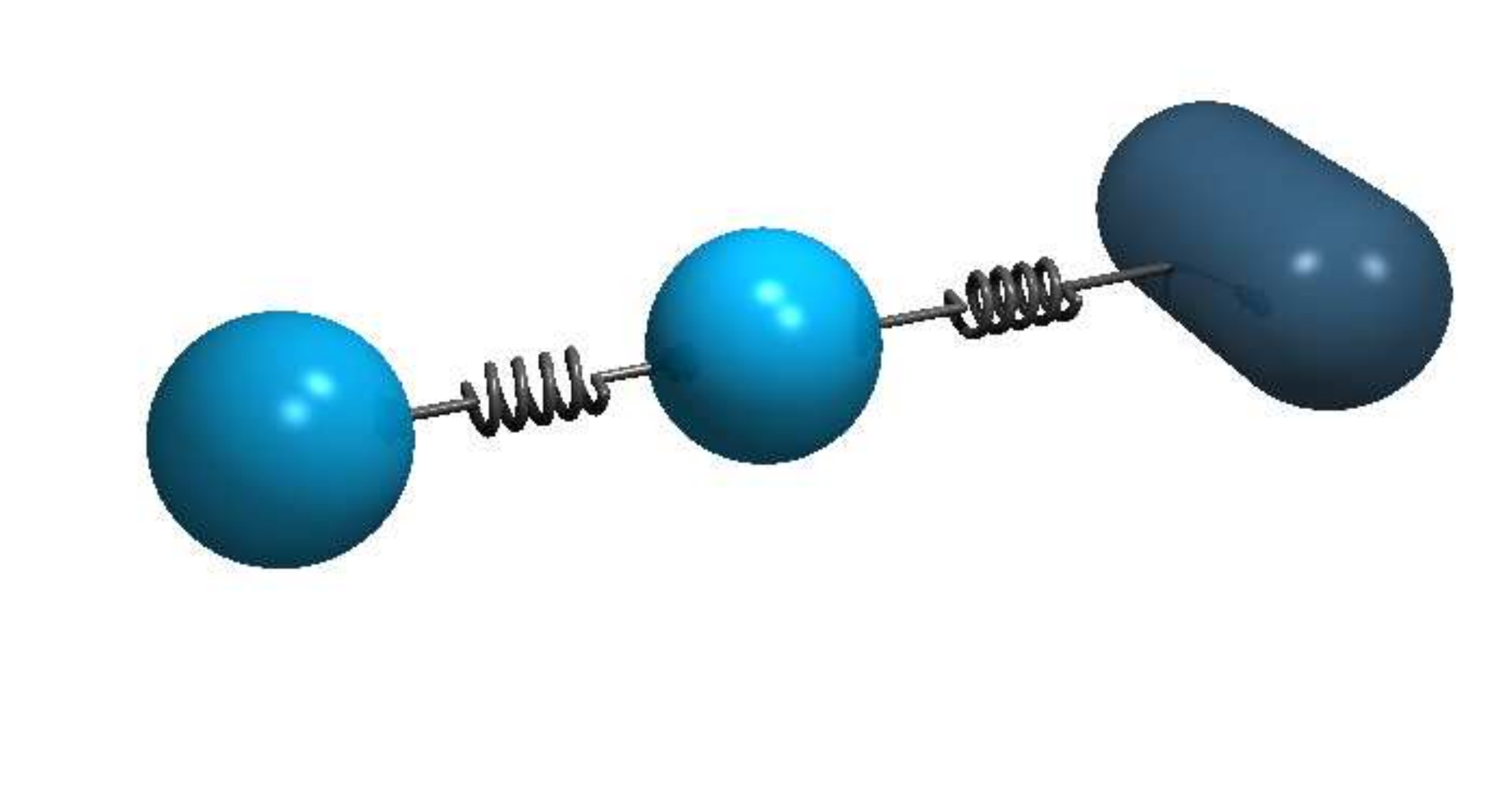}}} & 0.0041 & 2.43 & 3.65 & 0.84\\
\hline
{(g)}&\parbox[2cm][1cm][c]{2cm}{\centering{\includegraphics[height=0.8cm]{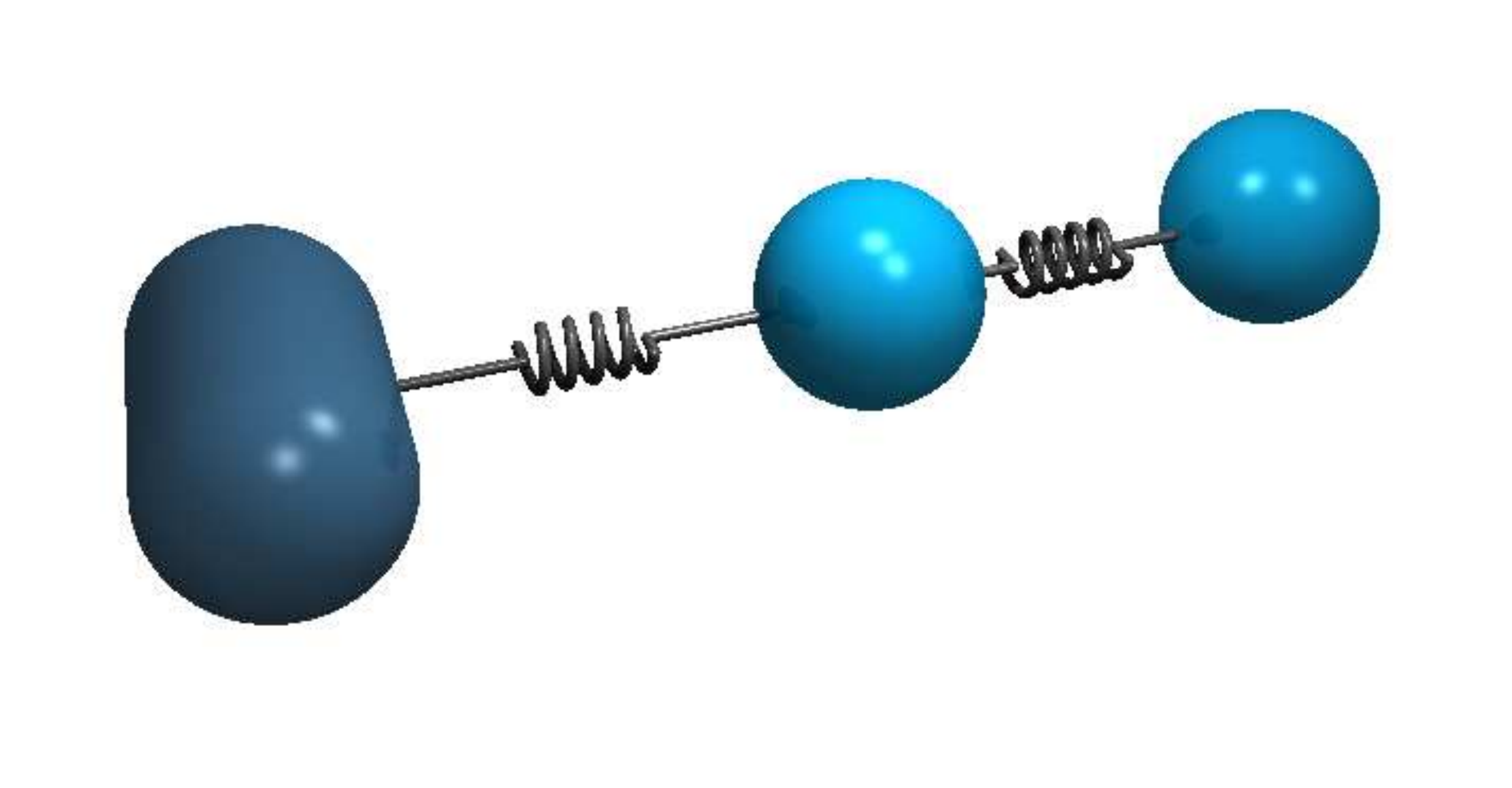}}} & 0.0041 & 2.17 & 3.80 & 0.91 \\
\hline
\end{tabular}
\caption{\footnotesize{The different parameters for the swimmers with perpendicular capsules and long springs. $\omega$ equals 0.00022. All the quantities are given in terms of their values on the lattice.}}
\label{tab:caps_family2_params}
\end{table}

\begin{table}
\centering
\begin{tabular}{|cc|c|c|c|c|}
\hline
\multicolumn{2}{|c|}{Design} & $K$ & $d_1$ & $d_2$ & $\sin(\varphi_1-\varphi_2)$ \\  
\hline
{(p)}&\parbox[2cm][1cm][c]{2cm}{\centering{\includegraphics[height=0.8cm]{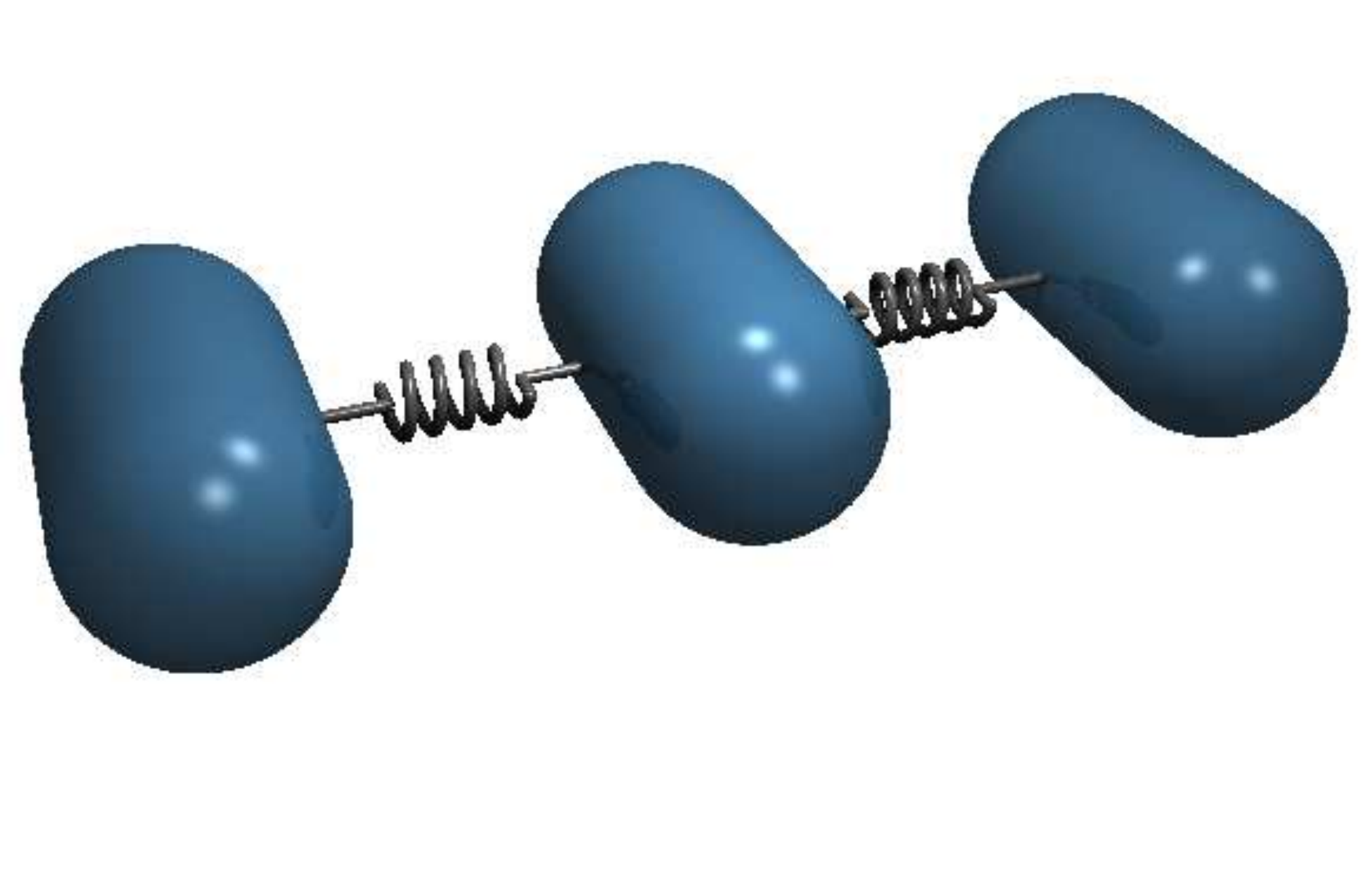}}}  & 0.0068 & 1.92 & 2.59  & 0.90\\
\hline
{(o)}&\parbox[2cm][1cm][c]{2cm}{\centering{\includegraphics[height=0.8cm]{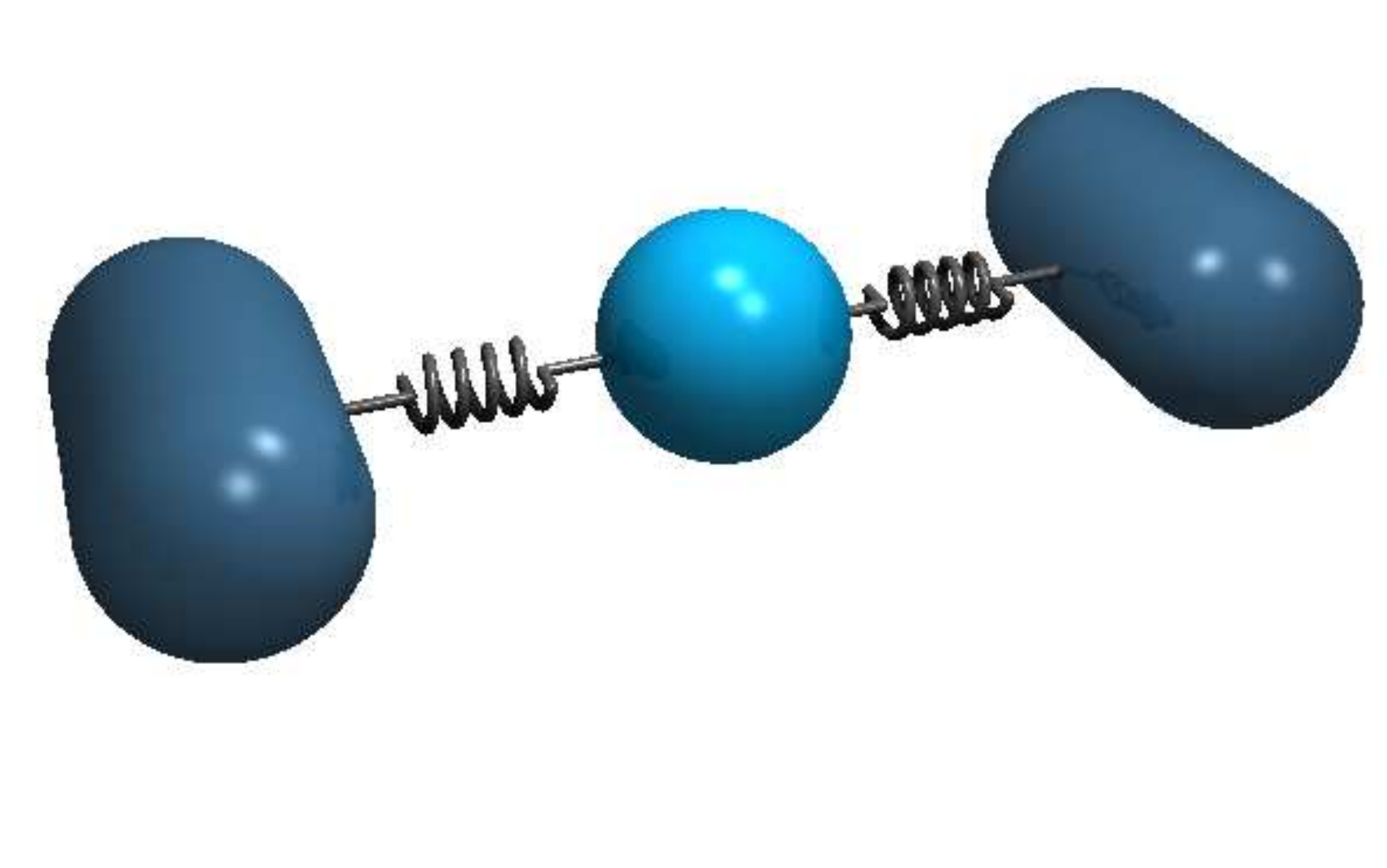}}} & 0.0058  & 2.07 & 3.33 & 0.76 \\
\hline
{(m)}&\parbox[2cm][1cm][c]{2cm}{\centering{\includegraphics[height=0.8cm]{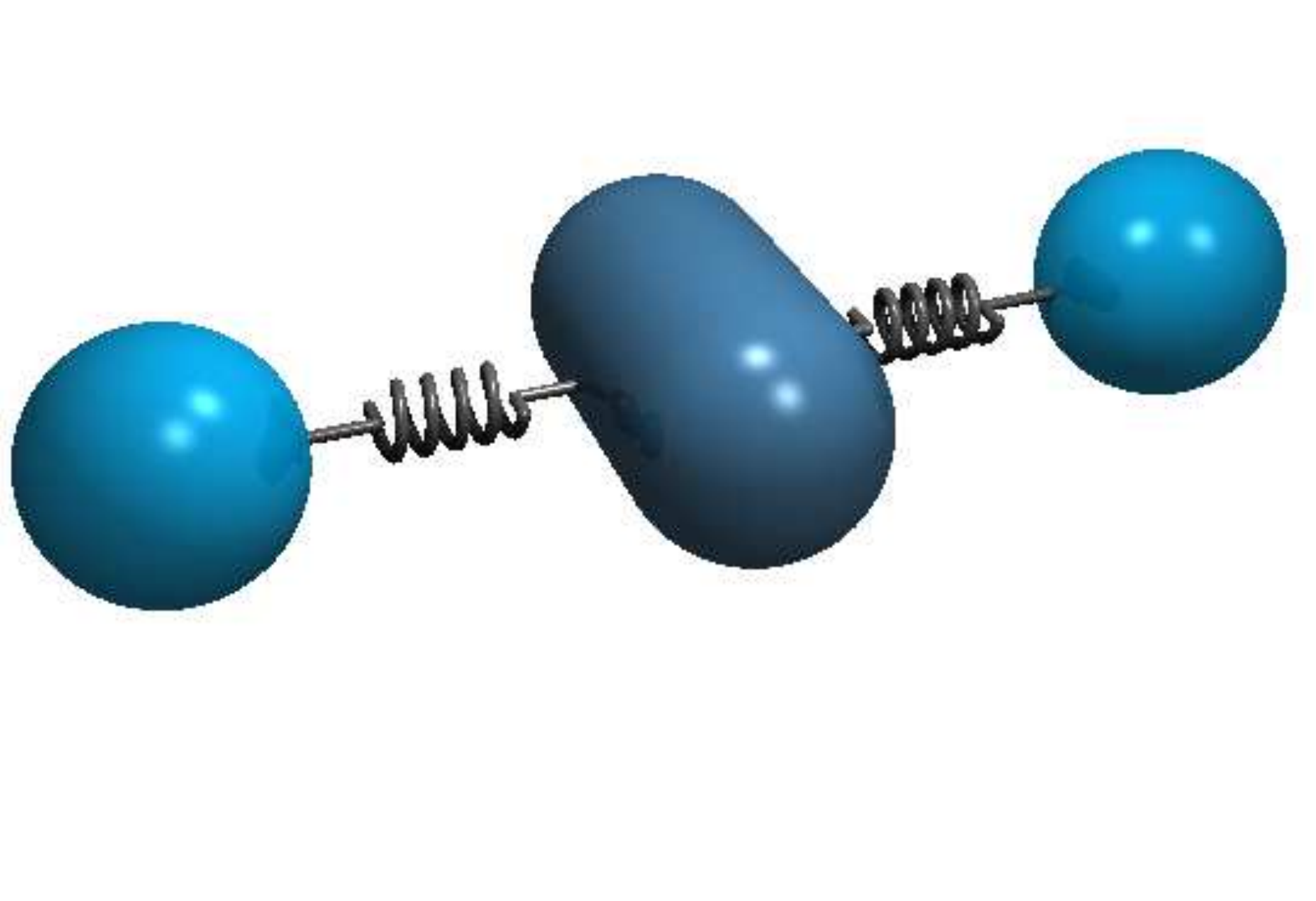}}}  & 0.0050 & 2.56 & 3.24 & 0.98 \\
\hline
{(n)}&\parbox[2cm][1cm][c]{2cm}{\centering{\includegraphics[height=0.8cm]{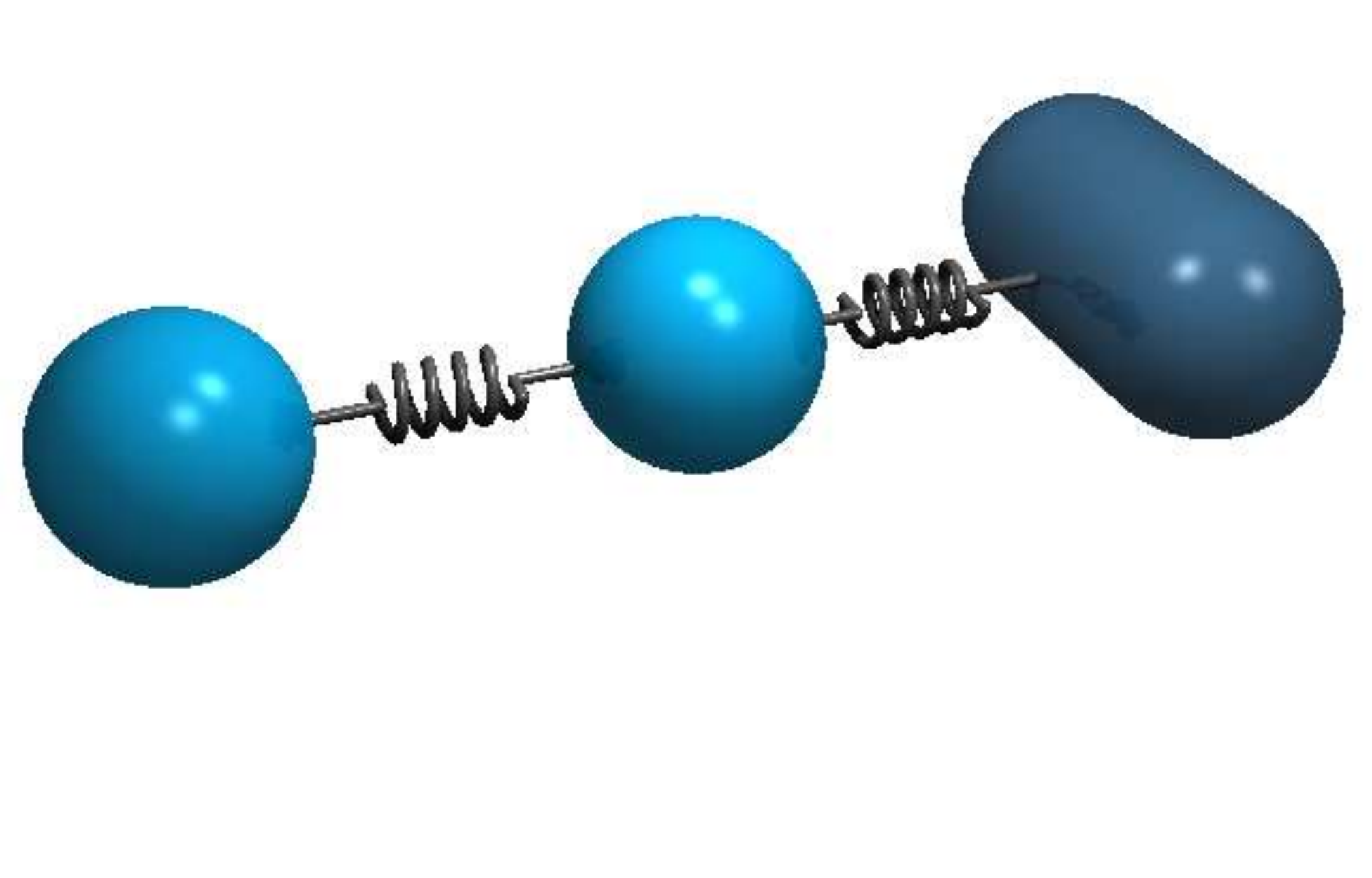}}} & 0.0050 & 2.54 & 3.39 & 0.81\\
\hline
{(l)}&\parbox[2cm][1cm][c]{2cm}{\centering{\includegraphics[height=0.8cm]{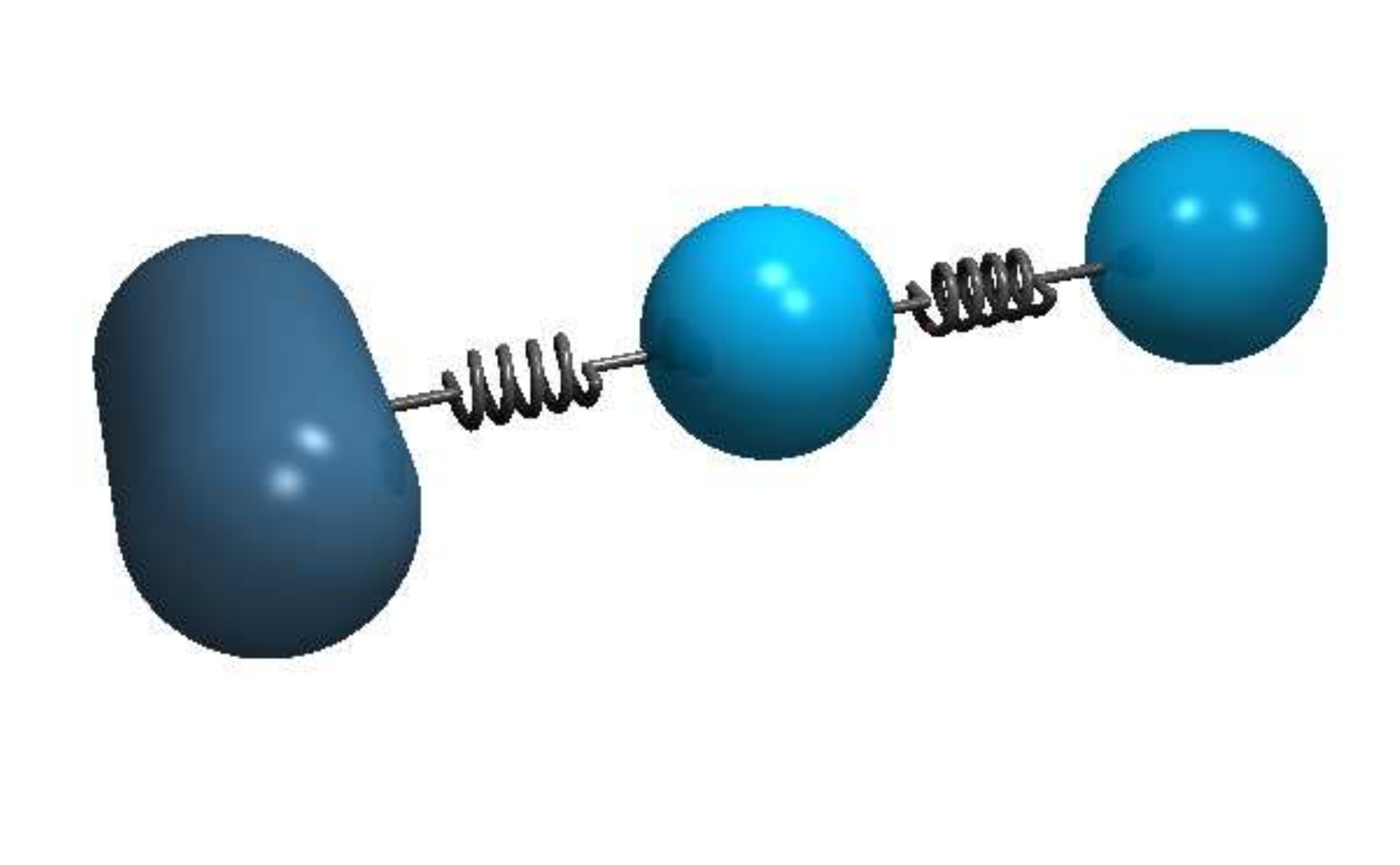}}} & 0.0050 & 2.00 & 3.75 & 0.92 \\
\hline
\end{tabular}
\caption{\footnotesize{The different parameters for the swimmers with perpendicular capsules and short springs. $\omega$ equals 0.00022. All the quantities are given in terms of their values on the lattice.}}
\label{tab:caps_family3_params}
\end{table}